\documentclass[pdftex,letterpaper,11pt]{article}
\pdfoutput=1
\usepackage[dvips]{graphicx,color}
\usepackage{wrapfig}
\usepackage[labelfont=bf,skip=1pt]{caption}
\usepackage{rotating}
\usepackage{subcaption}
\usepackage{array}
\usepackage{tabularx}
\usepackage{epstopdf}
\usepackage{enumerate}
\usepackage[para]{threeparttable}
\usepackage{multirow}
\usepackage{amsmath}
\usepackage{latexsym}
\usepackage{amssymb}
\usepackage[square,authoryear]{natbib}
\usepackage{color}
\usepackage[rgb,dvipsnames]{xcolor}
\definecolor{grey}{gray}{0.7}
\usepackage{hyperref}
\usepackage{fancyhdr}
\hypersetup{colorlinks=true, urlcolor=blue, citecolor=black, linkcolor=black}
\usepackage{extramarks}
\usepackage{tocloft}      %% for formatting Table of Contents
\usepackage{titlesec}
\usepackage[top=1.0in, bottom=1.0in, left=0.875in, right=0.875in]{geometry}
\usepackage{setspace}
\usepackage{indentfirst}
\newcommand{\shorttitle}{Quantified Energy Dissipation}
\fancypagestyle{plain}{
  \fancyhf{}
  \fancyhead[EOL]{\large \shorttitle}
  \fancyhead[EOR]{\large \leftmark}
  \fancyfoot[EOL]{\textit{L.B. Wilson III et al.}}
  \fancyfoot[EOC]{}
  \fancyfoot[EOR]{\thepage}
}
\newcommand{\footnoteremember}[2]{\footnote{#2}\newcounter{#1}\setcounter{#1}{\value{footnote}}}
\newcommand{\footnoterecall}[1]{\footnotemark[\value{#1}]}
\begin{document}

%%  Vertical spacing before/after each
%%
%%  -> Define spacing for floats
\titlespacing{\enumerate}{0pt}{*0.001}{*0.001}
\titlespacing{\figure}{0pt}{*0.01}{*0.01}
\titlespacing{\table}{0pt}{*0.005}{*0.005}
\titlespacing{\threeparttable}{0pt}{*0.005}{*0.005}
\titlespacing{\tablenotes}{0pt}{*0.01}{*0.01}
\titlespacing{\wrapfigure}{0pt}{*0.01}{*0.01}
%%  -> Define spacing for sections
\titlespacing{\section}{0pt}{*0.100}{*0.100}
\titlespacing{\subsection}{0pt}{*0.050}{*0.050}
\titlespacing{\subsubsection}{0pt}{*0.075}{*0.075}
\titlespacing{\paragraph}{0pt}{*0.075}{*0.075}
%%
%%  Format captions
%%
\captionsetup[table]{font=large,position=above,justification=centering}
\captionsetup[figure]{font=footnotesize,position=below,justification=RaggedRight}
%%
%%  Table of Contents:  Define # of levels labeled in TOCs
%%
\setcounter{tocdepth}{3}
%%  Page Format and Numbering
\pagestyle{plain}
\pagenumbering{arabic}
%%
%%  Title Page
%%
\title{\bf Quantified Energy Dissipation Rates:  Electromagnetic Wave Observations in the Terrestrial Bow Shock}
\author{L.B. Wilson III\footnoteremember{1}{NASA Goddard Space Flight Center, Greenbelt, Maryland, USA.}, D.G. Sibeck\footnoterecall{1}, \\ A.W. Breneman\footnoteremember{2}{School of Physics and Astronomy, University of Minnesota, Minneapolis, Minnesota, USA.}, \\ O. Le Contel\footnoteremember{3}{Laboratoire de Physique des Plasmas, Ecole Polytechnique, Palaiseau, France.}, \\ C. Cully\footnoteremember{4}{Department of Physics and Astronomy, University of Calgary, Calgary, Alberta, Canada.}, \\ D.L. Turner\footnoteremember{5}{Institute of Geophysics and Planetary Physics/Earth and Space Sciences, UCLA, Los Angeles, California, USA.}, and V. Angelopoulos\footnoterecall{5}}
\date{\today}
\maketitle
%%  Shut off page numbering here
\thispagestyle{empty}

%%  Abstract
\begin{abstract}
  We present the first quantified measure of the rate of energy dissipated per unit volume by high frequency electromagnetic waves in the transition region of the Earth's collisionless bow shock using data from the THEMIS spacecraft.  Every THEMIS shock crossing examined with available wave burst data showed both low frequency ($<$10 Hz) magnetosonic-whistler waves and high frequency ($\gtrsim$10 Hz) electromagnetic and electrostatic waves throughout the entire transition region and into the magnetosheath.  The waves in both frequency ranges had large amplitudes, but the higher frequency waves, which are the focus of this study, showed larger contributions to both the Poynting flux and the energy dissipation rates.  The higher frequency waves were identified as combinations of ion-acoustic waves, electron cyclotron drift instability driven waves, electrostatic solitary waves, and whistler mode waves.  These waves were found to have:  (1) amplitudes capable of exceeding $\delta$B $\sim$ 10 nT and $\delta$E $\sim$ 300 mV/m, though more typical values were $\delta$B $\sim$ 0.1-1.0 nT and $\delta$E $\sim$ 10-50 mV/m; (2) energy fluxes in excess of 2000 $\mu$W m$^{-2}$; (3) resistivities $>$ 9000 $\Omega$ m; and (4) energy dissipation rates $>$ 3 $\mu$W m$^{-3}$.  The dissipation rates were found to be in excess of four orders of magnitude greater than was necessary to explain the increase in entropy across the shocks.  Thus, the waves need only be, at times, $<$ 0.01\% efficient to balance the nonlinear wave steepening that produces the shocks.  Therefore, these results show for the first time that high frequency electromagnetic and electrostatic waves have the capacity to regulate the global structure of collisionless shocks.
\end{abstract}

%%  Comments: 41 pages, 14 PDF figures, submitting to Nature \\
%%  Subj-class: physics.space-ph \\

\clearpage
%%  Table of Contents
\tableofcontents
%%  Shut off page numbering here
\thispagestyle{empty}
%%  This must come AFTER the \maketitle command, which sets the page style to PLAIN
\clearpage

\thispagestyle{fancy}   %%  use fancy headers&footers
\pagenumbering{arabic}  %%  start page numbering here
\setcounter{page}{1}

\titlespacing{\title}{0pt}{*0.01}{*0.01}
\titlespacing{\author}{0pt}{*0.01}{*0.01}
\titlespacing{\section}{0pt}{*0.1}{*0.1}
\titlespacing{\subsection}{0pt}{*0.05}{*0.05}
\titlespacing{\subsubsection}{0pt}{*0.05}{*0.05}
\titlespacing{\enumerate}{0pt}{*0.01}{*0.01}
\titlespacing{\figure}{0pt}{*0.01}{*0.01}
\titlespacing{\wrapfigure}{0pt}{*0.01}{*0.01}

%% => Change footnote marks
\renewcommand{\thefootnote}{\fnsymbol{footnote}}
%%----------------------------------------------------------------------------------------
%%  Section:  Introduction
%%----------------------------------------------------------------------------------------
\numberwithin{equation}{section}
\section{Introduction}  \label{sec:introduction}
%%  Intro paragraph
\indent  Shock waves, in their simplest form, are discontinuities that result from the balance between nonlinear wave steepening and energy loss.  The energy loss must transform the incident bulk flow kinetic energy to some other form like random kinetic energy (i.e., heat).  If this energy transformation is irreversible, then the process is said to have dissipated energy \citep{petschek58, fishman60a}.  The distinction between reversible and irreversible is important because the initiation of a shock discontinuity from a nonlinearly steepened wave requires that the transformation be irreversible \citep{shu92a}.  In a collision dominated media like the Earth's atmosphere, energy dissipation is accomplished through binary particle collisions.  In a collisionless media like the solar wind, particle collisions occur too infrequently to significantly alter the incident bulk flow kinetic energy in the shock ramp.  Yet, collisionless shock waves are known to be efficient mechanisms by which charged particles can be heated and/or accelerated.  In addition, the basic theories for possible energy dissipation mechanisms in these shocks have remained relatively unchanged for over 40 years \citep[e.g.,][]{sagdeev66, coroniti70b, tidman71a, wu84a, treumann09a}.  However, the relative importance of each possible dissipation mechanism has not been quantified for low Mach number shocks.

%%  Introduce energy dissipation mechanisms
\indent  The four possible mechanisms by which a collisionless shock could transfer energy are:  wave dispersion \citep[e.g.,][]{mellott84a, krasnoselskikh02a, sundkvist12a}, wave-particle interactions \citep[e.g.,][]{sagdeev66, gary81c}, particle reflection \citep[e.g.,][]{edmiston84, kennel87a, bale05a}, and macroscopic quasi-static field effects \citep[e.g.,][]{scudder86a, scudder86b, scudder86c}.  Theory predicts that the dominant type of energy dissipation depends strongly upon fast mode Mach number, M${\scriptstyle_{f}}$, shock normal angle, $\theta{\scriptstyle_{Bn}}$, and the ratio of particle to magnetic pressures called the plasma beta, $\beta$ \citep[e.g.,][]{kennel85a}.  At low M${\scriptstyle_{f}}$, theory suggests that the dominant mechanisms are wave dispersion and/or wave-particle interactions \citep{kennel85a}.  Above some critical Mach number, M${\scriptstyle_{cr}}$, the shock requires additional energy dissipation in the form of particle reflection to limit wave steepening \citep{edmiston84}.  Numerous studies have argued that macroscopic quasi-static fields (the fourth mechanism) govern the dynamics of charged particles in low Mach number collisionless shock waves \citep[e.g.,][]{scudder86c, hull00a, hull01a}.  Theory and observation support the first three mechanisms, but we believe that the fourth mechanism has been given too much importance due to low resolution observations.

%%  Cross-Shock Potential
\indent  Recent studies have argued that quasi-static fields dominate over higher frequency waves \citep[e.g.,][]{dimmock12a, mitchell12a}.  However, previous observations of quasi-static electric fields were often under-sampled and/or the instrument saturated (e.g., Cluster saturates near $\sim$40 mV/m).  The under-sampled electric fields often show a ``spiky'' signature that is assumed to be the quasi-static electric field called the cross-shock potential \citep[e.g.,][]{mozer13a, sundkvist13a}.  These low frequency electric field observations require a reference frame transformation to remove convective field effects.  However, the observation of nonlinear ($\delta$B/B $>$ 1) electromagnetic fluctuations near the ramps of collisionless shocks \citep[e.g.,][]{wilsoniii12c, sundkvist13a} and evidence of shock reformation/non-stationarity \citep[e.g.,][]{lobzin07a, mazelle10a} raise doubts about the ability to remove convective field effects.  Moreover, observations \citep[e.g.,][]{wilsoniii07a, wilsoniii10a} show that higher frequency wave fields can be much larger ($\gtrsim$200 mV/m) than the quasi-static electric fields typically observed ($\sim$few 10's of mV/m) \citep[e.g.,][]{dimmock12a, mozer13a}.

%%  Need for irreversibility
\indent  If the quasi-static electric fields govern the dynamics of charged particles across the shock, then, ignoring other mechanisms, the transition would be thermodynamically reversible.  Of the four possible energy dissipation mechanisms in collisionless shocks, only wave-particle interactions can directly produce an irreversible change in energy.  Recently, \citet[][]{parks12a} observed that the entropy flux density increases across the bow shock, which argues against a purely reversible transition.  Previous studies that examined particle distributions in detail showed that the downstream distributions could not be explained by only assuming acceleration due to a quasi-static electric field across the shock ramp \citep[e.g.,][]{scudder86c, schwartz11a}.  To account for the discrepancy, they invoked wave-particle interactions as a possible mechanism to explain the difference between their observations and their predictions.  Note that the debate is not whether quasi-static electric fields exist within collisionless shock ramps, rather, the relative importance of these quasi-static fields versus the high frequency wave fields.  However, no study has quantified the relative importance of wave-particle interactions (with respect to the other dissipation mechanisms) in the energy budget of collisionless shock waves.

%%  Introduce energy dissipation mechanisms
\indent  Wave-particle interactions transfer energy between the electromagnetic fields (driven by instabilities) and charged particles, through effective collisions, with a net result that is analogous to an effective friction or drag force.  These effective collisions can act to irreversibly reduce the relative drift between electrons and ions that give rise to the currents producing the magnetic ramp in collisionless shock waves \citep[e.g.,][]{sagdeev66, gary81c}.  Recent observations \citep[e.g.,][]{wilsoniii07a, wilsoniii10a, wilsoniii12c, wilsoniii13a} and Vlasov simulations using realistic mass ratios \citep[e.g.,][]{petkaki06a, petkaki08a, yoon06a, yoon07a} have indirectly supported this theory.  Furthermore, recent PIC simulations have found that wave-particle interactions can modify the macroscopic structure of collisionless shock waves \citep[e.g.,][]{matsukiyo06b, riquelme11a, comisel11a}.  Therefore, a quantified estimate of the energy dissipation rate due to wave-particle interactions could help resolve the debate over the dominant energy dissipation mechanism in low Mach number collisionless shock waves.

%%  Summary and Outline
\indent  In this paper we describe the first quantified estimates showing that wave-particle interactions can provide more than enough energy dissipation to explain the observed increase in the specific entropy density across collisionless shock waves.  These results provide the first quantified observational evidence that the microphysics of collisionless shock waves can dominate their macroscopic behavior.  

%%  Summary and Outline:  Main
\indent  The paper is outlined as follows:  Section \ref{sec:DataSetsMethod} describes the data sets and methodology; Section \ref{sec:Observations} shows example bow shock crossings and low frequency waveforms; Section \ref{subsec:WaveModes} discusses the high frequency wave types observed and their relevance;  Section \ref{sec:EnergyDissipation} presents our evidence for wave energy dissipation; and Section \ref{sec:Discussion} discusses our conclusions.  

%%  Summary and Outline:  Main
\indent  We have included multiple appendices for supplemental material to provide the reader with more detailed discussions of our analysis techniques.  The appendices are outlined as follows:  Appendix \ref{app:Methodology} presents some of our analysis techniques; Appendix \ref{subapp:THEMISWaveObservations} presents the high frequency waves observed by THEMIS and their relevance; Appendix \ref{subapp:WaveFieldProperties} presents a statistical summary of the wave amplitudes; Appendix \ref{app:WindSTEREO} presents example waveform captures from the Wind and STEREO spacecraft; Appendix \ref{app:nifandbasis} introduces and defines the reference frame transformations and coordinate basis rotations used in our analysis; Appendix \ref{app:thermodynamics} introduces our technique for estimating macroscopic energy dissipation rates; Appendix \ref{subapp:Implementation} introduces our method for estimating the current density; Appendix \ref{subapp:Approximations} provides justification for our assumptions when estimating the current density; Appendix \ref{app:EnergyDissipation} discusses how we estimate energy dissipation due to wave-particle interactions and its relation to macroscopic energy dissipation estimates; and Appendix \ref{app:velocitymomentcorr} illustrates how we remove secondary ion populations from the ion velocity distributions.

%%----------------------------------------------------------------------------------------
%%  Section:  Data Sets
%%----------------------------------------------------------------------------------------
\section{Data Sets}  \label{sec:DataSetsMethod}
\indent  In this section we first describe the relevant data sets used to examine the \textbf{\textcolor{Red}{8}} bow shock crossings presented herein.  We selected bow shock crossings based upon whether a single THEMIS spacecraft had both wave burst electromagnetic field data and particle burst velocity distributions during the shock transition.  The details of our analysis techniques are presented in the appendices.  Throughout this paper we use the following notation to distinguish observables measured at different cadences.  We define any quantity observed at $\leq$128 sps as a quasi-static (DC-coupled) quantity, Q${\scriptstyle_{o}}$, and any quantity observed at $\sim$8192 sps as a fluctuating (AC-coupled) quantity, $\delta$Q.

%%  FGM
\indent  We utilized the THEMIS fluxgate magnetometer (FGM) \citep{auster08a} for quasi-static (DC-coupled) vector magnetic field measurements.  The THEMIS FGM is capable of returning three quasi-static magnetic field components, B${\scriptstyle_{o,j}}$, at up to 128 samples per second (sps) for short durations and nearly continuous measurements at 4 sps.  The FGM results were used to define the ramp (transition) region of the shocks and for the Rankine-Hugoniot relation solutions (discussed below).  The FGM data was also used for estimating the current density (see Appendices \ref{app:nifandbasis} and \ref{app:CurrentDensity}).

%%  EFI and SCM
\indent  Waveform burst captures were obtained from the Search Coil Magnetometer (SCM) \citep{lecontel08a, roux08a} and the Electric Field Instrument (EFI) \citep[][]{bonnell08a, cully08b}.  The EFI(SCM) receiver returns three electric(magnetic) field components, $\delta$E${\scriptstyle_{j}}$($\delta$B${\scriptstyle_{j}}$) at a nominal sample rate of $\sim$16,384($\sim$8,192) sps, or a Nyquist frequency of $\sim$8,192($\sim$4,096) Hz.  All of the wave burst data presented herein have a soft high-pass filter above $\sim$10 Hz because the EFI was sampling in an AC-coupled mode for every crossing.  Even when in an AC-coupled mode, the data returned by the EFI can still be contaminated by spin-dependent photoelectron emissions and electrostatic wake effects \citep[][]{bonnell08a}.  Therefore, we removed any interval that showed remnant contamination by hand prior to further analysis.

%%  ESA
\indent  The THEMIS electrostatic analyzers (ESA) \citep{mcfadden08a, mcfadden08b} provide full 4$\pi$ steradian particle velocity distribution functions for both electrons and ions ranging from a few eV to over 25 keV every spin period ($\sim$3 s) in burst mode.  The electron(ion) instrument is called EESA(IESA).  The EESA burst velocity distributions contain 32 energy and 88 solid-angle bins.  The IESA burst velocity distributions have similar energy/angular resolutions and cadence to EESA.  The ESA particle velocity distribution functions will be used to calculate the following particle velocity moments:  plasma number density, N${\scriptstyle_{i}}$; bulk flow velocity, \textbf{V}${\scriptstyle_{bulk}}$; average ion temperature, T${\scriptstyle_{i}}$; and average electron temperature, T${\scriptstyle_{e}}$.

%%  Particle velocity moments
\indent  Particle velocity moments from each instrument will be used to calculate shock conservation relations (discussed below), reference frame transformations (see Appendix \ref{subapp:nif}), and coordinate basis rotations (see Appendix \ref{subapp:nifbasis}).  Removal of secondary ion species (see Appendix \ref{app:velocitymomentcorr}) reflected from the shock, e.g., gyrating and/or gyrophase bunched ion distributions \citep[e.g.,][]{gurgiolo81a, meziane97}, is necessary in each event to approximate the velocity moments of the undisturbed upstream solar wind.

%%========================================================================================
%%  Table:  Shock Parameters
%%========================================================================================
\begin{table}[htb]
  \centering
  \caption{Shock Parameters and Rankine-Hugoniot Solutions}
  \label{tab:shockparams}
    \begin{tabular}{| l | c | c | c | c | c | c |}
      \hline
      \textbf{Date} & \textbf{Probe} & $\mid$V${\scriptstyle_{shn}}$$\mid$ & $\mid$U${\scriptstyle_{shn}}$$\mid$ & $\theta{\scriptstyle_{Bn}}$ & M${\scriptstyle_{f}}$ & N${\scriptstyle_{i2}}$/N${\scriptstyle_{i1}}$  \\
      (YYYY-MM-DD)  &  & (km/s) & (km/s) & (deg) &  &  \\
      \hline
      %%  Use Std. Dev. of Mean for Vshn and Ushn
      %%    -> b/c their uncertainties are dependent upon N
      %%      => our estimate for each would improve with larger N
      {\small 2009-07-13 [1st Crossing]} & B & 53 $\pm$  2 & 275 $\pm$ 2 & 43$^{\circ}$ $\pm$ 5$^{\circ}$ & 3.07 $\pm$ 0.10 & 6.7 $\pm$ 0.6   \\
      {\small 2009-07-21 [1st Crossing]} & C & 24 $\pm$  7 & 200 $\pm$ 2 & 51$^{\circ}$ $\pm$ 6$^{\circ}$ & 2.06 $\pm$ 0.12 & 3.6 $\pm$ 0.5   \\
      {\small 2009-07-23 [1st Crossing]} & C & 65 $\pm$  7 & 425 $\pm$ 2 & 83$^{\circ}$ $\pm$ 3$^{\circ}$ & 3.04 $\pm$ 0.04 & 4.1 $\pm$ 0.3   \\
      {\small 2009-07-23 [2nd Crossing]} & C & 13 $\pm$  7 & 504 $\pm$ 2 & 88$^{\circ}$ $\pm$ 2$^{\circ}$ & 3.62 $\pm$ 0.05 & 3.7 $\pm$ 0.2   \\
      {\small 2009-07-23 [3rd Crossing]} & C & 38 $\pm$ 10 & 417 $\pm$ 1 & 54$^{\circ}$ $\pm$ 4$^{\circ}$ & 3.11 $\pm$ 0.06 & 2.8 $\pm$ 0.4   \\
      {\small 2009-09-26 [1st Crossing]} & A & 29 $\pm$  8 & 339 $\pm$ 1 & 60$^{\circ}$ $\pm$ 9$^{\circ}$ & 4.83 $\pm$ 0.26 & 4.2 $\pm$ 0.8   \\
      {\small 2011-10-24 [1st Crossing]} & E & 44 $\pm$  9 & 361 $\pm$ 2 & 84$^{\circ}$ $\pm$ 5$^{\circ}$ & 2.22 $\pm$ 0.04 & 3.0 $\pm$ 0.3   \\
      {\small 2011-10-24 [2nd Crossing]} & E & 32 $\pm$  5 & 365 $\pm$ 2 & 88$^{\circ}$ $\pm$ 2$^{\circ}$ & 2.33 $\pm$ 0.01 & 4.8 $\pm$ 0.3   \\
      \hline
    \end{tabular}
\end{table}
%%========================================================================================
%%  Table:  Shock Parameters
%%========================================================================================

%%  Rankine-Hugoniot Solutions
\indent  We numerically solve the Rankine-Hugoniot relations \citep[e.g.,][]{vinas86a, koval08a} for each bow shock crossing in Table \ref{tab:shockparams} to estimate the shock normal vector ($\hat{\textbf{n}}$), the shock normal velocity in the spacecraft frame (V${\scriptstyle_{shn}}$), the shock normal velocity in the shock rest frame (U${\scriptstyle_{shn}}$), the shock normal angle ($\theta{\scriptstyle_{Bn}}$), the fast mode Mach number (M${\scriptstyle_{f}}$), and the shock compression ratio (N${\scriptstyle_{i2}}$/N${\scriptstyle_{i1}}$).  We use these parameters to characterize the macroscopic properties of the shock (see Appendices \ref{app:nifandbasis} and \ref{app:CurrentDensity}).  

%%  Rankine-Hugoniot Results
\indent  The results of the Rankine-Hugoniot analysis can be seen in Table \ref{tab:shockparams}.  We have examined \textbf{\textcolor{Red}{8}} THEMIS bow shock crossings, all of which were supercritical (for more details, see our analysis techniques and methodology in Appendix \ref{app:Methodology}).

%\clearpage
%%----------------------------------------------------------------------------------------
%%  Section:  Observations
%%----------------------------------------------------------------------------------------
\section{Observations}  \label{sec:Observations}
%%++++++++++++++++++++++++++++++++++++++++++++++++++++++++++++++++++++++++++++++++++++++++
%% Image:  Example Bow Shock Crossing [fgl and moments]
%%++++++++++++++++++++++++++++++++++++++++++++++++++++++++++++++++++++++++++++++++++++++++
\begin{wrapfigure}{r}{0.405\textwidth}
  \vspace{-15pt}
  \centering
    {\includegraphics[trim = 0mm 0mm 0mm 0mm, clip, width=0.415\textwidth]
    {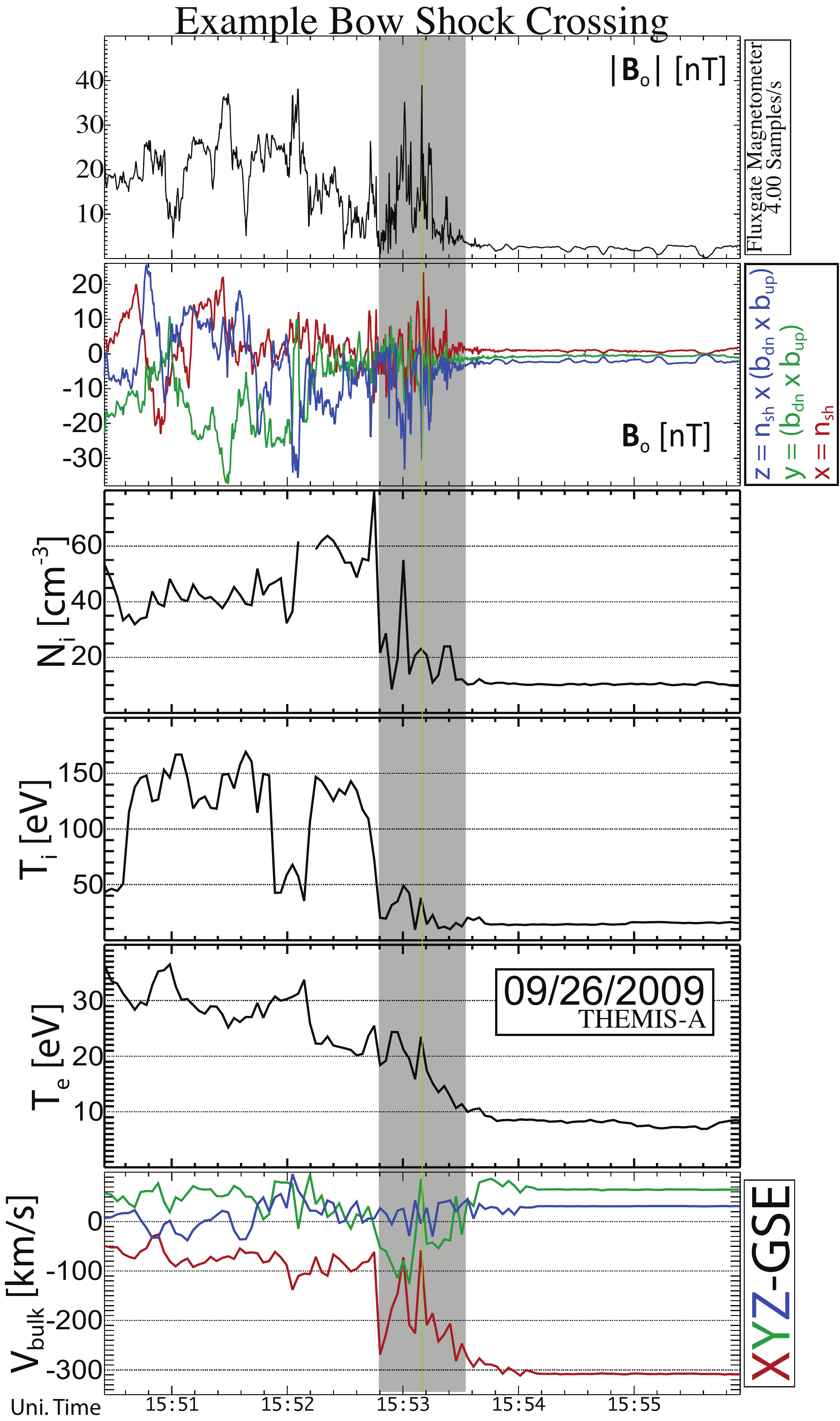}}
    %%  Define new caption setting options
    \captionsetup{width=0.395\textwidth,font=footnotesize,labelfont=bf}
    %%  Define caption
   \caption[Example Bow Shock Crossing]{This figure shows a $\sim$330 s window centered on $\sim$15:53:10.080 UT (green vertical line) with magnetic field and plasma parameters, observed by THEMIS-A on 2009-09-26, for a typical supercritical oblique bow shock crossing.  The top two panels show $\mid$\textbf{B}${\scriptstyle_{o}}$$\mid$ [nT] and \textbf{B}${\scriptstyle_{o}}$ [NCB, nT], respectively, observed with the FGM at $\sim$4 sps.  The rest of the panels show, in order, N${\scriptstyle_{i}}$ [cm$^{-3}$], T${\scriptstyle_{i}}$ [eV], T${\scriptstyle_{e}}$ [eV], and V${\scriptstyle_{bulk}}$ [GSE, km/s].  The gray shaded region indicates the timespan for Figure \ref{fig:exampleBSCfgh}.}
  \label{fig:exampleBSCfglMoms}
  \vspace{-10pt}
\end{wrapfigure}
%%++++++++++++++++++++++++++++++++++++++++++++++++++++++++++++++++++++++++++++++++++++++++
%% Image:  Example Bow Shock Crossing [fgl and moments]
%%++++++++++++++++++++++++++++++++++++++++++++++++++++++++++++++++++++++++++++++++++++++++

\indent  In this section we present examples of bow shock crossings and characteristic examples of the types of electromagnetic fluctuations observed.

\indent  We have examined \textbf{\textcolor{Red}{8}} bow shock crossings with the THEMIS spacecraft.  For every crossing, we have removed the secondary ion populations (see Appendix \ref{app:velocitymomentcorr}) and electric field spikes due to photoelectron emissions and electrostatic wake effects.  We filtered the electromagnetic fields above $\sim$10 Hz to remove DC-coupled quasi-static fluctuations from the EFI and SCM observations and convective electric fields from the EFI.  A cursory comparison (not shown) between the electric fields observed at $\sim$128 sps and those at $\sim$8,192 sps show that the high frequency electric fields consistently dominate the low frequency components ($\delta$E $\gg$ E${\scriptstyle_{o}}$).  Therefore, we did not focus on these lower frequency electric fields.

%%----------------------------------------------------------------------------------------
%%  Subsection:  Overview and Examples
%%----------------------------------------------------------------------------------------
\subsection{Overview and Examples}  \label{subsec:Overview}
\indent  Figure \ref{fig:exampleBSCfglMoms} shows an example bow shock crossing observed on 2009-09-26 by THEMIS-A.  The top two panels show the $\mid$\textbf{B}${\scriptstyle_{o}}$$\mid$ and its components in the normal incidence frame (NIF) coordinate basis (NCB, see Appendix \ref{app:nifandbasis} for definition) observed by the FGM at $\sim$4 sps.  The next four panels show particle velocity moments for the ions and electrons.  We removed secondary ions (see Appendix \ref{app:velocitymomentcorr}) due to shock reflection from the upstream velocity distributions prior to calculating the ion moments in Figure \ref{fig:exampleBSCfglMoms}.  The change in particle velocity moments in Figure \ref{fig:exampleBSCfglMoms} reflect the supercritical nature of this shock (see Tables \ref{tab:shockparams} and \ref{tab:machparams}).  One can see that the shock causes significant plasma compression (i.e., N${\scriptstyle_{i2}}$/N${\scriptstyle_{i1}}$ $\sim$ 4), strong ion heating (T${\scriptstyle_{i2}}$/T${\scriptstyle_{i1}}$ $>$ 6), and strong electron heating (T${\scriptstyle_{e2}}$/T${\scriptstyle_{e1}}$ $\gtrsim$ 3).  

%%  Discuss Figure 1 further
\indent  The structure of this bow shock is consistent with previous observations of supercritical bow shocks.  The large variability observed in $\mid$\textbf{B}${\scriptstyle_{o}}$$\mid$ and \textbf{B}${\scriptstyle_{o}}$ could be explained by sudden expansions and contractions of the bow shock or non-stationary shock reformation \citep[e.g.,][]{krasnoselskikh02a, lobzin07a}.  The large fluctuations in \textbf{B}${\scriptstyle_{o}}$ correlate with significant deflections in \textbf{V}${\scriptstyle_{bulk}}$ and changes in N${\scriptstyle_{i}}$.  The details of the large fluctuations in \textbf{B}${\scriptstyle_{o}}$ are discussed further below.

%%  Delayed Ti increase
\indent  Notice that the increase in T${\scriptstyle_{i}}$ is slightly delayed with respect to T${\scriptstyle_{e}}$.  This is partly due to our use of only the core in determining T${\scriptstyle_{i}}$ (see Appendix \ref{app:velocitymomentcorr}) and partly a real phenomena.  The flow is deflected and the electrons begin to thermalize before T${\scriptstyle_{i}}$ shows any significant change.  Therefore, the bulk flow reduction appears to be compensated by reflected ions and electron heating.  The two phenomena are not unrelated and will be discussed in greater detail later.

\begin{wrapfigure}{l}{0.425\textwidth}
  \vspace{-10pt}
  \centering
    {\includegraphics[trim = 0mm 0mm 0mm 0mm, clip, width=0.435\textwidth]
    {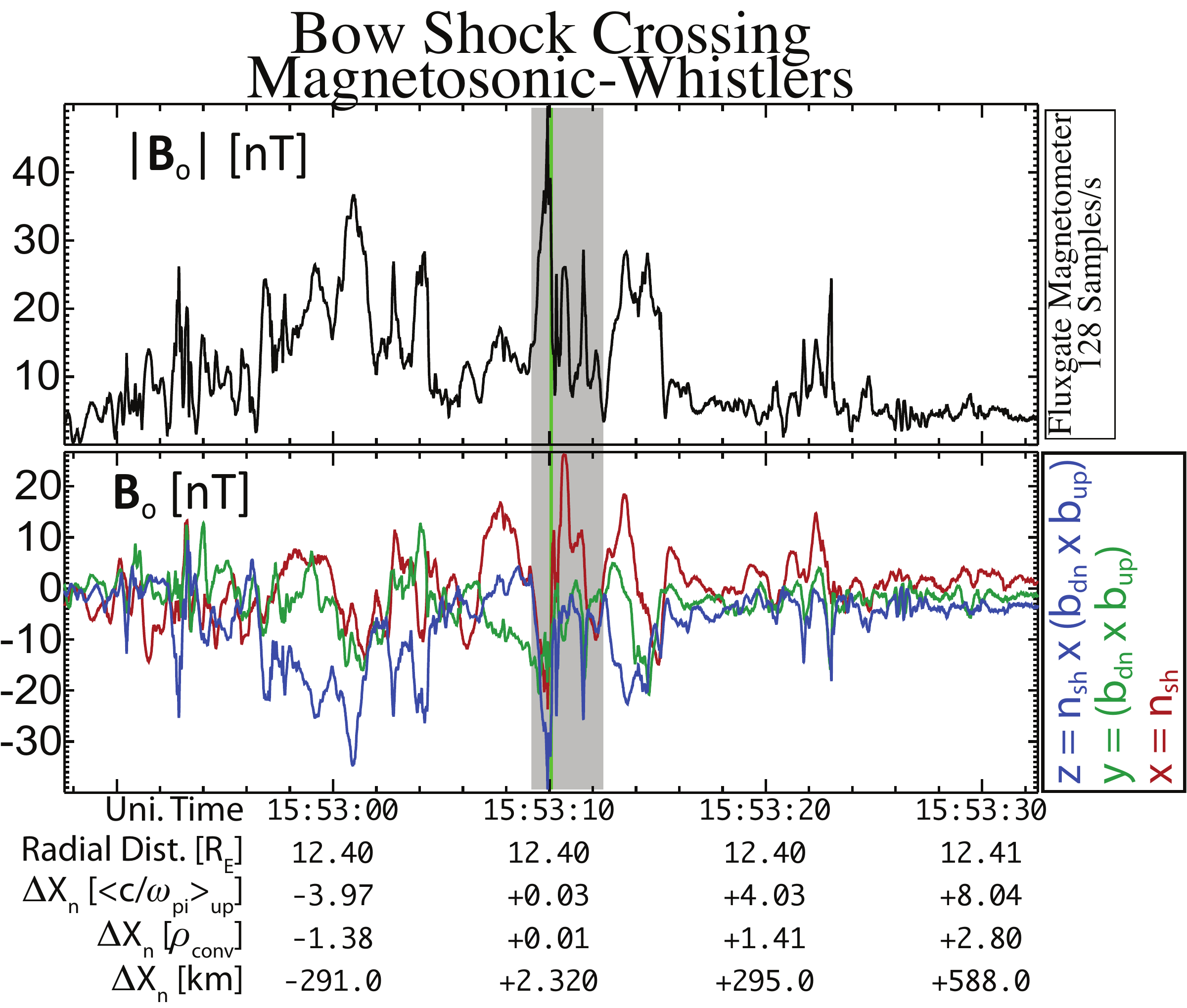}}
    %%  Define new caption setting options
    \captionsetup{width=0.415\textwidth,font=footnotesize,labelfont=bf}
    %%  Define caption
    \caption[FGH Example Bow Shock Crossing]{This figure illustrates how survey data can under-sample large amplitude fluctuations in typical bow shock crossings.  It shows a $\sim$45 s window of $\mid$\textbf{B}${\scriptstyle_{o}}$$\mid$ and its NCB components [nT] sampled at $\sim$128 sps by the FGM.  The tick marks below the plot are, from top-to-bottom, the following:  (1) Time [UT], (2) radial distance from center of Earth [R${\scriptstyle_{E}}$], (3)-(5) distance from center of shock ramp in units of $\langle c / \omega{\scriptstyle_{pe}} \rangle{\scriptstyle_{up}}$, $\rho{\scriptstyle_{conv}}$, and km.  The gray shaded region indicates the timespan for Figure \ref{fig:exampleEBSjEdjBSC}.}
    \label{fig:exampleBSCfgh}
    \vspace{-10pt}
\end{wrapfigure}

%%  Introduce Figure 2
\indent  The gray shaded region in Figure \ref{fig:exampleBSCfglMoms} shows the time range for Figure \ref{fig:exampleBSCfgh}.  Figure \ref{fig:exampleBSCfgh} shows $\sim$45 s of higher time resolution ($\sim$128 sps) magnetic field data observed by the FGM instrument.  The tick marks below the plot show the universal time (UT), radial distance from center of Earth (R${\scriptstyle_{E}}$), and the distance from center of shock ramp ($\sim$15:53:10.080 UT) in three units:  upstream average ion inertial lengths ($\langle c / \omega{\scriptstyle_{pe}} \rangle{\scriptstyle_{up}}$), convected ion gyroradii ($\rho{\scriptstyle_{conv}}$ $=$ U${\scriptstyle_{shn}}$/$\langle \Omega{\scriptstyle_{ci}} \rangle{\scriptstyle_{down}}$), and km.

%%  Discuss Figure 2 further
\indent  Notice that the amplitudes of both $\mid$\textbf{B}${\scriptstyle_{o}}$$\mid$ and its components are larger than observed by the FGM at $\sim$4 sps shown in Figure \ref{fig:exampleBSCfglMoms}.  The comparison illustrates that the largest amplitude fluctuations are not well resolved in the $\sim$4 sps FGM data shown in Figure \ref{fig:exampleBSCfglMoms}.  The fluctuations in Figure \ref{fig:exampleBSCfgh}, which occur on spatial scales much smaller than ion scales, illustrate the highly dynamic nature of the supercritical bow shock and are commonly observed in the bow shock crossings with high time resolution magnetometers.  Many of these magnetic pulsations have $\delta$B/B${\scriptstyle_{o}}$ $\gtrsim$ 4 and their gradient scale lengths along the shock normal vector are on electron scales.  These electromagnetic fluctuations are identified as magnetosonic-whistler mode waves.

%%++++++++++++++++++++++++++++++++++++++++++++++++++++++++++++++++++++++++++++++++++++++++
%% Image:  Example Bow Shock Crossing [fgh only]
%%++++++++++++++++++++++++++++++++++++++++++++++++++++++++++++++++++++++++++++++++++++++++
%%++++++++++++++++++++++++++++++++++++++++++++++++++++++++++++++++++++++++++++++++++++++++
%% Image:  Example Bow Shock Crossing [fgh only]
%%++++++++++++++++++++++++++++++++++++++++++++++++++++++++++++++++++++++++++++++++++++++++

%%  Discuss magnetosonic-whistlers
\indent  Nearly all of the bow shock crossings examined herein show large amplitude compressive magnetic fluctuations upstream of the shock ramp consistent with magnetosonic-whistler precursors \citep[e.g.,][and references therein]{wilsoniii09a}.  These fluctuations show enhanced power for f${\scriptstyle_{ci}}$ $<$ f${\scriptstyle_{sc}}$ $\lesssim$ f${\scriptstyle_{lh}}$, where f${\scriptstyle_{sc}}$ is the spacecraft frame frequency.  They are right-hand polarized (with respect to \textbf{B}${\scriptstyle_{o}}$) electromagnetic compressive fluctuations with magnetic fluctuations in phase with density fluctuations.  Theory suggests that they are driven by dispersion \citep[e.g.,][]{kennel85a, krasnoselskikh02a} and/or reflected ions \citep[e.g.,][]{wu83a, riquelme11a, comisel11a}.  Observations have shown evidence to support both dispersion \citep[e.g.,][]{sundkvist12a} and instabilities \citep[e.g.,][]{wilsoniii12c} as the source of these waves.  At highly oblique angles, these waves stochastically accelerate electrons parallel to \textbf{B}${\scriptstyle_{o}}$ and heat ions perpendicular to \textbf{B}${\scriptstyle_{o}}$ \citep[e.g.,][]{wu83a, cairns05a}, which has been supported by observations \citep[e.g.,][]{wilsoniii12c}.

%%  Discuss magnetosonic-whistler amplitudes
\indent  Magnetosonic-whistlers are primarily observed with the FGM instrument because we filter the SCM data to $\geq$10 Hz to match the EFI data when AC-coupled.  Figure \ref{fig:exampleBSCfgh} shows that these fluctuations can have amplitudes $\sim$2-4 times the upstream average field strength.  Such large amplitude fluctuations raise doubts about the capacity for electrons to remain magnetized as they move through this region \citep[e.g.,][]{mozer13a, sundkvist13a}.  These fluctuations are rarely observed in the filtered SCM data shown herein and they are not the focus of this study.

%%  Introduce Figure 3
\indent  The gray shaded region in Figure \ref{fig:exampleBSCfgh} shows the $\sim$3.3 s window for the time range in Figure \ref{fig:exampleEBSjEdjBSC}.  Figure \ref{fig:exampleEBSjEdjBSC} shows examples of the electromagnetic fluctuations that are the focus of this paper.  The top two panels show the $\mid$\textbf{B}${\scriptstyle_{o}}$$\mid$ and its NCB components (see Appendix \ref{app:nifandbasis} for definition) observed by the FGM at $\sim$128 sps.  The third and fourth panels shows $\delta$\textbf{E} and $\delta$\textbf{B}, respectively, in the spacecraft frame (SCF) of reference and in a field-aligned coordinates (FACs) basis (definition found in the inset box).  Both are sampled at $\sim$8192 sps with a soft high-pass filter above $\sim$10 Hz.  The fifth panel shows $\mid$$\delta \tilde{\textbf{S}}$$\mid$ (see Appendix \ref{app:Methodology} for definition).  The 6th panel shows two components of of the current density, \textbf{j}${\scriptstyle_{o}}$, in NCB (see Appendix \ref{app:nifandbasis} for definition).  The seventh panel shows the amount of ohmic dissipation, or $\left( - \textbf{j}{\scriptstyle_{o}} \cdot \delta \textbf{E} \right)$, at the $\delta$\textbf{E} time steps (red) with corresponding trend line (cyan).  $\delta$\textbf{E} was transformed into the NIF and rotated into the NCB prior to projecting onto \textbf{j}${\scriptstyle_{o}}$.  We will discuss this figure in more detail below.

%%----------------------------------------------------------------------------------------
%%  Subsection:  Separation of Frequencies
%%----------------------------------------------------------------------------------------
\subsection{Separation of Frequencies}  \label{subsec:SeparationFrequencies}
%%  Intro
\indent  In this section we discuss the differences between the fluctuations in $\delta$\textbf{B} and $\delta$\textbf{E} and their dependence upon frequency.  First we discuss the lower frequency magnetosonic-whistlers and then we will discuss the higher frequency electrostatic and electromagnetic modes that are the focus of this work.  We discuss these two separate frequency and spatial scales within the context of their relative contributions to the energy dissipation due to the work done on the particles by the electromagnetic fields, $\left( - \textbf{j} \cdot \textbf{E} \right)$.

%%++++++++++++++++++++++++++++++++++++++++++++++++++++++++++++++++++++++++++++++++++++++++
%% Image:  Example BSC with fields, currents, and S
%%++++++++++++++++++++++++++++++++++++++++++++++++++++++++++++++++++++++++++++++++++++++++
\begin{figure}[!htb]
  \centering
    {\includegraphics[trim = 0mm 0mm 0mm 0mm, clip, height=17cm]{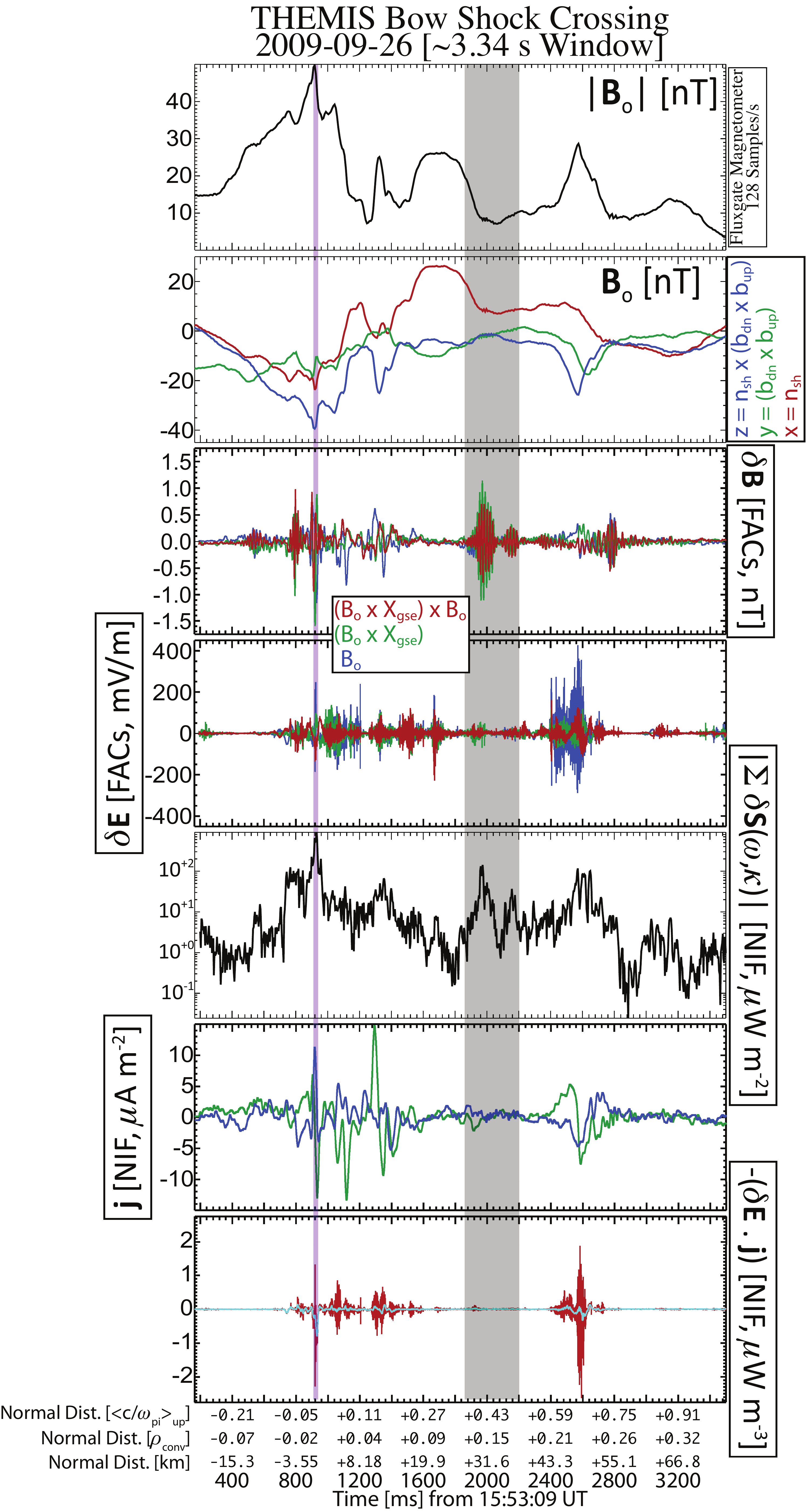}}
    \captionsetup{width=12cm}   %%  adjust caption width
    \caption[Example BSC with fields, currents, and S]{A $\sim$3.3 s window showing an example of the energy budget in a typical bow shock crossing.  The panels are, from top-to-bottom, the following:  (1) $\mid$\textbf{B}${\scriptstyle_{o}}$$\mid$ [nT]; (2) \textbf{B}${\scriptstyle_{o}}$ [NCB, nT]; (3) $\delta$\textbf{B} [FACs, nT]; (4) $\delta$\textbf{E} [FACs, mV/m]; (5) $\mid$$\delta \tilde{\textbf{S}}$$\mid$ [NIF, $\mu$W m$^{-2}$]; (6) \textbf{j}${\scriptstyle_{o}}$ [NCB, $\mu$A m$^{-2}$]; and (7) $\left(- \textbf{j}{\scriptstyle_{o}} \cdot \delta \textbf{E} \right)$ [NCB, $\mu$W m$^{-3}$].  The gray shaded region shows the timespan for the example whistler mode wave shown in Figure \ref{fig:examplewhistler}.  The magenta shaded region shows the timespan for the example electrostatic solitary waves shown in Figure \ref{fig:exampleESWs}.}
    \label{fig:exampleEBSjEdjBSC}
\end{figure}
%%++++++++++++++++++++++++++++++++++++++++++++++++++++++++++++++++++++++++++++++++++++++++
%% Image:  Example BSC with fields, currents, and S
%%++++++++++++++++++++++++++++++++++++++++++++++++++++++++++++++++++++++++++++++++++++++++

%%  Discuss magnetosonic-whistlers
\indent  For magnetosonic-whistlers, the largest values of E${\scriptstyle_{o}}$ and B${\scriptstyle_{o}}$ occur with spacecraft frame frequencies (f${\scriptstyle_{sc}}$) between the ion cyclotron (f${\scriptstyle_{ci}}$) and lower hybrid resonance (f${\scriptstyle_{lh}}$) frequencies or f${\scriptstyle_{ci}}$ $<$ f${\scriptstyle_{sc}}$ $\lesssim$ f${\scriptstyle_{lh}}$, which is $\sim$0.01-10 Hz for typical solar wind conditions.  These fluctuations can be as large as B${\scriptstyle_{o}}$ $\gtrsim$ 30 nT and E${\scriptstyle_{o}}$ $\gtrsim$ 40 mV/m, but they are typically smaller with B${\scriptstyle_{o}}$ $\sim$ 1-10 nT and E${\scriptstyle_{o}}$ $\sim$ 5-20 mV/m.  Magnetosonic-whistlers also typically have smaller \textbf{S}${\scriptstyle_{o}}$ contributions than the higher frequency waves because they have much smaller electric fields, or E${\scriptstyle_{o}}$ $\ll$ $\delta$E.

%%  Discuss Figure 3:  Dissipation
\indent  Notice in Figure \ref{fig:exampleEBSjEdjBSC} that the energy dissipation due to the work done on the particles by the electromagnetic fields or $\left( - \textbf{j}{\scriptstyle_{o}} \cdot \delta \textbf{E} \right)$ has peak magnitudes near the sharpest gradients in \textbf{B}${\scriptstyle_{o}}$ or the largest values of \textbf{j}${\scriptstyle_{o}}$.  These large \textbf{j}${\scriptstyle_{o}}$ are typically due to magnetosonic-whistlers and are typically dominated by frequencies $\lesssim$10 Hz.  However, the contribution to $\left\lvert \textbf{j} \cdot \textbf{E} \right\rvert$ by magnetosonic-whistlers tends to be at least an order of magnitude smaller than the higher frequency waves (discussed below) due to their smaller $\delta$E.  The peaks in $\left( - \textbf{j}{\scriptstyle_{o}} \cdot \delta \textbf{E} \right)$ near these large gradients are due to the higher frequency waves.  Previous observations have shown that higher frequency electromagnetic \citep[e.g.,][]{hull12a, wilsoniii13a} and electrostatic \citep[e.g.,][]{wilsoniii07a} fluctuations occur simultaneously with large amplitude magnetosonic-whistler precursor waves.  We also find from our THEMIS observations that these higher frequency fluctuations tend to occur simultaneously with large amplitude magnetosonic-whistlers.  

%%  Discuss Figure 3:  current driven instabilities
\indent  The occurrence of higher frequency waves near the largest values of \textbf{j}${\scriptstyle_{o}}$ is consistent with the idea that current-driven instabilities are responsible for dissipating the necessary energy to regulate the nonlinear steepening of an electromagnetic wave \citep[e.g.,][]{sagdeev66, gary81c, bale05a, treumann09a}.  Magnetosonic-whistlers are also capable of accelerating and reflecting particles when nonlinear and steepened \citep[e.g.,][]{wilsoniii13b}.  Therefore, we believe they provide an important source of free energy for the higher frequency waves (the focus of this study) and they can act as a conduit for energy/momentum exchange between fields and particles.

%%  Discuss Figure 3:  Fluctuations in general
\indent  The data shown in Figure \ref{fig:exampleEBSjEdjBSC} are not unusual for bow shock crossings.  Every bow shock crossing we have examined with available waveform data from Wind, STEREO, or THEMIS shows high frequency ($\gtrsim$10 Hz) large amplitude fluctuations in both $\delta$\textbf{B} and $\delta$\textbf{E}.  These higher frequency waves are composed of two categories:  (1) electromagnetic fluctuations; and (2) electrostatic (i.e., $\left( \textbf{k} \times \delta \textbf{E} \right)$ $\sim$ 0) fluctuations.

%%  Fluctuations:  Electromagnetic
\indent  The high frequency waves dominated by electromagnetic components show peak values of $\delta$B between f${\scriptstyle_{lh}}$ and the electron cyclotron frequency, f${\scriptstyle_{ce}}$, or f${\scriptstyle_{lh}}$ $\ll$ f${\scriptstyle_{sc}}$ $\leq$ f${\scriptstyle_{ce}}$.  These fluctuations can have large amplitudes with $\delta$B and $\delta$E up to $\sim$2 nT and $\sim$30 mV/m, respectively, but they typically have $\delta$B $\lesssim$ 0.5 nT and $\delta$E $\lesssim$ 10 mV/m.  They can produce significant contributions to $\delta$\textbf{S} but their contributions to $\left\lvert \textbf{j}{\scriptstyle_{o}} \cdot \delta \textbf{E} \right\rvert$ tend to be small compared to the electrostatic fluctuations discussed below and they are observed less often.  We also note that though these high frequency electromagnetic fluctuations can have large $\delta$B, they are always smaller than the lower frequency magnetosonic-whistlers.  Thus, as one would expect, the magnetic field spectrum shows a decreasing trend with increasing frequency.

%%  Fluctuations:  Electrostatic
\indent  The electric fields, however, can show an inverted spectrum in the presence of high frequency electrostatic waves.  Ignoring the high frequency electromagnetic fluctuations, there is typically a large frequency gap between the low frequency magnetosonic-whistlers and the higher frequency electrostatic waves.  Their peak values of $\delta$E typically occur between the ion (f${\scriptstyle_{pi}}$) and electron plasma frequency (f${\scriptstyle_{pe}}$), or f${\scriptstyle_{pi}}$ $\leq$ f${\scriptstyle_{sc}}$ $\lesssim$ f${\scriptstyle_{pe}}$.  These fluctuations have the largest contribution to $\delta$E and dominate the entire power spectrum.  They can have amplitudes $\delta$E $\gtrsim$ 300 mV/m, but they typically have $\delta$E $\sim$ 10-50 mV/m.  These electrostatic fluctuations typically produce the largest contributions to $\delta$\textbf{S} and $\left\lvert \textbf{j}{\scriptstyle_{o}} \cdot \delta \textbf{E} \right\rvert$ because of their incredibly large $\delta$E.  

%%  Fluctuations:  Reference to later discussion
\indent  We discuss both the high frequency electromagnetic and electric fluctuations in more detail in Section \ref{subsec:WaveModes} and Appendix \ref{app:HFWaveObservations}.

%\clearpage
%%----------------------------------------------------------------------------------------
%%  Subsection:  High Frequency Waves
%%----------------------------------------------------------------------------------------
\subsection{High Frequency Waves}  \label{subsec:WaveModes}

\indent  In this section, first we will introduce and discuss the various high frequency wave types observed.  We leave the detailed examples and discussion of these high frequency waves to the appendices, which are outlined as follows:  in Appendix \ref{subapp:THEMISWaveObservations} we present some example waveforms observed by THEMIS and discuss their relevance; in Appendix \ref{subapp:WaveFieldProperties} we summarize the statistics of the wave properties for all high frequency waves observed by THEMIS; and finally in Appendix \ref{app:WindSTEREO} we show example waveforms observed by the Wind and STEREO spacecraft for comparison.

%%  Summary of wave types
\indent  All of the bow shock crossings examined had large amplitude fluctuations in $\delta$B and $\delta$E.  Nearly all of the bow shock crossings examined herein show any combination of the following electromagnetic and electrostatic fluctuations, in no particular order:  (1) magnetosonic-whistler precursors \citep[e.g.,][and references therein]{wilsoniii09a}; (2) high frequency whistler mode waves \citep[e.g.,][]{hull12a, wilsoniii13a}; (3) trains of electrostatic solitary waves (ESWs) or electron phase space holes \citep[e.g.,][]{bale98b, bale02a}; (4) ion-acoustic waves (IAWs) \citep[e.g.,][]{wilsoniii07a}; and/or (5) nonlinear electrostatic fluctuations consistent with those examined by \citet[][]{hull06a} and \citet[][]{wilsoniii10a}.  Though magnetosonic-whistlers are large and have been shown to be important \citep[e.g.,][]{sundkvist12a, wilsoniii12c}, we will not focus on them herein.

%%  Relate names to previous discussion of EM and ES fluctuations
\indent  In Section \ref{subsec:SeparationFrequencies}, we discussed some properties of high frequency electromagnetic and electrostatic fluctuations.  The high frequency electromagnetic waves are whistler mode waves.  The high frequency electrostatic waves are composed of combinations of ECDI, IAWs, and trains of ESWs.  As we previously discussed, the electrostatic fluctuations produce the largest contributions to $\delta$\textbf{S} and $\left\lvert \textbf{j}{\scriptstyle_{o}} \cdot \delta \textbf{E} \right\rvert$.  This is significant because theory predicts that these high frequency electrostatic waves can provide the dominant form of energy dissipation for collisionless shocks \citep[e.g.,][]{sagdeev66, coroniti70b, tidman71a, wu84a, treumann09a}.  In the following, we discuss recent simulation results that are consistent with our observations and conclusions.

%%  Consistently Observed ECDI, IAWs, and ESWs
\indent  ECDI, IAWs, and trains of ESWs (similar to those shown in Figures \ref{fig:exampleecdi}-\ref{fig:exampleESWs}) are observed in nearly every bow shock crossing we have examined with not only THEMIS, but Wind \citep[e.g.,][]{wilsoniii10f} and STEREO \citep[e.g.,][]{breneman11c} as well (see Appendix \ref{app:HFWaveObservations}).  The amplitude of these fluctuations range from $\sim$10's of mV/m to $>$300 mV/m.  These modes are observed semi-continuously from the foot through the magnetosheath.  This is in contrast with some simulations which only show waves near the front(upstream) edge of the foot or shock ramp \citep[e.g.,][]{matsukiyo06b}.  However, \citet[][]{muschietti13a} noted that because these simulations (and theirs) were performed in the electron rest frame, the waves were limited to the shock foot.  If the effects of convection were included, they suggest that the waves could exist everywhere in the shock transition region, consistent with our observations.

%%  The ECDI and [IAWs and ESWs]
\indent  The study by \citet[][]{muschietti13a} focused on the effects of the ECDI in a perpendicular shock.  We have shown that the waves predominantly observed include the ECDI as well as IAWs and ESWs.  While we observed ECDI in each crossing, the majority of the electrostatic waves were more consistent with IAWs and ESWs.  \citet[][]{muschietti13a} ran an example simulation to compare the evolution of the ECDI and IAWs.  They found that at late times in their simulations, the ECDI and IAWs had very similar power spectrums (ignoring the peaks due to the Bernstein modes in the ECDI).  The only differences were in the wave polarization and their respective effects on the particle distributions.  The IAWs in their simulation began to form electron phase space holes at later times.  

%%  How to explain all 3
\indent  The similarity in the power spectrums for the ECDI and IAWs found by \citet[][]{muschietti13a} should be expected since the ECDI is a series of electron Bernstein modes coupled to Dopler-shifted IAWs.  However, it adds difficulty to the unique identification of each mode.  Moreover, \citet[][]{muschietti13a} found that the higher harmonics damped out leaving only the fundamental after sufficient time.  The result was a well defined peak near f${\scriptstyle_{ce}}$ and a broad, weaker spectrum at higher frequencies.  At this point, the ECDI power spectrum looks very similar to the IAW power spectrum.  Thus, it is not surprising that we identify fluctuations consistent with both the ECDI and IAWs.  

%%  How to explain all 3 (Continued)
\indent  As previously shown, ESWs can either couple to, or directly cause, the growth of IAWs \citep[e.g.,][]{dyrud06} or whistler mode waves \citep[e.g.,][]{lu08}.  Both of these modes may also be indirectly driven unstable by the large amplitude magnetosonic-whistler waves observed throughout the transition region.  The large magnetic fluctuations due magnetosonic-whistlers can produce strong localized currents that can excite current-driven instabilities like IAWs.  Magnetosonic-whistlers can also compress the plasma to produce a temperature anisotropy instability that may explain the origin of the high frequency whistler mode waves.  Therefore, our THEMIS, Wind, and STEREO observations of combinations of ECDI, ESWs, and IAWs throughout the entire shock transition region (from foot through the magnetosheath) are consistent with simulations \citep[e.g.,][]{muschietti13a} and previous observations \citep[e.g.,][]{wilsoniii07a, wilsoniii10a, wilsoniii13a}.

%%----------------------------------------------------------------------------------------
%%  Section:  Energy Dissipation
%%----------------------------------------------------------------------------------------
\section{Energy Dissipation}  \label{sec:EnergyDissipation}

\indent  In this section, we will show conclusive evidence that wave-particle interactions can provide enough energy dissipation to balance the nonlinear wave steepening producing the shock.

%%  Discuss ubiquity of waves
\indent  We have examined \textbf{\textcolor{Red}{8}} bow shock crossings with the THEMIS spacecraft, where the shock parameters can be seen in Tables \ref{tab:shockparams} and \ref{tab:machparams}.  In every THEMIS event examined, we observed large amplitude electromagnetic and electrostatic fluctuations in and around the shock ramps, with the largest amplitudes found near the sharpest magnetic field gradients.  The long duration of the THEMIS waveform captures compared to those observed by Wind and STEREO show that these fluctuations can remain enhanced for $>$10 seconds.  Figures \ref{fig:exampleWind} and \ref{fig:exampleSTEREO} support our conclusion that large amplitude electromagnetic fluctuations are an ubiquitous phenomena in the collisionless bow shock transition region.  Previous studies \citep[e.g.,][]{wilsoniii07a} at interplanetary shocks came to a similar conclusion.  We also note that the observation of these waves is not limited to quasi-perpendicular geometry, as seen in Table \ref{tab:shockparams} and previously reported by \citet[][]{wilsoniii07a}.  These results add to the mounting evidence that electromagnetic waves play an important role in the macroscopic redistribution of energy in collisionless shocks.  Therefore, we decided to quantify the relative contribution of these electromagnetic waves in the global energy budget of the collisionless bow shock.

%%++++++++++++++++++++++++++++++++++++++++++++++++++++++++++++++++++++++++++++++++++++++++
%% Image:  (*eta* |j|^2)/∆∑ vs. |(∂E . j)|/∆∑
%%++++++++++++++++++++++++++++++++++++++++++++++++++++++++++++++++++++++++++++++++++++++++
\begin{figure}[!htb]
  \centering
    {\includegraphics[trim = 0mm 0mm 0mm 0mm, clip, width=12cm]{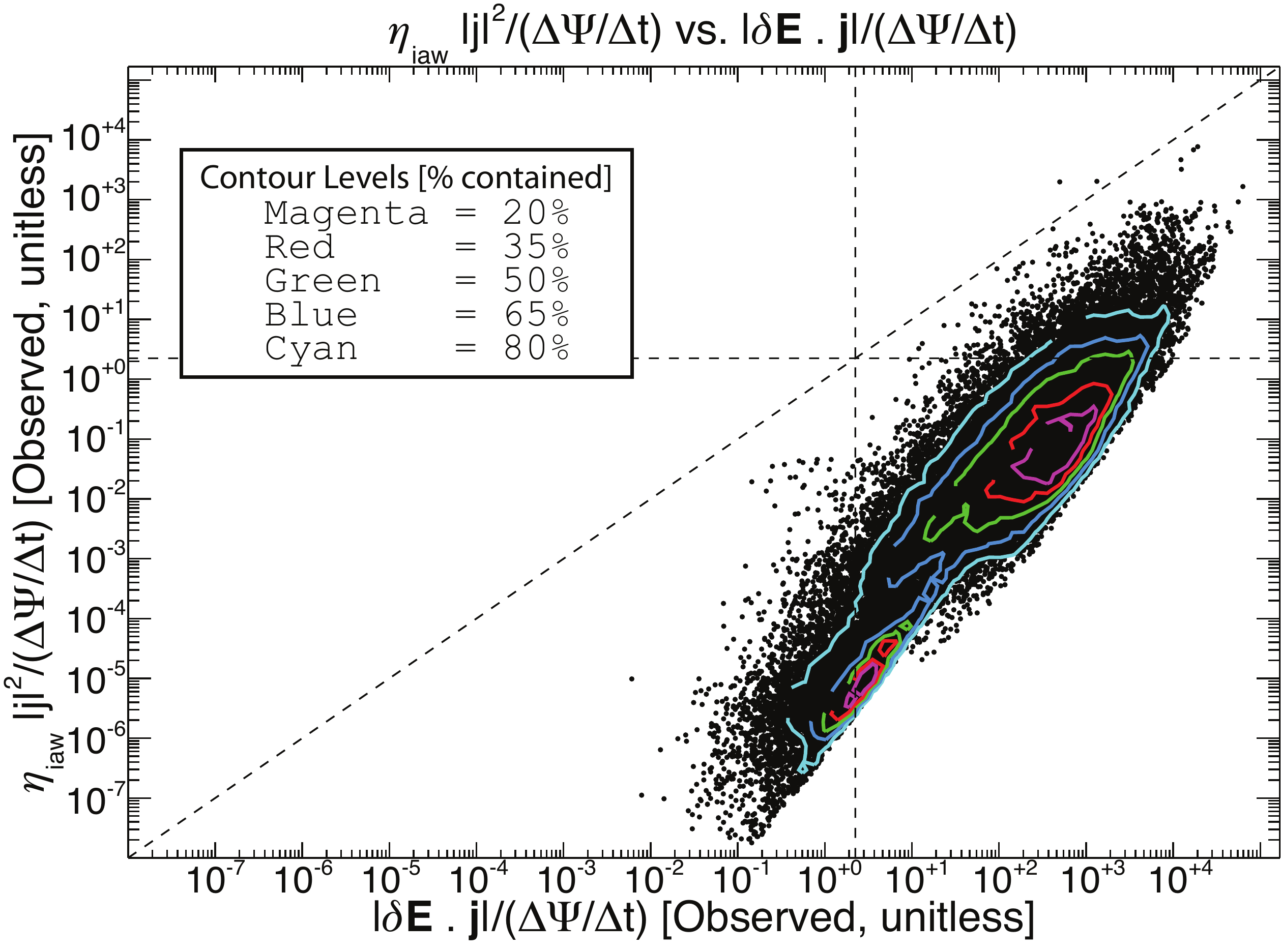}}
    \captionsetup{width=12cm}   %%  adjust caption width
    \caption[Comparison of Ohmic Dissipation Rates]{Plot showing that the wave energy dissipation rate often exceeds the dissipation rate necessary to produce the observed increase in entropy.  The plot shows $\mathcal{Y}{\scriptstyle_{\Psi}}$ ($\equiv$ $\eta{\scriptstyle_{iaw}} \left\lvert \textbf{j}{\scriptstyle_{o}} \right\rvert^{2}$/$\dot{\Psi}$) versus $\mathcal{R}{\scriptstyle_{\Psi}}$ ($\equiv$ $\left\lvert \textbf{j}{\scriptstyle_{o}} \cdot \delta \textbf{E} \right\rvert$/$\dot{\Psi}$) for all events examined herein.  The contours show the percentage of points contained within each.  The inset box defines the corresponding percentage.}
    \label{fig:comparedissrate}
\end{figure}
%%++++++++++++++++++++++++++++++++++++++++++++++++++++++++++++++++++++++++++++++++++++++++
%% Image:  (*eta* |j|^2)/∆∑ vs. |(∂E . j)|/∆∑
%%++++++++++++++++++++++++++++++++++++++++++++++++++++++++++++++++++++++++++++++++++++++++

%%  Introduce (*eta* |j|^2)/∆∑ vs. |(∂E . j)|/∆∑
\indent  We performed this test by calculating the ratio of the dissipation rate of the waves to the dissipation rates necessary to explain the observed increase in entropy, which we defined as $\dot{\Psi}$ $\equiv$ ($\rho$T)$\Delta \mathfrak{s}$/$\Delta$t.  We calculated this ratio using two slightly different methods for reasons explained in Appendix \ref{app:EnergyDissipation}.  The first ratio we defined as $\mathcal{Y}{\scriptstyle_{\Psi}}$ $\equiv$ $\eta{\scriptstyle_{iaw}} \left\lvert \textbf{j}{\scriptstyle_{o}} \right\rvert^{2}$/$\dot{\Psi}$ and the second as $\mathcal{R}{\scriptstyle_{\Psi}}$ $\equiv$ $\left\lvert \textbf{j}{\scriptstyle_{o}} \cdot \delta \textbf{E} \right\rvert$/$\dot{\Psi}$.  Physically, $\mathcal{Y}{\scriptstyle_{\Psi}}$ and $\mathcal{R}{\scriptstyle_{\Psi}}$ are ratios of the rate of work per unit volume done by the waves on the plasma to the rate of energy dissipation per unit volume necessary to produce the increase in entropy.  Therefore, if either ratio is $\geq$ 1, then the magnitude of the rate of work per unit volume done on the particles by the fluctuating electric fields exceeds the dissipation rate necessary to produce the observed increase in entropy.  Meaning, if either ratio is $\geq$ 1 then the waves can provide more than enough energy dissipation to explain the shock dynamics.

%%  Discuss calculation of (*eta* |j|^2)/∆∑ vs. |(∂E . j)|/∆∑
\indent  Previous observations found a high relative occurrence of IAWs in the ramp regions of interplanetary collisionless shock waves \citep[e.g.,][]{wilsoniii07a}.  Therefore, we used the IAW dispersion relation to give our estimate of $\eta{\scriptstyle_{iaw}}$ for $\mathcal{Y}{\scriptstyle_{\Psi}}$.  The assumptions used to estimate \textbf{j}${\scriptstyle_{o}}$ are only valid near the shock ramp, so we only included wave bursts within roughly $\pm$10 seconds of the center of the shock ramp.  We linearly interpolated \textbf{j}${\scriptstyle_{o}}$ to the same time steps as $\delta$\textbf{E} prior to calculating $\mathcal{Y}{\scriptstyle_{\Psi}}$ and $\mathcal{R}{\scriptstyle_{\Psi}}$.  The details of and justifications for up-sampling \textbf{j}${\scriptstyle_{o}}$ are given in Appendix \ref{app:CurrentDensity}.

%%  Introduce Figure:  (*eta* |j|^2)/∆∑ vs. |(∂E . j)|/∆∑
\indent  Figure \ref{fig:comparedissrate} plots $\mathcal{Y}{\scriptstyle_{\Psi}}$ versus $\mathcal{R}{\scriptstyle_{\Psi}}$, calculated from the high frequency waves, for all the bow shock crossings observed.  The contours define regions containing a percentage of the total number of points shown, where the contour levels are defined in the inset box in the upper left-hand corner.  The vertical(horizontal) dashed line shows where $\mathcal{R}{\scriptstyle_{\Psi}}$($\mathcal{Y}{\scriptstyle_{\Psi}}$) $=$ 1.  The diagonal dashed line shows where $\mathcal{R}{\scriptstyle_{\Psi}}$ $=$ $\mathcal{Y}{\scriptstyle_{\Psi}}$.

%%  Discuss Figure:  (*eta* |j|^2)/∆∑ vs. |(∂E . j)|/∆∑
\indent  In every THEMIS bow shock crossing, we found $>$100 data points (on FGM time-steps) that satisfy $\mathcal{R}{\scriptstyle_{\Psi}}$ $\geq$ 1 and $\mathcal{Y}{\scriptstyle_{\Psi}}$ $\geq$ 1.  Note that this reduces to roughly $>$50 data points for $\mathcal{R}{\scriptstyle_{\Lambda}}$ or $\mathcal{Y}{\scriptstyle_{\Lambda}}$ (see Appendix \ref{app:EnergyDissipation} for more details).  The fact that every event has $>$100 individual time steps satisfying $\mathcal{R}{\scriptstyle_{\Psi}}$($\mathcal{Y}{\scriptstyle_{\Psi}}$) $\geq$ 1 implies that the waves can support the global dynamics of the shock structure for at least $\sim$0.8 seconds or scale lengths ranging from $\sim$10-50 km.  One can see from Figure \ref{fig:exampleEBSjEdjBSC} that this scale length is much larger than the typical gradient scale lengths observed in \textbf{B}${\scriptstyle_{o}}$ and/or \textbf{j}${\scriptstyle_{o}}$.  In addition, an examination of the DC-coupled electric field measurements (sampled at $\sim$128 sps) shows that their contribution to the wave energy dissipation, $\left\lvert \textbf{j}{\scriptstyle_{o}} \cdot \textbf{E}{\scriptstyle_{o}} \right\rvert$, was typically an order of magnitude lower than from the higher frequency AC-coupled measurements.  Therefore, we argue that the waves have the capacity to dominate the global energy budget of collisionless shock waves.

%%  Comparison between (*eta* |j|^2)/∆∑ and |(∂E . j)|/∆∑
\indent  The comparison between $\mathcal{R}{\scriptstyle_{\Psi}}$ and $\mathcal{Y}{\scriptstyle_{\Psi}}$ in Figure \ref{fig:comparedissrate} shows that the our assumption $\delta$\textbf{E} $\approx$ $\left( \overleftrightarrow{\mathbb{\eta}}{\scriptstyle_{_{iaw}}} \cdot \thickspace \textbf{j}{\scriptstyle_{o}} \right)$ for $\mathcal{Y}{\scriptstyle_{\Psi}}$ is too small by up to four orders of magnitude.  This agrees with a similar comparison found in Vlasov simulations \citep[e.g.,][]{petkaki06a, petkaki08a, yoon06a, yoon07a}, where the analytical estimates for $\eta{\scriptstyle_{iaw}}$ were found to be up to 2-3 orders of magnitude too small.  Even so, this plot shows that every shock crossing has multiple points with $\mathcal{Y}{\scriptstyle_{\Psi}}$ $\geq$ 1.  Therefore, $\eta{\scriptstyle_{iaw}} \left\lvert \textbf{j}{\scriptstyle_{o}} \right\rvert^{2}$ can be used as a lower bound for an estimate of the wave energy dissipation rate in a collisionless shock.

%%----------------------------------------------------------------------------------------
%%  Section:  Discussion
%%----------------------------------------------------------------------------------------
\section{Discussion}  \label{sec:Discussion}
%%  Summary
\indent  We present the first quantified measure of the energy dissipation rate, due to wave-particle interactions, in collisionless shocks.  The work presented herein can be summarized by the following points:
\begin{enumerate}
  \setlength{\itemsep}{0pt}    %% tighten up spacing between items
  \setlength{\parskip}{0pt}    %% tighten up spacing between items
  \item  Every bow shock crossing examined with available wave burst data from THEMIS showed large amplitude $\delta$\textbf{E} and $\delta$\textbf{B} high frequency ($\gtrsim$10 Hz) waves throughout the entire transition region and into the magnetosheath.  These high frequency ($\gtrsim$10 Hz) wave amplitudes can exceed $\delta$B $\sim$ 10 nT and $\delta$E $\sim$ 300 mV/m, though they are typically observed at $\delta$B $\sim$ 0.1-1.0 nT and $\delta$E $\sim$ 10-50 mV/m.
  \item  The high frequency ($\gtrsim$10 Hz) waves were composed of multiple modes, identified as ion-acoustic waves (IAWs), electron cyclotron drift instability (ECDI), trains of electrostatic solitary waves (ESWs), and electromagnetic whistler mode waves.  The low frequency ($<$10 Hz) observations were dominated by magnetosonic-whistler waves.
  \item  The high frequency ($\gtrsim$10 Hz) waves were found to have $\mid$$\delta \textbf{S}$$\mid$ in excess of 2000 $\mu$W m$^{-2}$, though typical values were between $\sim$ 1-10 $\mu$W m$^{-2}$.  They could produce resistivities $\eta{\scriptstyle_{iaw}}$ $>$ 9000 $\Omega$ m and energy dissipation rates $\left\lvert \textbf{j}{\scriptstyle_{o}} \cdot \delta \textbf{E} \right\rvert$ $>$ 3 $\mu$W m$^{-3}$.
  \item  The cursory examination of Wind and STEREO bow shock crossings showed similar wave modes and comparable electric field amplitudes.  Both spacecraft observe large amplitude ($\delta$E $\geq$ 100 mV/m) waves throughout the shock transition region and magnetosheath when ever wave burst data was available.  Thus, we conclude that these large amplitude waves are ubiquitous throughout the entire shock transition region and the magnetosheath.
  \item  The low frequency ($<$10 Hz) magnetosonic-whistler waves were also an ubiquitous mode upstream of the shock ramps, but we did not focus on them.  While their magnetic amplitudes can be very large (B${\scriptstyle_{o}}$ $>$ 30 nT), their contribution to the wave energy dissipation was typically an order of magnitude less than from the higher frequency modes, or $\left\lvert \textbf{j}{\scriptstyle_{o}} \cdot \textbf{E}{\scriptstyle_{o}} \right\rvert$ $\ll$ $\left\lvert \textbf{j}{\scriptstyle_{o}} \cdot \delta \textbf{E} \right\rvert$.
  \item  When we compared the wave energy dissipation rates, $\left\lvert \textbf{j}{\scriptstyle_{o}} \cdot \delta \textbf{E} \right\rvert$ and $\eta{\scriptstyle_{iaw}} \left\lvert \textbf{j}{\scriptstyle_{o}} \right\rvert^{2}$, to the values necessary to produce the observed increase in entropy, $\dot{\Psi}$ ($\equiv$ $\rho T \Delta \mathfrak{s} / \Delta t$), we found that every event has a majority of data points where $\mathcal{R}{\scriptstyle_{\Psi}}$ ($\equiv$ $\left\lvert \textbf{j}{\scriptstyle_{o}} \cdot \delta \textbf{E} \right\rvert$/$\dot{\Psi}$) $>$ 1 or $\mathcal{Y}{\scriptstyle_{\Psi}}$ ($\equiv$ $\eta{\scriptstyle_{iaw}} \left\lvert \textbf{j}{\scriptstyle_{o}} \right\rvert^{2}$/$\dot{\Psi}$) $>$ 1.  Moreover, the wave dissipation rates can greatly exceed the dissipation rates necessary to balance the nonlinear steepening of the shock ramp.  These results have the following implications:
  \begin{enumerate}
    \setlength{\itemsep}{0pt}    %% tighten up spacing between items
    \setlength{\parskip}{0pt}    %% tighten up spacing between items
    \item  The waves can provide more than enough energy dissipation to produce the observed increase in entropy across the shock ramp.
    \item  Therefore, the waves can provide enough energy dissipation to balance the nonlinear wave steepening leading to the shock itself.
    \item  More importantly, this implies that the efficiency of the wave energy dissipation need only be $\lesssim$ 0.01\% to regulate the global shock dynamics.
  \end{enumerate}
  \item  We performed an example calculation for the growth rates of the ECDI in every event to verify that the instability could reach sufficient amplitude before convecting into the downstream.  Our estimates showed that the waves could saturate in much less time than is necessary for them to convect across the shock foot alone.  Therefore, we can conclude that the waves can be driven by an instability upstream of the shock ramp and still provide sufficient energy dissipation before convecting downstream.
\end{enumerate}
These observations are the first results that quantitatively show that wave-particle interactions have the capacity to control the global dynamics of collisionless shock waves.  These results support recent observations \citep[e.g.,][]{wilsoniii07a, wilsoniii10a, wilsoniii12c} and simulations \citep[e.g.,][]{matsukiyo06b, comisel11a, muschietti13a} that suggest microphysical processes can dominate the global structure of low Mach number collisionless shocks.

%%  Efficiency of fields
%%  Physical Significance:  (*eta* |j|^2)/∆∑ and |(∂E . j)|/∆∑ > 1
\indent  Observations with $\mathcal{R}{\scriptstyle_{\Psi}}$ $>$ 1 or $\mathcal{Y}{\scriptstyle_{\Psi}}$ $>$ 1 are highly suggestive of the relative importance of wave-particle-driven energy dissipation.  The physical significance of observing $\mathcal{R}{\scriptstyle_{\Psi}}$ $\geq$ 1 or $\mathcal{Y}{\scriptstyle_{\Psi}}$ $\geq$ 1 implies that the waves are capable of providing more energy dissipation than is necessary to produce the observed increase in entropy.  This means that there is more work done by the wave electric fields on the particles in a unit volume than is necessary to balance the nonlinear wave steepening that leads to the shock formation.  More importantly, this implies that the waves need not be 100\% efficient when exchanging energy/moment with the particles to explain the observed increase in entropy.  In fact, the data corresponding to $\mathcal{R}{\scriptstyle_{\Psi}}$ in Figure \ref{fig:comparedissrate} show that the electromagnetic fluctuations need only be, at times, $<$ 0.01\% efficient to mediate the global shock transition.

%%  Other Dissipation Mechanisms
\indent  We emphasize the efficiency of the wave dissipation because we observe two other energy loss mechanisms in many of these shock waves.  For instance, all the THEMIS events satisfy M${\scriptstyle_{f}}$/M${\scriptstyle_{cr}}$ $>$ 1 (i.e., supercritical), implying ion reflection, which we observe in the analyzed shocks.  In addition, many of the THEMIS shock crossings observed have magnetosonic-whistler precursors, which can carry energy away from the shock ramp.  Therefore, at least some of the incident bulk flow kinetic energy may be lost through mechanisms other than wave-particle interactions.  Note that reflection is not, alone, an irreversible process.  However, we have quantitatively shown that these high frequency waves have the capacity to be the dominant energy dissipation mechanism.

%%  Irreversibility:  Reflected Ions
\indent  While particle reflection can be an energy loss mechanisms that collisionless shock waves use to balance nonlinear wave steepening, it cannot transform the incident bulk flow kinetic energy into thermal energy irreversibly through direct means.  For instance, if the reflected ions drive an instability that then stochastically scatters incident particles, then they can indirectly transform energy irreversibly through the instability.  As discussed in Section \ref{subsec:WaveModes}, we believe the observed high frequency waves are driven by unstable particle distributions.  The most likely source of free energy for many of the observed modes is due to the relative drift between the reflected ions and incident electrons and/or ions.

%%  Irreversibility:  Wave Dispersion
\indent  We now discuss whether the dispersive radiation of a magnetosonic-whistler precursor wave, due to nonlinear wave steepening, is an irreversible process.  The dispersive properties of magnetosonic-whistler waves result in the higher(smaller) frequency(wavelength) propagating faster than the lower frequencies.  Therefore, the shock ramp can appear to be a train of compressive magnetosonic-whistler waves with the highest frequency waves observed the farthest from the shock ramp.  One might think that the spatial spreading of frequency components is not directly irreversible.  However, one does not expect a radiated wave to return to its source without external influences.  In addition, if the radiated waves carry momentum/energy into the upstream (i.e., their group velocity exceeds the shock velocity), then they are directly removing momentum/energy from the shock.  Thus, until the radiated waves interact with the upstream medium and impart their momentum/energy to the plasma, it is not immediately obvious that this process is directly irreversible.

%%  Irreversibility:  Wave Dispersion (2)
\indent  If these waves are dispersively radiated and they carry energy away from the shock ramp, then they can indirectly transform energy (irreversibly) by either stochastically scattering particles directly or exciting waves that scatter the particles.  In the latter case, their large magnetic fluctuations can produce strong localized currents that drive electrostatic instabilities (e.g., IAWs) or they can compress the plasma to produce temperature anisotropy instabilities (e.g., whistler mode waves).  Previous observations have found that higher frequency electrostatic \citep[e.g.,][]{wilsoniii07a} and electromagnetic \citep[e.g.,][]{hull12a, wilsoniii13a} waves occur simultaneously with large amplitude magnetosonic-whistler waves.  Note that previous studies have found that the magnetosonic-whistlers can be generated either by dispersion \citep[e.g.,][]{sundkvist12a} or instabilities \citep[e.g.,][]{wilsoniii12c}.  

%%  Irreversibility:  Discussion
\indent  Note that in both of these scenarios, particle reflection or dispersive radiation, the end state is an irreversible transformation of energy at a microscopic scale through wave-particle interactions.  Whether the high frequency waves we observe throughout the transition region were driven by the free energy from the reflected ions or by the localized currents in the magnetosonic-whistlers is not the focus of this study.  The over abundance of potential energy dissipation that these high frequency waves can produce is the most important result in our study.  Theory and simulation have shown that the observed modes are capable of efficiently exchanging energy/momentum between particle species leading to an irreversible transformation of energy.

Therefore, we conclude that the observed wave modes have the capacity to dominate the global dynamics of collisionless shock waves.  \\

%%  \indent  \textbf{\textcolor{Magenta}{Need discussion about types of waves, how they dissipate energy, how they exchange momentum between particle species, what particles they interact with, etc....}}  \\

%%----------------------------------------------------------------------------------------
%%  Acknowledgements
%%----------------------------------------------------------------------------------------
{\normalsize \noindent \textbf{Acknowledgements}}  \\
%\begin{acknowledgments}
\indent  We would like to thank A.F.- Vi{\~n}as, D. Sundkvist, V.V. Krasnoselskikh, D. Bryant, D.A. Roberts, R. Lysak, and M.L. Goldstein for useful discussions of the fundamental physics involved in our study.  
%\end{acknowledgments}

%%----------------------------------------------------------------------------------------
%%  Bibliography
%%----------------------------------------------------------------------------------------
\singlespacing
%% => Shut off section counters
\setcounter{secnumdepth}{-1} 
\addtocontents{toc}{}	% Tighten spacing in Table of Contents
\section{\emph{References}} \label{sec:references} 
\renewcommand{\refname}{}  %%  shut off automatic section name
\renewcommand{\bibsep}{5pt}	% Tighten spacing between references

%%  \bibliography{my_bib_maker}

\clearpage
\appendix
\setcounter{secnumdepth}{5}
%%----------------------------------------------------------------------------------------
%%  Appendix:  Analysis Techniques
%%----------------------------------------------------------------------------------------
\section{Analysis Techniques}  \label{app:Methodology}
\indent  In this appendix we outline some of our analysis techniques.

%%  Downsample EFI to calculate S
\indent  For events where the EFI was sampling at $\sim$16,384 sps while the SCM was sampling at $\sim$8,192 sps, we down-sampled the EFI to the SCM time steps, after calibration.  Then these two fields, $\delta$\textbf{E} and $\delta$\textbf{B}, were used to estimate the Poynting flux, $\delta$\textbf{S} [$=$ ($\delta$\textbf{E}$\times \delta$\textbf{B})/$\mu{\scriptstyle_{o}}$], in temporal and frequency space.  $\delta$\textbf{S} was calculated after both $\delta$\textbf{E} and $\delta$\textbf{B} were filtered and after each was transformed or rotated into the desired reference frame or coordinate basis, respectively.

%%  Frequency space estimate of S
\indent  The frequency space estimate, $\mid$$\delta \tilde{\textbf{S}}$$\mid$, involves summing over the Fourier transformed time and frequency bins of $\delta$\textbf{S} to estimate the magnitude of the electromagnetic energy flux.  We calculated $\mid$$\delta \tilde{\textbf{S}}$$\mid$ through the following steps:  (1) create a Hanning window with N${\scriptstyle_{fft}}$ elements and multiply by $\delta$\textbf{B}${\scriptstyle_{i:k}}$ and $\delta$\textbf{E}${\scriptstyle_{i:k}}$, i.e., where $i$ and $k$ are the first and last abscissa of any given N${\scriptstyle_{fft}}$-element increment; (2) calculate the Fourier transform (FFT) of these products, $\delta \tilde{\textbf{E}}{\scriptstyle_{i:k}}$ and $\delta \tilde{\textbf{B}}{\scriptstyle_{i:k}}$, respectively; (3) calculate $\delta \tilde{\mathfrak{P}}{\scriptstyle_{i:k}}$ $=$ $\delta \tilde{\textbf{E}}{\scriptstyle_{i:k}} \times \delta \tilde{\textbf{B}}{\scriptstyle_{i:k}}^{*}$/(2$\mu{\scriptstyle_{o}}$) in frequency space; (4) calculate $\delta \tilde{\mathfrak{S}}{\scriptstyle_{m}}^{2}$ $=$ $\sum_{i}^{k}$ $\mid$$\delta \tilde{\mathfrak{P}}{\scriptstyle_{i:k}}$$\mid^{2}$; and (5) finally, $\mid$$\delta \tilde{\textbf{S}}$$\mid$ $=$ $\delta \tilde{\mathfrak{S}}{\scriptstyle_{m}}$ $\Delta$f, where $\Delta$f is the bandwidth for any given N${\scriptstyle_{fft}}$-element increment.

%%  Purpose of calculating S
\indent  The purpose of calculating $\mid$$\delta \tilde{\textbf{S}}$$\mid$ was to examine the upper bound on $\delta$\textbf{S}.  This calculation decomposes the signal into frequency space before performing the cross-product, which ensures that matching frequency components were multiplied together.  However, $\mid$$\delta \tilde{\textbf{S}}$$\mid$ will return positive definite values even if $\delta$\textbf{S} is periodically varying.

%%  Shock Parameter Uncertainties
\indent  Before moving on, we will explain the uncertainties shown in Table \ref{tab:shockparams}.  For each shock crossing, we selected a time range for the upstream and downstream regions with an equal number, $N$, of particle moment time-steps, t${\scriptstyle_{i}}$.  The time ranges were selected ``by-eye'' and were chosen by attempting to minimize changes in the particle velocity moments and magnetic field vectors to the Rankine-Hugoniot relations.  The Rankine-Hugoniot relations are numerically minimized for each input set of particle velocity moments and magnetic field vectors, \textbf{Y}${\scriptstyle_{ij}}$, where $i$ is the abscissa for the time-steps and $j$ defines the plasma parameter (e.g., number density).  The uncertainties for V${\scriptstyle_{shn}}$ and U${\scriptstyle_{shn}}$ depend upon $N$ and tend to decrease with increasing $N$.  Therefore the uncertainties for V${\scriptstyle_{shn}}$ and U${\scriptstyle_{shn}}$ are given by the standard deviation of the mean or $\sigma{\scriptstyle_{x}}/\sqrt{N}$.  The rest of the parameter uncertainties are given by the standard deviation or $\sigma{\scriptstyle_{x}}$.

%%========================================================================================
%%  Table:  Mach Parameters
%%========================================================================================
\begin{table}[htb]
  \centering
  \caption{Critical Mach Number Ratios}
  \label{tab:machparams}
    \begin{tabular}{| l | c | c | c | c |}
      \hline
      \textbf{Date} & M${\scriptstyle_{f}}$/M${\scriptstyle_{cr}}$ & M${\scriptstyle_{f}}$/M${\scriptstyle_{w}}$ & M${\scriptstyle_{f}}$/M${\scriptstyle_{gr}}$ & M${\scriptstyle_{f}}$/M${\scriptstyle_{nw}}$   \\
      \hline
      {\small 2009-07-13 [1st Crossing]} & 2.97 $\pm$ 0.10 & 0.20 $\pm$ 0.02 & 0.15 $\pm$ 0.01 & 0.14 $\pm$ 0.01   \\
      {\small 2009-07-21 [1st Crossing]} & 1.45 $\pm$ 0.13 & 0.15 $\pm$ 0.02 & 0.12 $\pm$ 0.01 & 0.11 $\pm$ 0.01   \\
      {\small 2009-07-23 [1st Crossing]} & 2.50 $\pm$ 0.06 & 1.24 $\pm$ 0.48 & 0.95 $\pm$ 0.37 & 0.88 $\pm$ 0.34   \\
      {\small 2009-07-23 [2nd Crossing]} & 2.98 $\pm$ 0.07 & 5.24 $\pm$ 5.19 & 4.04 $\pm$ 4.00 & 3.71 $\pm$ 3.67   \\
      {\small 2009-07-23 [3rd Crossing]} & 2.61 $\pm$ 0.07 & 0.25 $\pm$ 0.02 & 0.19 $\pm$ 0.02 & 0.18 $\pm$ 0.02   \\
      {\small 2009-09-26 [1st Crossing]} & 4.61 $\pm$ 0.26 & 0.45 $\pm$ 0.13 & 0.35 $\pm$ 0.10 & 0.32 $\pm$ 0.09   \\
      {\small 2011-10-24 [1st Crossing]} & 2.02 $\pm$ 0.04 & 1.00 $\pm$ 0.75 & 0.77 $\pm$ 0.58 & 0.71 $\pm$ 0.53   \\
      {\small 2011-10-24 [2nd Crossing]} & 1.75 $\pm$ 0.01 & 2.58 $\pm$ 1.48 & 1.99 $\pm$ 1.14 & 1.83 $\pm$ 1.05   \\
      \hline
    \end{tabular}
\end{table}
%%========================================================================================
%%  Table:  Mach Parameters
%%========================================================================================

%%  Critical Mach Numbers
\indent  We have examined \textbf{\textcolor{Red}{8}} bow shock crossings so far.  For each crossing, we calculated M${\scriptstyle_{f}}$ and then determined whether the shock was supercritical or not.  The first critical Mach number, M${\scriptstyle_{cr}}$, determines the theoretical value of M${\scriptstyle_{f}}$, above which, the shock can no longer dissipate sufficient energy using resistivity or dispersion alone \citep[][]{edmiston84}.  There are three whistler critical Mach numbers \citep[][]{krasnoselskikh02a} and are defined as:  M${\scriptstyle_{w}}$ corresponds to the maximum Mach number at which a linear whistler can phase stand with respect to the shock front;  M${\scriptstyle_{gr}}$ is the maximum Mach number which would allow a whistler wave to carry energy into the upstream;  and M${\scriptstyle_{nw}}$ is the maximum Mach number for which a stationary shock front solution can be found, above which, the wave breaks.  The Mach ratios can be seen in Table \ref{tab:machparams}.

%%  Explain Mach # Table
\indent  The values shown in Table \ref{tab:machparams} were calculated using the \textbf{Y}${\scriptstyle_{ij}}$ inputs.  We calculated each Mach number with associated standard deviations, giving us M${\scriptstyle_{k}}$ $\pm$ $\sigma{\scriptstyle_{k}}$.  We then calculated the ratio of M${\scriptstyle_{f}}$ to each of the critical Mach numbers discussed above and used the standard technique for the propagation of uncertainties [i.e., $\delta$q/$\mid$q$\mid$ $=$ (($\delta$x/x)$^{2}$ $+$ ... $+$ ($\delta$z/z)$^{2}$)$^{1/2}$, where q $=$ q(x,...,z)].  As one can see, every shock examined herein satisfies M${\scriptstyle_{f}}$/M${\scriptstyle_{cr}}$ $>$ 1, and is therefore supercritical.

%%  Poynting's theorem
\indent  From Poynting's theorem (see Appendix \ref{app:EnergyDissipation}, Equation \ref{eq:quantenergydiss_0}), we can see that $\left\lvert \textbf{j} \cdot \textbf{E} \right\rvert$ ($\sim$ $\left\lvert \textbf{j}{\scriptstyle_{o}} \cdot \delta \textbf{E} \right\rvert$) is an energy sink/loss (see Appendix \ref{app:CurrentDensity} for estimation of \textbf{j}${\scriptstyle_{o}}$), called ohmic dissipation.  Therefore, we will use this as our first estimate of wave energy dissipation.  After some assumptions, we can rewrite this term as $\eta{\scriptstyle_{iaw}} \left\lvert \textbf{j}{\scriptstyle_{o}} \right\rvert^{2}$, which will be our second estimate of energy dissipation.  We use two methods to estimate the wave dissipation rate because each method relies upon assumptions and each method has advantages/disadvantages (see Appendix \ref{app:EnergyDissipation}).  

%%  Thermodynamics
\indent  The most important analysis in this paper is our quantitative comparison between the macroscopic and wave energy (microscopic) dissipation rates.  To do this, we calculated the ratio of the wave energy dissipation rate to the energy dissipation rate necessary to produce the observed increase in specific entropy density per unit time, or ($\rho$T)$\Delta \mathfrak{s}$/$\Delta$t ($\equiv$ $\dot{\Psi}$).  These ratios, $\mathcal{Y}{\scriptstyle_{\Psi}}$ $\equiv$ $\eta{\scriptstyle_{iaw}} \left\lvert \textbf{j}{\scriptstyle_{o}} \right\rvert^{2}$/$\dot{\Psi}$ and $\mathcal{R}{\scriptstyle_{\Psi}}$ $\equiv$ $\left\lvert \textbf{j}{\scriptstyle_{o}} \cdot \delta \textbf{E} \right\rvert$/$\dot{\Psi}$, were calculated to estimate the wave energy (microscopic) dissipation rate (see Equations \ref{eq:quantenergydiss_3a} and \ref{eq:quantenergydiss_3b}) relative to the macroscopic energy dissipation rate.

%%  |Jo.Eo| << |Jo.∂E|
\indent  Though we did examine the low frequency ($\leq$10 Hz) fields, we found that they consistently showed E${\scriptstyle_{o}}$ $\ll$ $\delta$E and $\left\lvert \textbf{j}{\scriptstyle_{o}} \cdot \textbf{E}{\scriptstyle_{o}} \right\rvert$ $\ll$ $\left\lvert \textbf{j}{\scriptstyle_{o}} \cdot \delta \textbf{E} \right\rvert$.  Therefore, we have assumed that the ratios $\mathcal{Y}{\scriptstyle_{\Psi}}$ and $\mathcal{R}{\scriptstyle_{\Psi}}$ are dominated by the high frequency ($\gtrsim$10 Hz) waves that we focus on in this study.

%%----------------------------------------------------------------------------------------
%%  Appendix:  High Frequency Wave Observations
%%----------------------------------------------------------------------------------------
\section{High Frequency Wave Observations}  \label{app:HFWaveObservations}

%%----------------------------------------------------------------------------------------
%%  Subsection:  THEMIS Waves: Examples
%%----------------------------------------------------------------------------------------
\subsection{THEMIS Waves: Examples}  \label{subapp:THEMISWaveObservations}

%%  Outline Wave Modes
\indent  The first instability we will discuss involves the nonlinear electrostatic fluctuations.  We will then discuss observations of ESWs, high frequency whistler mode waves, and finally ion-acoustic waves (IAWs).

%%++++++++++++++++++++++++++++++++++++++++++++++++++++++++++++++++++++++++++++++++++++++++
%% Image:  ECDI Example
%%++++++++++++++++++++++++++++++++++++++++++++++++++++++++++++++++++++++++++++++++++++++++
\begin{figure}[!htb]
  \centering
    \vspace{-0pt}
    {\includegraphics[trim = 0mm 0mm 0mm 0mm, clip, width=11cm]{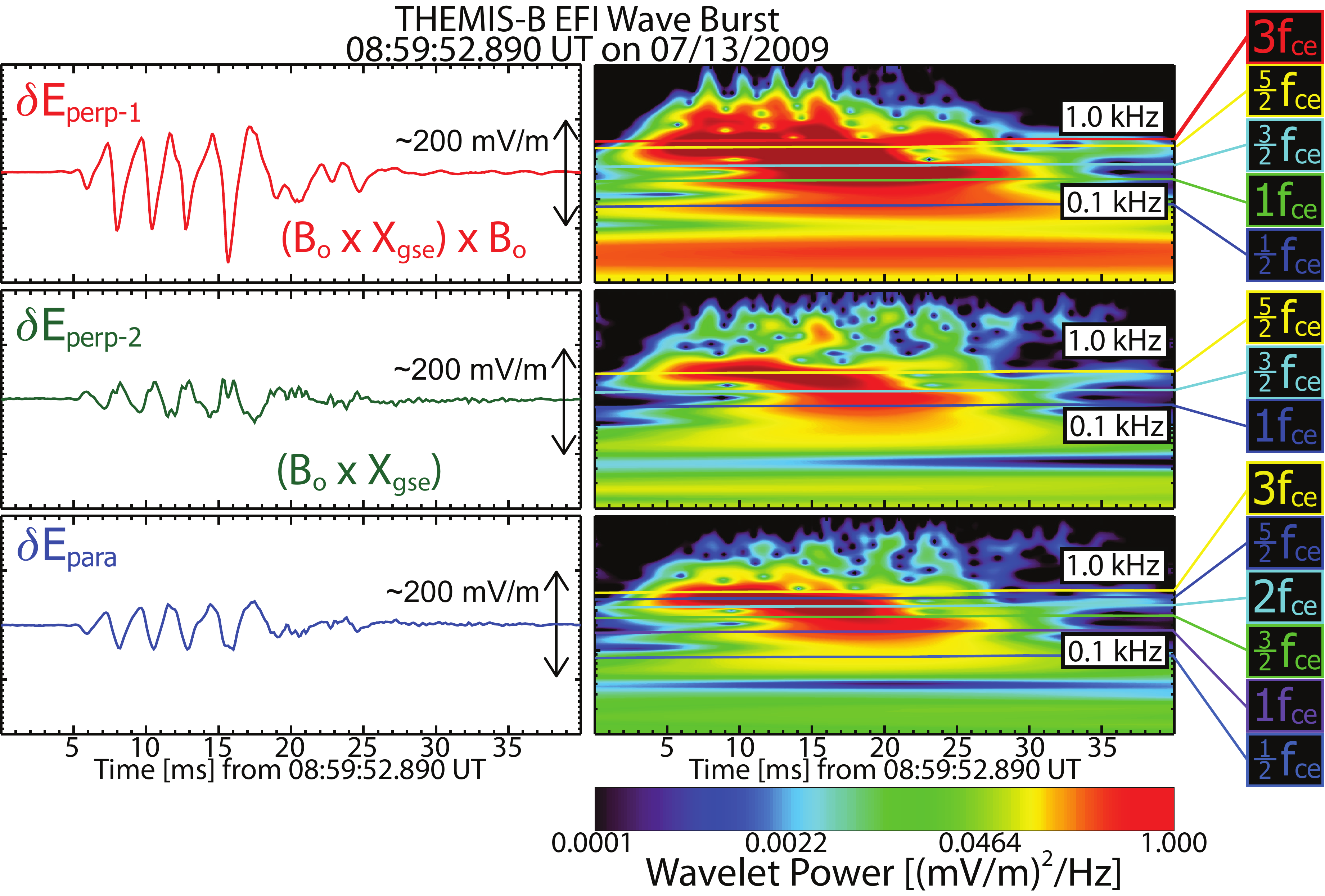}}
    \captionsetup{width=12cm}   %%  adjust caption width
    \caption[ECDI Example]{An example of an ECDI waveform and corresponding wavelet transforms observed by the EFI instrument in wave burst mode for a 40 ms time window near the shock ramp on 2009-07-13.  The right-hand column shows the corresponding wavelet transforms \citep[][]{torrence98b} with the frequency on a logarithmic scale from $\sim$10-4000 Hz.}
    \label{fig:exampleecdi}
    \vspace{-10pt}
\end{figure}
%%++++++++++++++++++++++++++++++++++++++++++++++++++++++++++++++++++++++++++++++++++++++++
%% Image:  ECDI Example
%%++++++++++++++++++++++++++++++++++++++++++++++++++++++++++++++++++++++++++++++++++++++++

\indent  \citet[][]{hull06a} described these nonlinear fluctuations as electrostatic IAWs while \citet[][]{wilsoniii10a} argued they were consistent with the electron cyclotron drift instability (ECDI) \citep[e.g.,][]{muschietti13a}.  The ECDI fluctuations are consistent with a mixture of Doppler shifted IAWs and electron cyclotron harmonics at integer and half-integer harmonics of f${\scriptstyle_{ce}}$.  They are driven unstable by the relative drift between incident electrons and shock-reflected ions \citep[e.g.,][]{matsukiyo06b, muschietti13a}.  In this paper, we show evidence to support our hypothesis that these fluctuations are consistent with the ECDI.

%%==================================
%%  ECDI:  Intro Example Figure
%%==================================
\indent  Figure \ref{fig:exampleecdi} shows an example of an ECDI waveform observed by THEMIS-B near the bow shock ramp on 2009-07-13.  The left-hand column shows the three electric field components in field-aligned coordinates (FACs) with corresponding amplitudes defined by the black arrows at the right-hand side of each panel.  Each wavelet is shown on the same color-scale range defined by the color bar at the bottom of the right-hand column.  Overlaid on the wavelets are color-coded lines showing integer and half-integer harmonics of f${\scriptstyle_{ce}}$.

%%  ECDI:  Discuss Example
\indent  The fluctuations have waveform and frequency spectrum characteristics similar to previous observations \citep[e.g.,][]{hull06a, wilsoniii10a}.  However, this wave shows properties consistent with the ECDI-driven waves described by \citet[][]{wilsoniii10a}, not simple Doppler shifted IAWs \citep[e.g., see Figure 10 in][]{wilsoniii10a}.  The shared properties include:  (1) asymmetric oscillation of $\delta$\textbf{E} about a mean value [i.e., may imply a net potential drop]; (2) significant amplitudes perpendicular to \textbf{B}${\scriptstyle_{o}}$; (3) significant amplitudes parallel to the shock normal vector (not shown); and (4) power focused at integer and half-integer harmonics of f${\scriptstyle_{ce}}$.

%%  ECDI:  Importance
\indent  ECDI-driven waves are important for shock physics because they are capable of resonantly interacting with the bulk of the ion distribution and preferentially heating the electrons perpendicular to \textbf{B}${\scriptstyle_{o}}$ \citep[][]{forslund70b, forslund72a, lampe72a}.  More recent work has shown that the ECDI can produce a suprathermal tail on the ion distribution and strongly heat the electrons \citep[][]{muschietti13a}.  This is accomplished through the following process:  (1) the ECDI is excited at multiple harmonics of f${\scriptstyle_{ce}}$ by removing bulk kinetic energy from the reflected ions; (2) these electrostatic fluctuations interact with the electrons and trap some of the reflected ions; (3) the wave amplitudes decrease, where higher harmonics experience more damping than lower harmonics; thus allowing (4) the waves to exchange energy and momentum between particle species irreversibly.

%%++++++++++++++++++++++++++++++++++++++++++++++++++++++++++++++++++++++++++++++++++++++++
%% Image:  Whistler Example
%%++++++++++++++++++++++++++++++++++++++++++++++++++++++++++++++++++++++++++++++++++++++++
\begin{figure}[!htb]
  \centering
    \vspace{-10pt}
    {\includegraphics[trim = 0mm 0mm 0mm 0mm, clip, width=10cm]{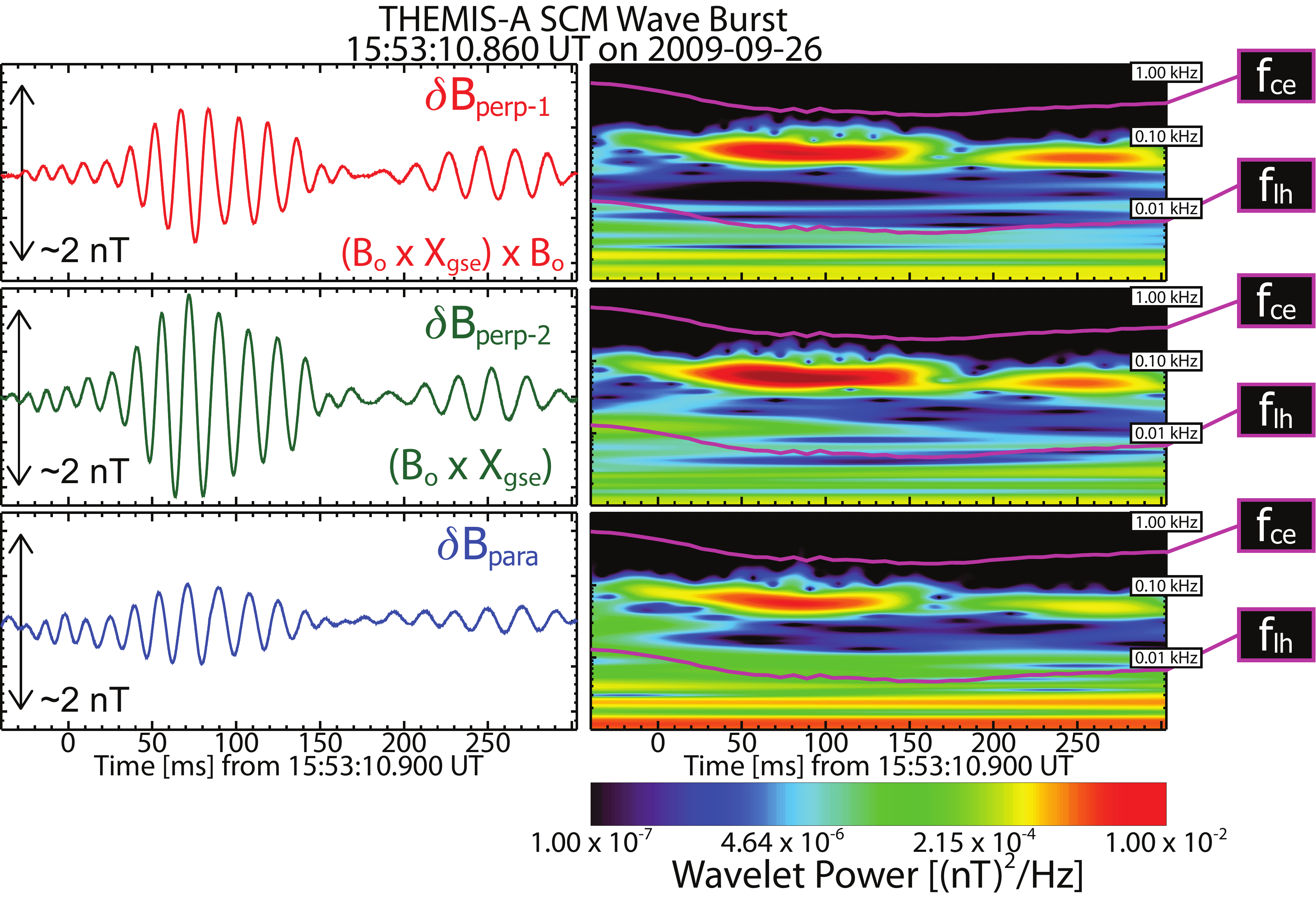}}
    \captionsetup{width=13cm}   %%  adjust caption width
    \caption[Whistler Example]{An example of a typical high frequency whistler mode wave observed by the SCM instrument in wave burst mode for a $\sim$340 ms time window near the shock ramp on 2009-09-26 (gray shaded timespan in Figure \ref{fig:exampleEBSjEdjBSC}).  The format is similar to that of Figure \ref{fig:exampleecdi} except that this shows three magnetic field components, the frequency ranges from $\sim$1-1000 Hz, and only f${\scriptstyle_{lh}}$ and f${\scriptstyle_{ce}}$ are shown.}
    \label{fig:examplewhistler}
    \vspace{-10pt}
\end{figure}
%%++++++++++++++++++++++++++++++++++++++++++++++++++++++++++++++++++++++++++++++++++++++++
%% Image:  Whistler Example
%%++++++++++++++++++++++++++++++++++++++++++++++++++++++++++++++++++++++++++++++++++++++++

%%==================================
%%  Whistler:  Intro Example Figure
%%==================================
\indent  Figure \ref{fig:examplewhistler} shows an example of a typical high frequency (f${\scriptstyle_{lh}}$ $\ll$ f $<$ f${\scriptstyle_{ce}}$) electromagnetic whistler mode wave.  Previous observations of these modes found that they propagate at small angles relative to \textbf{B}${\scriptstyle_{o}}$ \citep[e.g.,][]{hull12a, wilsoniii13a}, as illustrated by the relatively small ratios of $\delta$B${\scriptstyle_{\parallel}}$/$\delta$B${\scriptstyle_{\perp}}$ in Figure \ref{fig:examplewhistler}.  The waves are often observed as short duration ($\sim$few 10's to 100's of ms) bursty wave packets and can have large amplitudes ($\delta$B $>$ few nT).  The corresponding electric field amplitudes for this example ranged from $\sim$few mV/m up to $\sim$30 mV/m.

%%  Whistler:  Importance
\indent  High frequency electromagnetic whistler mode waves are commonly observed in and around the ramp region \citep[e.g.,][]{hull12a} and downstream \citep[e.g.,][]{wilsoniii13a} of collisionless shocks.  The source of these modes is thought to be either an electron temperature anisotropy (T${\scriptstyle_{\perp}}$/T${\scriptstyle_{\parallel}}$ $>$ 1) instability  \citep[e.g.,][]{kennel66a} or an electron heat flux instability \citep[e.g.,][]{gary94a}.  These modes are important because they are thought to regulate the electron heat flux and electron halo temperature anisotropy in the solar wind.  At shocks, these modes are important because they can efficiently exchange energy/momentum between electrons and ions and can couple to multiple wave modes \citep{dyrud06}.

%%  Whistler:  Relevance
\indent  Therefore, whistler mode waves have multiple pathways to transform energy from electromagnetic to kinetic or vice versa.  Note, however, that these modes are observed less often and their contribution to $\left\lvert \textbf{j}{\scriptstyle_{o}} \cdot \delta \textbf{E} \right\rvert$ is often much smaller (e.g., see Figure \ref{fig:exampleEBSjEdjBSC}) than the electrostatic modes.  This is not to say these modes are unimportant because previous studies have shown that they can have a significant influence on the halo energy electrons \citep[e.g.,][]{gary94a, wilsoniii13a}.

\begin{figure}[!htb]
  \centering
    \vspace{-10pt}
    {\includegraphics[trim = 0mm 0mm 0mm 0mm, clip, width=10cm]{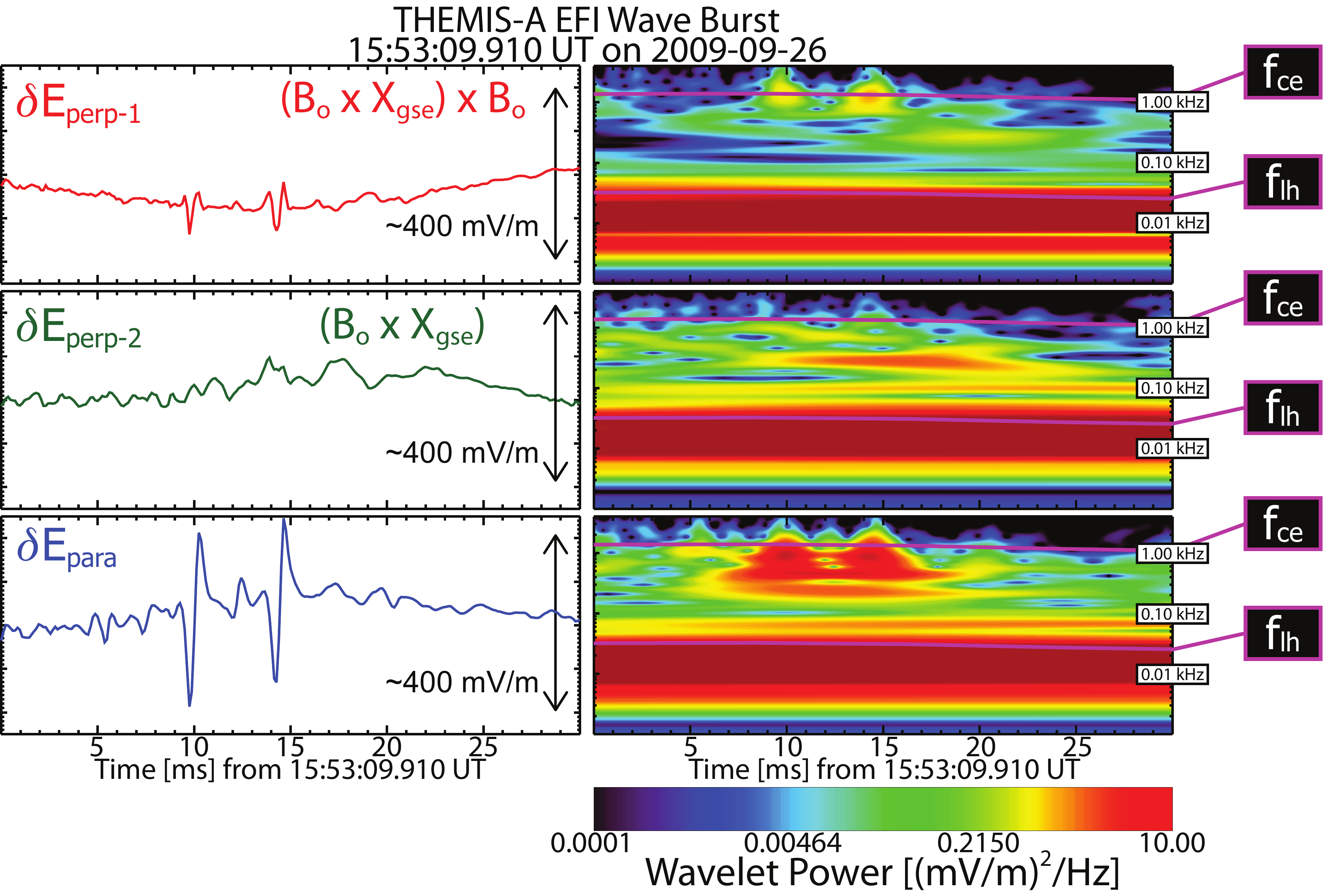}}
    \captionsetup{width=13cm}   %%  adjust caption width
    \caption[ESWs Example]{An example of two ESWs observed by the EFI instrument in wave burst mode for a $\sim$30 ms time window near the shock ramp on 2009-09-26 (magenta shaded timespan in Figure \ref{fig:exampleEBSjEdjBSC}).  The frequency axis of the wavelet plots range from $\sim$1-1000 Hz.  The format is similar to that of Figures \ref{fig:exampleecdi} and \ref{fig:examplewhistler}.}
    \label{fig:exampleESWs}
    \vspace{-10pt}
\end{figure}

%%==================================
%%  ESWs:  Intro Example Figure
%%==================================
\indent  Figure \ref{fig:exampleESWs} shows an example of two large amplitude ESWs observed at the peak of the shock overshoot (see magenta region in Figure \ref{fig:exampleEBSjEdjBSC}).  The low frequency ($\lesssim$40 Hz) signal, on which the ESWs bipolar signatures are superposed, is most likely artificial and should be ignored.  The fluctuations have waveform and frequency spectrum characteristics similar to previous observations of ESWs \citep[e.g.,][]{bale98b, wilsoniii07a, wilsoniii10a}.  ESWs are often too short in duration for many electric field detectors to resolve.  Even at $\sim$8,192 sps, these two examples are nearly under-sampled.  The same fluctuations observed at $\sim$16,384 sps (not shown), by comparison, show smooth and continuous bipolar pulses in $\delta$E${\scriptstyle_{\parallel}}$.

%%++++++++++++++++++++++++++++++++++++++++++++++++++++++++++++++++++++++++++++++++++++++++
%% Image:  ESWs Example
%%++++++++++++++++++++++++++++++++++++++++++++++++++++++++++++++++++++++++++++++++++++++++
%%++++++++++++++++++++++++++++++++++++++++++++++++++++++++++++++++++++++++++++++++++++++++
%% Image:  ESWs Example
%%++++++++++++++++++++++++++++++++++++++++++++++++++++++++++++++++++++++++++++++++++++++++

%%  ESWs Importance
\indent  ESWs can be driven unstable by electron beams \citep[e.g.,][]{cattell05a, ergun98d, franz05a}, modified two-stream instability (MTSI) \citep[e.g.,][]{matsukiyo06b}, etc.  ESWs are one of the more important modes because they can trap incident electrons \citep[e.g.,][]{dyrud06, lu08}, heat ions \citep[e.g.,][]{ergun98d}, and/or couple to (or directly cause) the growth of IAWs \citep[e.g.,][]{dyrud06}, whistler mode waves \citep{lu08}, and/or electron acoustic waves \citep[e.g.,][]{matsukiyo06b}.  Thus, solitary waves can directly heat and/or scatter particles or they can indirectly cause these effects through the generation of, or coupling to, secondary waves.  Because ESWs act like clumps of positive charge, they can efficiently scatter incident ions.  Previous observations have shown that a train of ESWs can increase T${\scriptstyle_{\perp, i}}$ by as much as the total initial thermal energy \citep{ergun98d}.

%%----------------------------------------------------------------------------------------
%%  Subsection:  THEMIS Waves: Properties
%%----------------------------------------------------------------------------------------
\subsection{THEMIS Waves: Properties}  \label{subapp:WaveFieldProperties}

\indent  In this appendix we present a summary of our statistical analysis of the wave amplitudes to help illustrate that they are important, large, and their relevance to shock physics.

%%  Intro Table:  Wave Property Statistics
\indent  From our THEMIS bow shock crossings, we have 731140 data points up-sampled to the $\delta$\textbf{E} time stamps (justification given in Appendix \ref{app:CurrentDensity}).  From those data points, we found the statistics shown in Table \ref{tab:WavePropertyStatistics}.  The columns are defined in the following order:  (1) parameter name and units; (2) minimum value; (3) maximum value; (4) mean or average value, $\langle$X$\rangle$; (5) standard deviation, $\sigma{\scriptstyle_{x}}$; (6) standard deviation of the mean, $\sigma{\scriptstyle_{x}}$/$\sqrt{N}$; and (7) the number of points used, N.  In Table \ref{tab:WavePropertyStatistics} we use the following definitions for brevity:  $\mathfrak{B}$ $\equiv$ $\mid$$\delta$\textbf{B}$\mid$/$\langle \mid$\textbf{B}${\scriptstyle_{o}}$$\mid \rangle{\scriptstyle_{up}}$, $\mathfrak{b}$ $\equiv$ $\mid$\textbf{B}${\scriptstyle_{o}}$$\mid$/$\langle \mid$\textbf{B}${\scriptstyle_{o}}$$\mid \rangle{\scriptstyle_{up}}$, $\mathcal{W}{\scriptstyle_{B}}$ $\equiv$ $\mid$$\delta$\textbf{B}$\mid^{2}$/(2$\mu{\scriptstyle_{o}}$), $\mathcal{W}{\scriptstyle_{E}}$ $\equiv$ $\varepsilon{\scriptstyle_{o}} \mid$$\delta$\textbf{E}$\mid^{2}$/2, $\mathfrak{E}$ $\equiv$ $\mid$$\delta$\textbf{E}$\mid$/(c $\langle \mid$\textbf{B}${\scriptstyle_{o}}$$\mid \rangle{\scriptstyle_{up}}$), $\mathcal{S}{\scriptstyle_{\eta}}$ $\equiv$ $\eta{\scriptstyle_{iaw}} \left\lvert \textbf{j}{\scriptstyle_{o}} \right\rvert^{2}$, $\mathfrak{R}{\scriptstyle_{\eta}}$ $\equiv$ $\left\lvert \textbf{j}{\scriptstyle_{o}} \cdot \delta \textbf{E} \right\rvert$, $\mathcal{R}{\scriptstyle_{\Psi}}$ $\equiv$ $\left\lvert \textbf{j}{\scriptstyle_{o}} \cdot \delta \textbf{E} \right\rvert$/$\dot{\Psi}$, and $\mathcal{Y}{\scriptstyle_{\Psi}}$ $\equiv$ $\eta{\scriptstyle_{iaw}} \left\lvert \textbf{j}{\scriptstyle_{o}} \right\rvert^{2}$/$\dot{\Psi}$ (defined in Appendix \ref{app:EnergyDissipation}).

%%========================================================================================
%%  Table:  Wave Property Statistics
%%========================================================================================
\begin{table}[htb]
  \centering
  \caption{Wave Property Statistics}
  \label{tab:WavePropertyStatistics}
    \begin{tabular}{| c | c | c | c | c | c | c |}
      \hline
      \textbf{Type} & Min. & Max. & $\langle$X$\rangle$ & $\sigma{\scriptstyle_{x}}$ & $\sigma{\scriptstyle_{x}}$/$\sqrt{N}$ & N  \\
      \hline
      $\mid$$\delta$\textbf{E}$\mid$ [mV m$^{-1}$]               & 4.16$\times$10$^{-02}$ & 2.99$\times$10$^{+02}$ & 1.34$\times$10$^{+01}$ & 1.85$\times$10$^{+01}$ & 2.16$\times$10$^{-02}$ & 730855  \\
      $\mid$$\delta$\textbf{B}$\mid$ [nT]                        & 5.60$\times$10$^{-04}$ & 1.00$\times$10$^{+01}$ & 2.53$\times$10$^{-01}$ & 4.53$\times$10$^{-01}$ & 5.30$\times$10$^{-04}$ & 731136  \\
      $\mid$$\delta$\textbf{S}$\mid$ [$\mu$W m$^{-2}$]           & 4.93$\times$10$^{-05}$ & 2.08$\times$10$^{+03}$ & 3.13$\times$10$^{+00}$ & 1.54$\times$10$^{+01}$ & 1.80$\times$10$^{-02}$ & 730812  \\
      $\mid$$\delta \tilde{\textbf{S}}$$\mid$ [$\mu$W m$^{-2}$]  & 7.26$\times$10$^{-03}$ & 2.89$\times$10$^{+03}$ & 1.95$\times$10$^{+01}$ & 7.65$\times$10$^{+01}$ & 8.96$\times$10$^{-02}$ & 728047  \\
      \hline
      $\mathfrak{B}$ [unitless]                                  & 4.74$\times$10$^{-05}$ & 1.45$\times$10$^{+00}$ & 2.77$\times$10$^{-02}$ & 4.18$\times$10$^{-02}$ & 4.88$\times$10$^{-05}$ & 731136  \\
      $\mathfrak{b}$ [unitless]                                  & 1.92$\times$10$^{-01}$ & 1.64$\times$10$^{+01}$ & 3.70$\times$10$^{+00}$ & 1.62$\times$10$^{+00}$ & 1.89$\times$10$^{-03}$ & 731140  \\
      $\mathfrak{E}$ [unitless]                                  & 3.66$\times$10$^{-05}$ & 3.29$\times$10$^{-01}$ & 7.06$\times$10$^{-03}$ & 1.29$\times$10$^{-02}$ & 1.51$\times$10$^{-05}$ & 730855  \\
      \hline
      $\mathcal{W}{\scriptstyle_{B}}$ [$\mu$W m$^{-3}$]          & 1.25$\times$10$^{-13}$ & 3.99$\times$10$^{-05}$ & 1.07$\times$10$^{-07}$ & 6.79$\times$10$^{-07}$ & 7.94$\times$10$^{-10}$ & 731136  \\
      $\mathcal{W}{\scriptstyle_{E}}$ [$\mu$W m$^{-3}$]          & 7.67$\times$10$^{-15}$ & 3.96$\times$10$^{-07}$ & 2.31$\times$10$^{-09}$ & 1.03$\times$10$^{-08}$ & 1.20$\times$10$^{-11}$ & 730855  \\
      \hline
      $\nu{\scriptstyle_{iaw}}$ [coll. s$^{-1}$]                 & 3.72$\times$10$^{-05}$ & 2.48$\times$10$^{+03}$ & 5.75$\times$10$^{+00}$ & 3.30$\times$10$^{+01}$ & 3.86$\times$10$^{-02}$ & 730756  \\
      $\eta{\scriptstyle_{iaw}}$ [$\Omega$ m]                    & 6.52$\times$10$^{-05}$ & 9.38$\times$10$^{+03}$ & 1.67$\times$10$^{+01}$ & 1.13$\times$10$^{+02}$ & 1.33$\times$10$^{-01}$ & 730756  \\
      \hline
      $\mathfrak{R}{\scriptstyle_{\eta}}$ [$\mu$W m$^{-3}$]      & 2.85$\times$10$^{-11}$ & 3.98$\times$10$^{+00}$ & 1.41$\times$10$^{-02}$ & 6.11$\times$10$^{-02}$ & 7.14$\times$10$^{-05}$ & 730855  \\
      $\mathcal{S}{\scriptstyle_{\eta}}$ [$\mu$W m$^{-3}$]       & 1.08$\times$10$^{-15}$ & 5.15$\times$10$^{-01}$ & 1.94$\times$10$^{-04}$ & 4.61$\times$10$^{-03}$ & 5.39$\times$10$^{-06}$ & 730756  \\
      \hline
      $\mathcal{Y}{\scriptstyle_{\Psi}}$ [unitless]              & 2.76$\times$10$^{-12}$ & 8.75$\times$10$^{+03}$ & 5.99$\times$10$^{+00}$ & 8.72$\times$10$^{+01}$ & 1.02$\times$10$^{-01}$ & 730756  \\
      $\mathcal{R}{\scriptstyle_{\Psi}}$ [unitless]              & 1.10$\times$10$^{-05}$ & 1.59$\times$10$^{+05}$ & 8.40$\times$10$^{+02}$ & 2.80$\times$10$^{+03}$ & 3.27$\times$10$^{+00}$ & 730855  \\
      \hline
    \end{tabular}
\end{table}
%%========================================================================================
%%  Table:  Wave Property Statistics
%%========================================================================================

%%  Discuss Table Results
\indent  One can see that there are a wide range of wave amplitudes with high frequency (i.e., AC-coupled) electric and magnetic fields exceeding 250 mV/m and 10 nT, respectively.  The magnetic fluctuations relative to an upstream averaged magnetic field magnitude, $\mid$\textbf{B}${\scriptstyle_{o}}$$\mid$/$\langle \mid$\textbf{B}${\scriptstyle_{o}}$$\mid \rangle{\scriptstyle_{up}}$ (FGM) can easily exceed $\sim$ 10 and $\mid$$\delta$\textbf{B}$\mid$/$\langle \mid$\textbf{B}${\scriptstyle_{o}}$$\mid \rangle{\scriptstyle_{up}}$ (SCM) can exceed $\sim$ 1.  The peak values for $\mid$$\delta \textbf{S}$$\mid$ and $\mid$$\delta \tilde{\textbf{S}}$$\mid$ can exceed 2000 $\mu$W m$^{-2}$.  These values are incredibly large, comparable to the energy fluxes needed to drive the terrestrial aurora \citep[e.g.,][]{angelopoulos02a, wygant00a} and larger than the largest values observed for whistler mode waves in the radiation belts \citep[e.g., see example in][]{wilsoniii11e}.  Such large values illustrate the relative importance of these high frequency waves in collisionless shock dissipation.

%%  Introduce Growth vs. Convection Times
\indent  To verify that these fluctuations can indeed result from instabilities, we will show example growth rate estimates for the ECDI \citep[e.g., see Equation 8 in][]{muschietti13a}.  We need to confirm that the waves can grow to a sufficient amplitude in less time than is necessary to convect across the shock foot.  The physically significant time scale is the lower hybrid resonance period, $\tau{\scriptstyle_{lh}}$ ($=$ 2$\pi$/$\omega{\scriptstyle_{lh}}$, where $\omega{\scriptstyle_{lh}}$ $=$ $\left( \Omega{\scriptstyle_{ci}} \Omega{\scriptstyle_{ce}} \right)^{1/2}$).  The growth rate of the ECDI for large wavelengths is $\gamma{\scriptstyle_{max}}$ $\sim$ $\left( M{\scriptstyle_{i}}/m{\scriptstyle_{e}} \right)^{3/4} \left( \alpha/8 \pi \right)^{1/4} \Omega{\scriptstyle_{ci}}$, where $\alpha$ is the ratio of ion beam to total ion density.  Let us we define a convection scale length, L${\scriptstyle_{conv}}$ $\equiv$ V${\scriptstyle_{e}}$$\tau{\scriptstyle_{lh}}$ (where V${\scriptstyle_{e}}$ is the incident electron speed in the shock frame), and a scale length for the shock foot, L${\scriptstyle_{f}}$ $\equiv$ U${\scriptstyle_{shn}}$/$\langle \Omega{\scriptstyle_{ci}} \rangle{\scriptstyle_{up}}$.  Then we can say that L${\scriptstyle_{conv}}$/L${\scriptstyle_{f}}$ $<$ 1 means that the waves could grow to sufficient amplitudes before convecting across the shock foot.

%%  Discuss Growth vs. Convection Times
\indent  For the THEMIS events examined herein, we found 0.0007 $\lesssim$ L${\scriptstyle_{conv}}$/L${\scriptstyle_{f}}$ $\lesssim$ 0.14.  Therefore, the waves should convect distances much less than the scale of the shock foot and thus should be observed at significant amplitudes throughout this region.  As we discussed in Appendix \ref{subapp:THEMISWaveObservations} and Appendix \ref{app:WindSTEREO}, we observe large amplitude waves throughout the entire transition region, consistent with the simulation results of \citet[][]{muschietti13a}.

%%----------------------------------------------------------------------------------------
%%  Subsection:  Wind and STEREO Examples
%%----------------------------------------------------------------------------------------
\subsection{Wind and STEREO Examples}  \label{app:WindSTEREO}

%%++++++++++++++++++++++++++++++++++++++++++++++++++++++++++++++++++++++++++++++++++++++++
%% Image:  Wind Example
%%++++++++++++++++++++++++++++++++++++++++++++++++++++++++++++++++++++++++++++++++++++++++
\begin{wrapfigure}{r}{0.405\textwidth}
  \centering
    \vspace{-10pt}
    {\includegraphics[trim = 0mm 0mm 0mm 0mm, clip, width=0.425\textwidth]
    {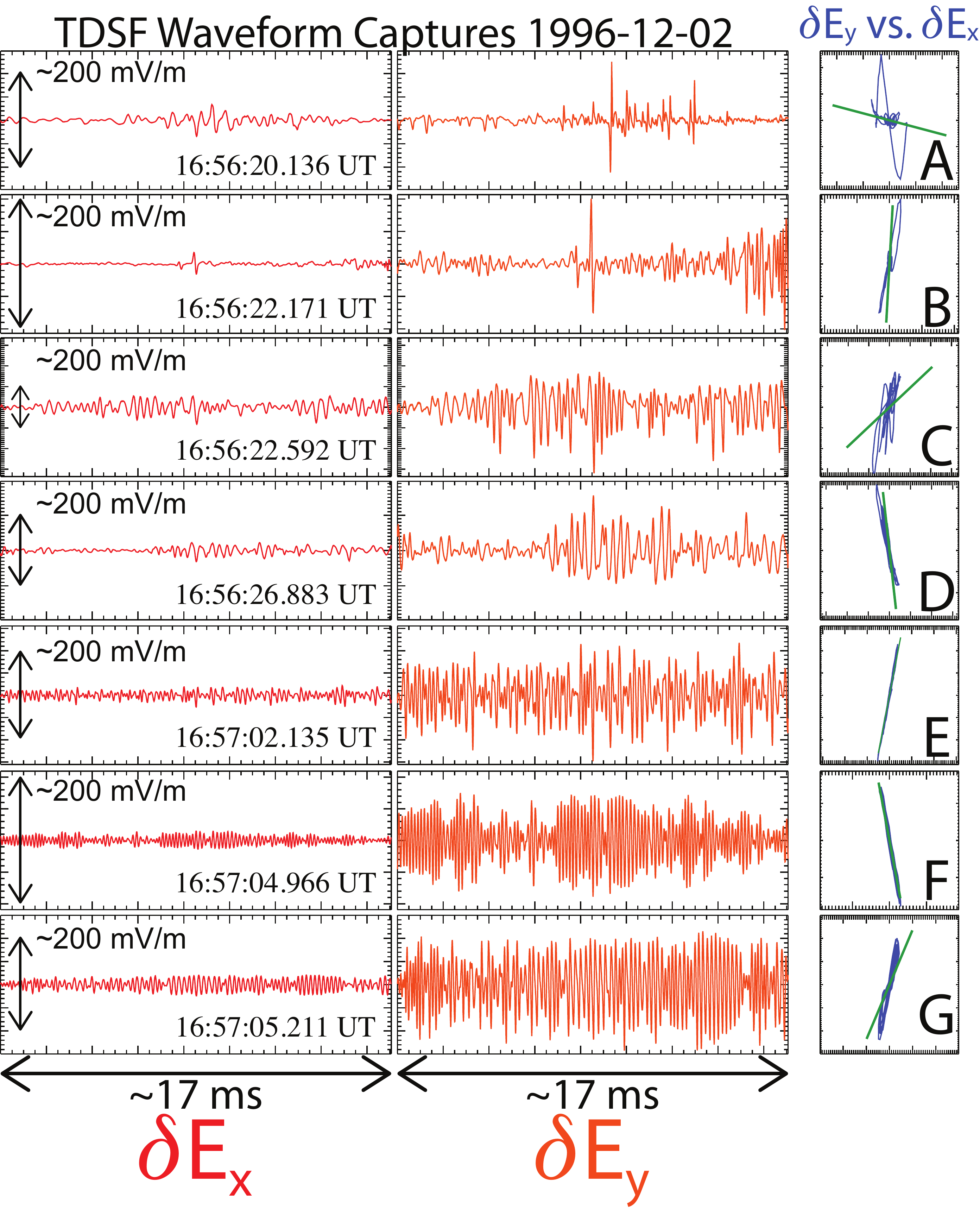}}
    %%  Define new caption setting options
    \captionsetup{width=0.395\textwidth,font=footnotesize,labelfont=bf}
    %%  Define caption
    \caption[Wind Example]{Seven waveform captures observed by the Wind/WAVES instrument during an inbound bow shock crossing on 1996-12-02.}
    \label{fig:exampleWind}
    \vspace{-10pt}
\end{wrapfigure}
%%++++++++++++++++++++++++++++++++++++++++++++++++++++++++++++++++++++++++++++++++++++++++
%% Image:  Wind Example
%%++++++++++++++++++++++++++++++++++++++++++++++++++++++++++++++++++++++++++++++++++++++++

%%  Intro Wind and STEREO
\indent  In this appendix we show a few waveform capture examples observed by the Wind/WAVES \citep[][]{bougeret95a} and STEREO S/WAVES \citep[][]{bale08a, bougeret08a} instruments.  The examples are characteristic of the observations from both spacecraft for every bow shock crossing examined by either Wind or STEREO.

%%  Wind Example Figure:  Introduce data
\indent  Figure \ref{fig:exampleWind} shows a series of seven waveform captures, $\sim$17.1 ms in duration, observed by the spin-stabilized Wind spacecraft during a bow shock crossing on 1996-12-02.  The electric fields are shown in instrument coordinates (non-rotating coordinates) sampled at $\sim$120,000 sps.  By instrument coordinates we mean that $\delta$E${\scriptstyle_{x(y)}}$ corresponds to data only from the X(Y)-antenna, $\sim$100(15) m tip-to-tip, without being rotated into a spacecraft or physically significant coordinate basis.  The purpose of showing the data in this manner prevents mixing of different noise levels from the different antenna since they have very different lengths.  

%%  Wind Example Figure:  Introduce Figure
\indent  The columns in Figure \ref{fig:exampleWind} show the following:  (left) $\delta$E${\scriptstyle_{x}}$; (middle) $\delta$E${\scriptstyle_{y}}$; and (right) $\delta$E${\scriptstyle_{y}}$ vs. $\delta$E${\scriptstyle_{x}}$ with \textbf{B}${\scriptstyle_{o}}$ projected onto the plane (green line).  The hodograms are uniformly scaled with the same range as the Y-axis range in the two left columns.  The black arrows in the left-hand column gives the relative scale of $\sim$200 mV/m for each waveform.  The data are shown in instrument coordinates with associated hodograms in the right-most column.  The hodograms show only the central 1/9th of the data.  This is to aid the reader in comparing the polarization to the projection of \textbf{B}${\scriptstyle_{o}}$.

%%  Wind Example Figure:  Discuss waveforms
\indent  Waveforms \textbf{A} and \textbf{B} in Figure \ref{fig:exampleWind} are consistent with trains of ESWs \citep[e.g.,][]{bale98b, bale02a}, waveform \textbf{C} is consistent with ECDI-driven waves \citep[e.g.,][]{wilsoniii10a}, and waveforms \textbf{D} and \textbf{G} are consistent with IAWs \citep[e.g.,][]{wilsoniii07a}.  One can see that every wave shown here has $\delta$E $\gtrsim$ 150 mV/m peak-to-peak.  These types of waves were observed in the shock ramp and throughout the magnetosheath of every Wind event we examined.  These large amplitude waves were observed in bow shock crossing when waveform capture data was available.  The waves are consistent with those we observed with the THEMIS spacecraft.

\begin{wrapfigure}{l}{0.425\textwidth}
  \vspace{-10pt}
  \centering
    {\includegraphics[trim = 0mm 0mm 0mm 0mm, clip, width=0.435\textwidth]
    {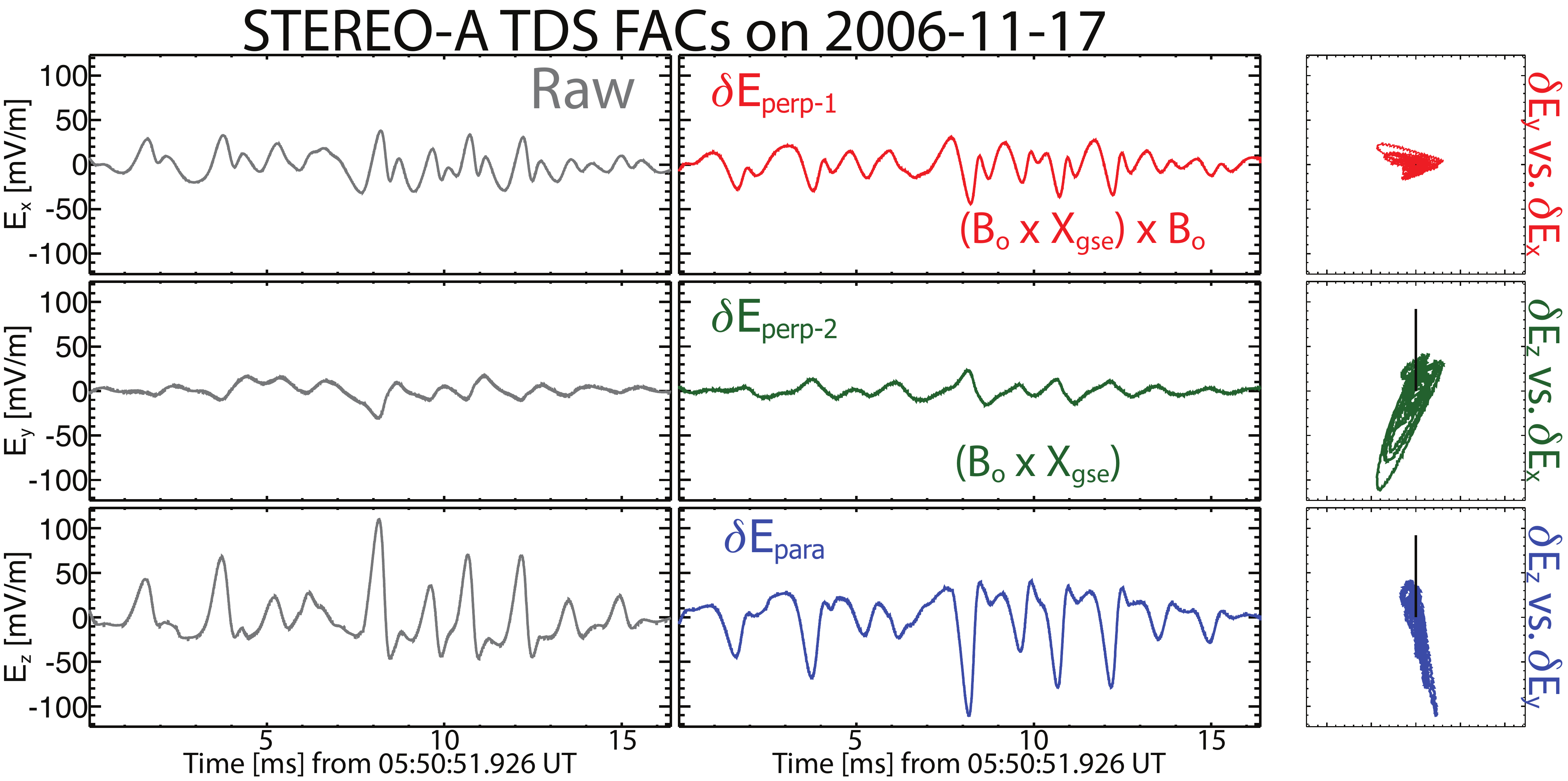}}
    %%  Define new caption setting options
    \captionsetup{width=0.405\textwidth,font=footnotesize,labelfont=bf}
    %%  Define caption
    \caption[STEREO Example]{An example waveform capture observed by the STEREO S/WAVES instrument during a bow shock crossing on 2006-11-17.}
    \label{fig:exampleSTEREO}
  \vspace{-10pt}
\end{wrapfigure}

%%  STEREO Example Figure:  Introduce Figure
\indent  Figure \ref{fig:exampleSTEREO} shows an example waveform capture, $\sim$16.4 ms in duration, sampled at $\sim$250,000 sps, observed by the STEREO-Ahead spacecraft during a bow shock crossing on 2006-11-17.  The data are shown in field-aligned coordinates (FACs), where the component directions are defined in the plot.  The columns show, in the following order:  (left) raw $\delta$E${\scriptstyle_{j}}$; (middle) corrected $\delta$E${\scriptstyle_{j}}$; and (right) $\delta$E${\scriptstyle_{j}}$ vs. $\delta$E${\scriptstyle_{i}}$ with the quasi-static magnetic field vector projected onto the plane (black line).  The hodograms are uniformly scaled with the same range as the Y-axis range in the two left columns.  One can see that the wave is polarized primarily along quasi-static magnetic field.

%%++++++++++++++++++++++++++++++++++++++++++++++++++++++++++++++++++++++++++++++++++++++++
%% Image:  STEREO Example
%%++++++++++++++++++++++++++++++++++++++++++++++++++++++++++++++++++++++++++++++++++++++++
%%++++++++++++++++++++++++++++++++++++++++++++++++++++++++++++++++++++++++++++++++++++++++
%% Image:  STEREO Example
%%++++++++++++++++++++++++++++++++++++++++++++++++++++++++++++++++++++++++++++++++++++++++

%%  STEREO Example Figure:  Discuss waveforms
\indent  These fluctuations are consistent with a train of ESWs.  The typical waveforms observed by STEREO \citep[e.g.,][]{breneman11c} are similar to those observed by Wind and those observed by THEMIS.  Again, in every bow shock crossing we examined with available waveform data, we observed waves with $\delta$E $\gtrsim$ 150 mV/m peak-to-peak.
%%  We found that the shorter antenna on Wind (i.e., the Y-antenna) typically showed larger amplitudes than the longer.  \textbf{\textcolor{Magenta}{The STEREO waveforms show comparable amplitudes on the X- and Y-antenna and slightly smaller for the Z-antenna, which suffers from a much shorter effective antenna length than the other two \citep[e.g.,][]{kellogg09a}.  Need to check this...}}

%%  Summary
\indent  Therefore, we conclude that large amplitude electrostatic waves are an ubiquitous phenomena in the Earth's collisionless bow shock.  The observations by Wind and STEREO add to the evidence supplied by our THEMIS observations that high frequency waves play an important role in the global energy dissipation budget of medium Mach number ($\sim$2-6) collisionless shocks.

%%----------------------------------------------------------------------------------------
%%  Appendix:  Normal Incidence Frame and Coordinate Basis
%%----------------------------------------------------------------------------------------
\section{Normal Incidence Frame and Coordinate Basis} \label{app:nifandbasis}
%%----------------------------------------------------------------------------------------
%%  Subsection:  Normal Incidence Frame
%%----------------------------------------------------------------------------------------
\subsection{Normal Incidence Frame}  \label{subapp:nif}
\indent  In this appendix, we will define our reference frame transformation into the Normal Incidence Frame (NIF) and coordinate basis rotations into the Normal incidence frame Coordinate Basis (NCB).  We will present the transformations/rotations in a generalized manner, but for the purposes of this manuscript the measurements are in the SpaceCraft Frame (SCF) and GSE coordinate basis.  We define the generalized basis as the Input Coordinate Basis (ICB).

\indent  We can define the velocity transformation from any arbitrary frame of reference (\textit{e.g.} SCF) to the shock frame of reference (SHF) as:
\begin{equation}
  \label{eq:fieldtrans_0}
  \textbf{V}{\scriptstyle_{sh}}^{rest} = \textbf{V}^{arb.} - \left( \textbf{V}{\scriptstyle_{sh}}^{arb.} \cdot \hat{\textbf{n}} \right) \hat{\textbf{n}}
\end{equation}
where $\hat{\textbf{n}}$ is the vector normal to the assumed planar shock front.  For an experimentalist's purposes, \textbf{V}$^{arb.}$ $\rightarrow$ \textbf{V}${\scriptstyle_{sw}}$, where \textbf{V}${\scriptstyle_{sw}}$ is the bulk flow velocity (e.g., solar wind velocity) in the SCF and ICB (e.g., GSE).  Let us define V${\scriptstyle_{sh,n}}$ as the shock normal speed in the SCF, determined from the numerical Rankine-Hugoniot solution techniques \citep[e.g.,][]{vinas86a, koval08a}.  Therefore, we can define the upstream incident bulk flow velocity in the SHF, which is given by:
\begin{equation}
  \label{eq:fieldtrans_2}
  \textbf{V}{\scriptstyle_{u}} = \textbf{V}{\scriptstyle_{sw}} - \left( V{\scriptstyle_{sh,n}} \hat{\textbf{n}} \right)  \text{  .}
\end{equation}
The transformation velocity from the SHF to the NIF is given by:
\begin{subequations}
  \begin{align}
    \textbf{V}^{NIF} & = \hat{\textbf{n}} \times \left( \textbf{V}{\scriptstyle_{u}} \times \hat{\textbf{n}} \right)  \label{eq:fieldtrans_3a}  \\
    & = \hat{\textbf{n}} \times \left( \textbf{V}{\scriptstyle_{sw}} \times \hat{\textbf{n}} \right)  \label{eq:fieldtrans_3b}
%%    \left\langle  \right\rangle
%%    \left[  \right]
%%    \left(  \right)
  \end{align}
\end{subequations}
therefore, the upstream flow velocity in the NIF and the GSE basis is given by:
\begin{equation}
  \label{eq:fieldtrans_4}
  \textbf{V}{\scriptstyle_{u}}^{NIF} = \textbf{V}{\scriptstyle_{u}} - \textbf{V}^{NIF}  \text{  .}
\end{equation}
\indent  Since the change in velocity between any SHF the local SCF satisfies $\mid$$\boldsymbol{\beta}$$\mid$ $\equiv$ $\mid$$\Delta$\textbf{V}$\mid$/c $\ll$ 1 for any shock within the heliosphere, the Lorentz transformations of the electric and magnetic fields \citep[page 558 of][]{jackson98a} can be given by:
\begin{subequations}
  \begin{align}
    \textbf{E}' & \approx \left( \textbf{E} + \boldsymbol{\beta} \times \textbf{B} \right)  \label{eq:fieldtrans_5b}  \\
    \textbf{B}' & \approx \textbf{B}  \text{  .}  \label{eq:fieldtrans_5d}
  \end{align}
\end{subequations}
Thus, the frame transformation velocity, $\Delta$\textbf{V}, between SCF and NIF is given by:
\begin{subequations}
  \begin{align}
    \Delta\textbf{V} & = \textbf{V}{\scriptstyle_{sw}} - \textbf{V}{\scriptstyle_{u}}^{NIF}  \label{eq:fieldtrans_6a}  \\
    & = \textbf{V}{\scriptstyle_{sw}} - \left[ \textbf{V}{\scriptstyle_{sw}} - \left( V{\scriptstyle_{sh,n}} \hat{\textbf{n}} \right) \right] + \textbf{V}^{NIF}  \label{eq:fieldtrans_6b}  \\
    & = \left( V{\scriptstyle_{sh,n}} \hat{\textbf{n}} \right) + \textbf{V}^{NIF}  \label{eq:fieldtrans_6c}
  \end{align}
\end{subequations}
which allows us to show that the electric field in the NIF, \textbf{E}$^{NIF}$, can be determined from the electric field observed in the SCF, \textbf{E}$^{SCF}$, through the following:
\begin{subequations}
  \begin{align}
    \textbf{E}^{NIF} & = \textbf{E}^{SCF} + \left( \Delta\textbf{V} \times \textbf{B} \right)  \label{eq:fieldtrans_7a}  \\
     & = \textbf{E}^{SCF} + \left[ \left( V{\scriptstyle_{sh,n}} \hat{\textbf{n}} + \textbf{V}^{NIF} \right) \times \textbf{B} \right]  \text{  .}  \label{eq:fieldtrans_7b}
  \end{align}
\end{subequations}
It should be noted that this reference frame transformation, $\left( \Delta\textbf{V} \times \textbf{B} \right)$, is rarely more than a few mV/m in magnitude.  For instance, for the 2009-09-26 event, $\mid$$\left( \Delta\textbf{V} \times \textbf{B} \right)$$\mid$ $\lesssim$ 2.5 mV/m.  Thus, these convective frame-dependent electric fields are relatively insignificant compared to the large fluctuations due to the observed waves.
%%----------------------------------------------------------------------------------------
%%  Subsection:  NIF Coordinate Basis
%%----------------------------------------------------------------------------------------
\subsection{NIF Coordinate Basis}  \label{subapp:nifbasis}
\indent  We can rotate into the Normal incidence frame Coordinate Basis (NCB) from the Input Coordinate Basis (ICB) by defining a rotation matrix, $\mathbb{A}$ \citep{scudder86a}, given by:
\begin{equation}
  \label{eq:nifbasis_1}
 \mathbb{A} = \left[
  \begin{array}{ c c c }
    n{\scriptstyle_{x}}     & n{\scriptstyle_{y}}     & n{\scriptstyle_{z}}      \\
    \beta{\scriptstyle_{x}} & \beta{\scriptstyle_{y}} & \beta{\scriptstyle_{z}}  \\
    \zeta{\scriptstyle_{x}} & \zeta{\scriptstyle_{y}} & \zeta{\scriptstyle_{z}}
  \end{array} \right]
\end{equation}
where $\hat{\textbf{n}}$ is the shock normal vector and $\boldsymbol{\beta}$ and $\boldsymbol{\zeta}$ are given by:
\begin{subequations}
  \begin{align}
    \hat{\textbf{y}} = \boldsymbol{\beta} & = \frac{ \textbf{B}{\scriptstyle_{2}} \times \textbf{B}{\scriptstyle_{1}} }{ \mid \textbf{B}{\scriptstyle_{1}} \times \textbf{B}{\scriptstyle_{2}} \mid }  \label{eq:nifbasis_2a}  \\
    \hat{\textbf{z}} = \boldsymbol{\zeta} & = \frac{ \hat{\textbf{n}} \times \boldsymbol{\beta} }{ \mid \hat{\textbf{n}} \times \boldsymbol{\beta} \mid }  \label{eq:nifbasis_2b}
  \end{align}
\end{subequations}
where \textbf{B}${\scriptstyle_{1(2)}}$ is the average upstream(downstream) magnetic field vector.  If the vectors $\hat{\textbf{n}}$, $\boldsymbol{\beta}$, and $\boldsymbol{\zeta}$ start in the ICB (e.g., GSE), then one would expect that $\mathbb{A}$ acting on $\hat{\textbf{n}}$, $\boldsymbol{\beta}$, or $\boldsymbol{\zeta}$ should give the corresponding NCB axis unit vector.  Meaning, we expect the following to be true:
\begin{subequations}
  \begin{align}
    \mathbb{A} \cdot \hat{\textbf{n}} & = \langle 1, 0, 0 \rangle  \label{eq:nifbasis_3a}  \\
    \mathbb{A} \cdot \boldsymbol{\beta} & = \langle 0, 1, 0 \rangle  \label{eq:nifbasis_3b}  \\
    \mathbb{A} \cdot \boldsymbol{\zeta} & = \langle 0, 0, 1 \rangle \text{  .}  \label{eq:nifbasis_3c}
  \end{align}
\end{subequations}
Thus, $\mathbb{A}$ should rotate any ICB vector into the NCB.

%%  Caveat
\indent  If the coordinate vectors used to create $\mathbb{A}$ are not orthogonal, then the correct rotation tensor is given by $\mathbb{R}$ $=$ ($\mathbb{A}^{T}$)$^{-1}$, or the inverse transpose of $\mathbb{A}$.  The need to perform the inverse transpose of $\mathbb{A}$ arises from the non-orthogonal nature of the NIF basis.  If the NIF were created from an orthogonal basis, then $\mathbb{A}$ would be an orthogonal matrix, which means $\mathbb{A}^{T}$ $=$ $\mathbb{A}^{-1}$.  For any invertible matrix, the following is true:  ($\mathbb{A}^{T}$)$^{-1}$ $=$ ($\mathbb{A}^{-1}$)$^{T}$.  Thus, an orthogonal NIF basis would imply $\mathbb{R}$ $=$ ($\mathbb{A}^{T}$)$^{-1}$ $=$ ($\mathbb{A}^{T}$)$^{T}$ $=$ $\mathbb{A}$.  In general, however, the NIF basis vectors are not orthogonal and thus $\mathbb{R}$ $\neq$ $\mathbb{A}$.

%%----------------------------------------------------------------------------------------
%%  Appendix:  Macroscopic Energy Dissipation
%%----------------------------------------------------------------------------------------
\section{Macroscopic Energy Dissipation} \label{app:thermodynamics}
\indent  In this appendix, we will discuss how we quantify the energy dissipated by the shock on a macroscopic, fluid scale.  These estimates are used as a proxy for the total amount of energy the shock needs to dissipate in order to produce the observed changes in entropy and/or enthalpy density.  We can then compare the macroscopic dissipation rates to wave dissipation rates (see Appendix \ref{app:EnergyDissipation}).  The comparison will allow us to determine whether electromagnetic waves can provide sufficient energy dissipation to mediate the shock transition.

\indent  Let us define $\Delta \mathcal{E}$ as the change in internal energy not including the energy needed to displace its surroundings.  Given that, we can write:
\begin{equation}
  \label{eq:thermodynamics_0}
  \Delta \mathcal{E} = \Delta Q + \Delta W{\scriptstyle_{mech}} + \Delta W{\scriptstyle_{ext}}
\end{equation}
where $\Delta$Q $\equiv$ \emph{heat} added to the system, $\Delta$W${\scriptstyle_{mech}}$ $\equiv$ mechanical work done on the system by the surroundings\footnote{this can actually be the converse, which simply results in a sign change}, and $\Delta$W${\scriptstyle_{ext}}$ $\equiv$ work done by any external or imposed forces on the system (e.g., introduce a current to a circuit element).  The work done to displace the surrounding medium is given by P$\Delta$V, where P is the scalar pressure (assume $\propto$ $\rho^{\gamma}$) and V is a volume element.  We have used the definitions $\rho$ $\equiv$ scalar mass density and $\gamma$ $\equiv$ polytrope index or ratio of specific heats.  We know the following:
\begin{equation}
  \label{eq:thermodynamics_1}
  \Delta \left( P V \right) = P \Delta V + V \Delta P
\end{equation}
and we know that $\Delta$Q $=$ T$\Delta$S (for a reversible process), where T $\equiv$ scalar temperature and S $\equiv$ scalar entropy.  We can now rewrite Equation \ref{eq:thermodynamics_0} as:
\begin{equation}
  \label{eq:thermodynamics_2}
  \Delta \mathcal{E} + \Delta \left( P V \right) = \Delta Q + V \Delta P + \Delta W{\scriptstyle_{ext}}
\end{equation}
where we now define $\Delta$($\mathcal{E}$ $+$ PV) $=$ $\Delta \mathcal{H}$ $\equiv$ enthalpy.  Enthalpy is the total internal energy of a system.  To move on, we first show that:
\begin{subequations}
  \begin{align}
    \Delta V & = \Delta \left( \frac{ m }{ \rho } \right) = - \left( \frac{ m }{ \rho^{2} } \right) \Delta \rho  \label{eq:thermodynamics_3a}  \\
    \Delta P & = \Delta \left( C{\scriptstyle_{o}} \rho^{\gamma} \right)  \label{eq:thermodynamics_3b}  \\
    & = C{\scriptstyle_{o}} \gamma \rho^{\gamma - 1} \Delta \rho  \label{eq:thermodynamics_3c}  \\
    & = \left( \frac{ \gamma P }{ \rho } \right) \Delta \rho  \label{eq:thermodynamics_3d}  \\
    & = C{\scriptstyle_{s}}^{2} \Delta \rho  \label{eq:thermodynamics_3e}
  \end{align}
\end{subequations}
where $m$ is the particle mass and C${\scriptstyle_{s}}$ $\equiv$ scalar speed of sound.  Thus, we have:
\begin{equation}
  \label{eq:thermodynamics_4}
  V \Delta P = \left( \frac{ m }{ \rho } \right) C{\scriptstyle_{s}}^{2} \Delta \rho  \text{  .}
\end{equation}
Recall that our equation for enthalpy was given by:
\begin{equation}
  \label{eq:thermodynamics_5}
  \Delta \mathcal{H} = T \Delta S + V \Delta P + \Delta W{\scriptstyle_{ext}}
\end{equation}
so that if we multiply both sides by ($\rho$/m), then we have:
\begin{subequations}
  \begin{align}
    \left( \frac{ \rho }{ m } \right) \Delta \mathcal{H} & = \left( \frac{ \rho T }{ m } \right) \Delta S + C{\scriptstyle_{s}}^{2} \Delta \rho + \left( \frac{ \rho }{ m } \right) \Delta W{\scriptstyle_{ext}}  \label{eq:thermodynamics_6a}  \\
    \Delta \mathfrak{h} & = \left( \rho T \right) \Delta \mathfrak{s} + C{\scriptstyle_{s}}^{2} \Delta \rho + \Delta \mathfrak{w}{\scriptstyle_{ext}}  \label{eq:thermodynamics_6b}
  \end{align}
\end{subequations}
where we note that $\Delta \mathfrak{h}$ has units of energy per unit volume [e.g., J m$^{-3}$] and can be referred to as the change in the specific enthalpy density.  The change in specific entropy, $\Delta \mathfrak{s}$, has the well known definition given by:
\begin{subequations}
  \begin{align}
    \Delta \mathfrak{s} & = C{\scriptstyle_{v}} \ln \left\lvert \frac{ P{\scriptstyle_{2}} }{ P{\scriptstyle_{1}} } \left( \frac{ \rho{\scriptstyle_{1}} }{ \rho{\scriptstyle_{2}} } \right)^{\gamma} \right\rvert  \label{eq:thermodynamics_7a}  \\
    C{\scriptstyle_{v}} & = \frac{ k{\scriptstyle_{B}} }{ m \left( \gamma - 1 \right) }  \label{eq:thermodynamics_7b}
  \end{align}
\end{subequations}
where the subscript 1(2) refers to the initial(final) state \citep[e.g.,][]{gurnett05}.  Note that $\Delta \mathfrak{s}$ has units of energy per degree Kelvin per unit mass [e.g., J $^{\circ}$K$^{-1}$ kg$^{-1}$].

\indent  If we divide both sides of Equation \ref{eq:thermodynamics_6b} by $\Delta$t, then we have:
\begin{subequations}
  \begin{align}
    \frac{ \Delta \mathfrak{h} }{ \Delta t } & = \left( \rho T \right) \frac{ \Delta \mathfrak{s} }{ \Delta t } + C{\scriptstyle_{s}}^{2} \frac{ \Delta \rho }{ \Delta t } + \frac{ \Delta \mathfrak{w}{\scriptstyle_{ext}} }{ \Delta t }  \label{eq:thermodynamics_8a}  \\
    \intertext{or:}
    \frac{ \Delta \mathfrak{h} }{ \Delta t } & = \dot{\Lambda} + \frac{ \Delta \mathfrak{w}{\scriptstyle_{ext}} }{ \Delta t }  \label{eq:thermodynamics_8b}
  \end{align}
\end{subequations}
which gives the rate of change of the total internal energy per unit volume per unit time [e.g., J m$^{-3}$ s$^{-1}$ or W m$^{-3}$].  The third term in Equation \ref{eq:thermodynamics_8a} defines the rate of work density done per unit time for adiabatic compression.  We are concerned with quantifying the amount of energy transformed irreversibly, which relates to the term involving $\Delta \mathfrak{s}$ satisfying $\Delta \mathfrak{s}$ $>$ 0.

%%----------------------------------------------------------------------------------------
%%  Appendix:  Current Density
%%----------------------------------------------------------------------------------------
\section{Current Density} \label{app:CurrentDensity}
\indent  In this appendix, we will discuss our estimates for spatial scales and current densities.  We need the current density to estimate the wave dissipation rate due to the work done by the wave electric fields on the plasma per unit volume.  The wave dissipation rate is compared to the macroscopic dissipation rate defined in Appendix \ref{app:thermodynamics} to determine if the waves can mediate the shock transition alone.
%%----------------------------------------------------------------------------------------
%%  Subsection:  Implementation
%%----------------------------------------------------------------------------------------
\subsection{Implementation}  \label{subapp:Implementation}
%%  ∆X
\indent  We wish to estimate the current density, \textbf{j}${\scriptstyle_{o}}$ ($=$ $\nabla \times$\textbf{B}${\scriptstyle_{o}}$/$\mu{\scriptstyle_{o}}$), which requires a spatial scale to substitute for $\nabla$.  We have already determined the shock normal speed in the spacecraft frame (SCF) of reference (V${\scriptstyle_{shn}}$ defined in Appendix \ref{app:nifandbasis}) from our Rankine-Hugoniot analysis.  Therefore, we can use the sample period from the FGM instrument, $\Delta t$, and V${\scriptstyle_{shn}}$, using the ``Taylor hypothesis,'' to calculate an associated spatial scale $\Delta$X${\scriptstyle_{n}}$ $\equiv$ V${\scriptstyle_{shn}}$ $\Delta$t.  It is important to note that this assumption only works for relatively large spatial scales and it assumes that all fluctuations are being convected at the same speed.  The high frequency waves we are interested in have spatial scales much too small to utilize multi-spacecraft observations.  Therefore, we are limited to single-spacecraft observations and our assumption of the connection between temporal and spatial scales at low frequencies near the shock ramp.

%%  j calculation --> |∂E . j|
\indent  We want to calculate \textbf{j}${\scriptstyle_{o}}$ in the Normal Incidence Frame (NIF) and the Normal incidence frame Coordinate Basis (NCB) \citep[e.g.,][]{scudder86a}.  In this basis, V${\scriptstyle_{shn}}$, and thus $\Delta$X${\scriptstyle_{n}}$, is along the $\hat{\textbf{x}}$-component.  We also know that $\nabla \cdot$\textbf{B}${\scriptstyle_{o}}$ $\approx$ $\hat{\textbf{n}} \cdot$\textbf{B}${\scriptstyle_{o}}$ $\approx$ 0, where $\hat{\textbf{n}}$ is the shock normal vector.  Note that in the following calculations, we assume that \textbf{j}${\scriptstyle_{o}}$ is the total current density and in the NIF and NCB.  Therefore, we can write:
\begin{subequations}
  \begin{align}
    \mu{\scriptstyle_{o}} \textbf{j}{\scriptstyle_{o}} & = \nabla \times \textbf{B}{\scriptstyle_{o}}  \label{eq:currentdensest_4a}  \\
    & = \left[ \hat{\textbf{x}} \left( \partial{\scriptstyle_{y}} B{\scriptstyle_{oz}} - \partial{\scriptstyle_{z}} B{\scriptstyle_{oy}} \right) + \hat{\textbf{y}} \left( \partial{\scriptstyle_{z}} B{\scriptstyle_{ox}} - \partial{\scriptstyle_{x}} B{\scriptstyle_{oz}} \right) + \hat{\textbf{z}} \left( \partial{\scriptstyle_{x}} B{\scriptstyle_{oy}} - \partial{\scriptstyle_{y}} B{\scriptstyle_{ox}} \right) \right]  \label{eq:currentdensest_4b}  \\
    & \approx \left[ \hat{\textbf{x}} \left( \partial{\scriptstyle_{y}} B{\scriptstyle_{oz}} - \partial{\scriptstyle_{z}} B{\scriptstyle_{oy}} \right) + \hat{\textbf{y}} \left( 0 - \partial{\scriptstyle_{x}} B{\scriptstyle_{oz}} \right) + \hat{\textbf{z}} \left( \partial{\scriptstyle_{x}} B{\scriptstyle_{oy}} - 0 \right) \right]  \label{eq:currentdensest_4c}  \\
    \intertext{and if we assume $\nabla \cdot$\textbf{j}${\scriptstyle_{o}}$ $\approx$ $\hat{\textbf{n}} \cdot$\textbf{j}${\scriptstyle_{o}}$ $\approx$ 0, then we have:}
    & \approx \left[ \hat{\textbf{x}} \left( 0 \right) + \hat{\textbf{y}} \left( 0 - \partial{\scriptstyle_{x}} B{\scriptstyle_{oz}} \right) + \hat{\textbf{z}} \left( \partial{\scriptstyle_{x}} B{\scriptstyle_{oy}} - 0 \right) \right]  \label{eq:currentdensest_4d}  \\
    & = \left[ - \hat{\textbf{y}} \partial{\scriptstyle_{x}} B{\scriptstyle_{oz}} + \hat{\textbf{z}} \partial{\scriptstyle_{x}} B{\scriptstyle_{oy}} \right]  \label{eq:currentdensest_4e}  \\
    \intertext{which we can rewrite as:}
    \textbf{j}{\scriptstyle_{o}} & \approx \frac{ 1 }{ \mu{\scriptstyle_{o}} } \left[ - \hat{\textbf{y}} \frac{ \Delta B{\scriptstyle_{oz}} }{ \Delta X{\scriptstyle_{n}} } + \hat{\textbf{z}} \frac{ \Delta B{\scriptstyle_{oy}} }{ \Delta X{\scriptstyle_{n}} } \right]  \text{  .}  \label{eq:currentdensest_4f}
  \end{align}
\end{subequations}

%%++++++++++++++++++++++++++++++++++++++++++++++++++++++++++++++++++++++++++++++++++++++++
%% Image:  Example BSC with fields, j, and ∂j
%%++++++++++++++++++++++++++++++++++++++++++++++++++++++++++++++++++++++++++++++++++++++++
\begin{figure}[!htb]
  \centering
    {\includegraphics[trim = 0mm 0mm 0mm 0mm, clip, width=12cm]
    {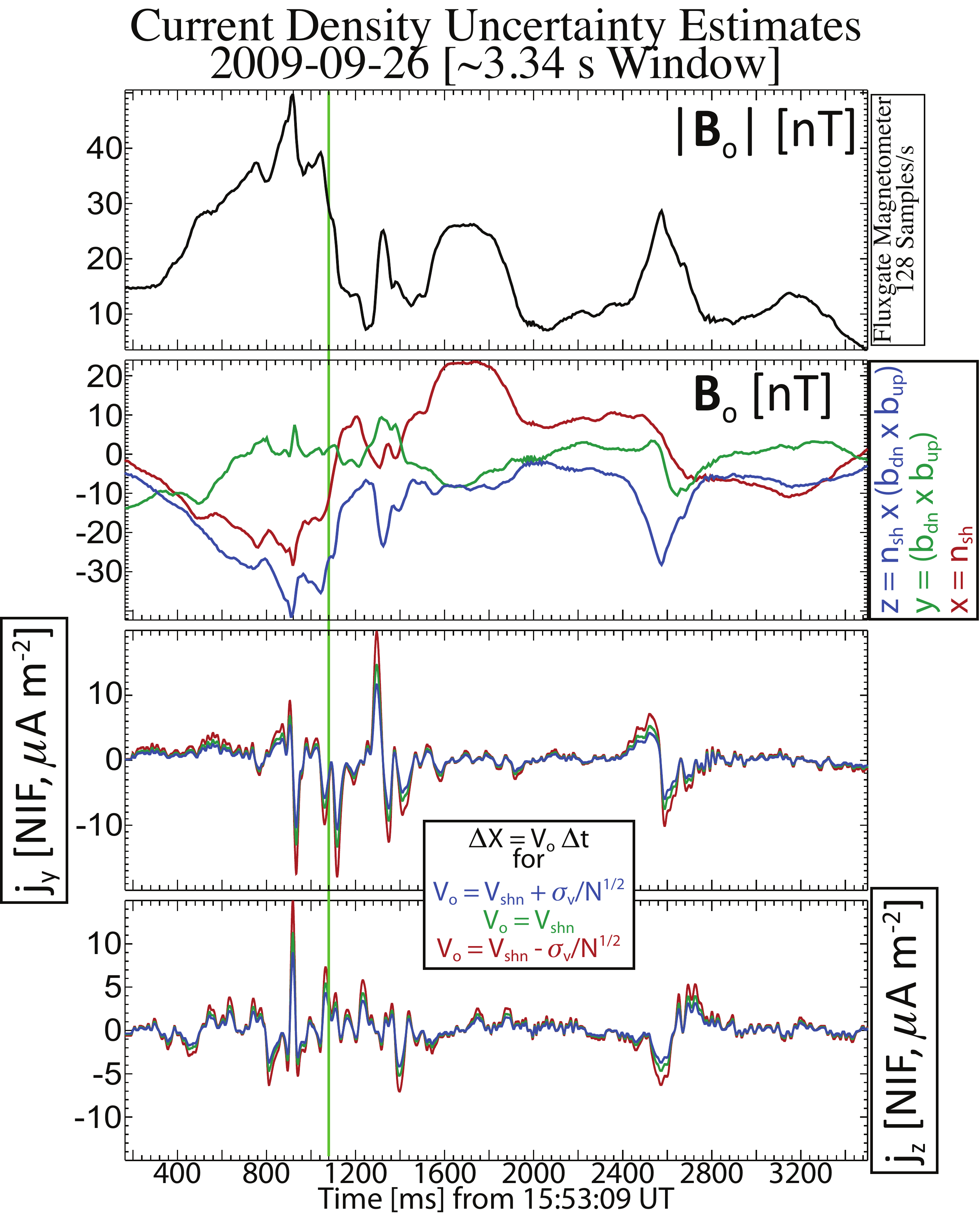}}
    \captionsetup{width=11cm}   %%  adjust caption width
    \caption[Example BSC with fields, currents, and uncertainties]{This example shows the same time range as Figure \ref{fig:exampleEBSjEdjBSC} and the top two panels are the same.  The third(fourth) panel shows three values of j${\scriptstyle_{y(z)}}$.  The color-coded lines show the value of V${\scriptstyle_{shn}}$ (see inset box) used to estimate $\Delta$X${\scriptstyle_{n}}$ for Equation \ref{eq:currentdensest_4f}.  The green vertical line indicates the center of the shock ramp at $\sim$15:53:10.080 UT.}
    \label{fig:exampleEBSdeltaj}
\end{figure}
%%++++++++++++++++++++++++++++++++++++++++++++++++++++++++++++++++++++++++++++++++++++++++
%% Image:  Example BSC with fields, j, and ∂j
%%++++++++++++++++++++++++++++++++++++++++++++++++++++++++++++++++++++++++++++++++++++++++

%%  purpose of j calculation --> |∂E . j|
\indent  We need \textbf{j}${\scriptstyle_{o}}$ to estimate an effective dissipation rate from Poynting's theorem.  We wish to compare the possible dissipation rates due to electromagnetic waves, $\left\lvert \textbf{j}{\scriptstyle_{o}} \cdot \delta \textbf{E} \right\rvert$ $\approx$ $\eta{\scriptstyle_{iaw}} \left\lvert \textbf{j}{\scriptstyle_{o}} \right\rvert^{2}$, to our macroscopic estimate from entropy production, ($\rho$T)$\Delta \mathfrak{s}$/$\Delta$t.  Since \textbf{j}${\scriptstyle_{o}}$ is on the same time step as the FGM, we first transform $\delta$\textbf{E} into the NIF and then rotate into the NCB.  Then we up-sampled \textbf{j}${\scriptstyle_{o}}$ to match the time steps for $\delta$\textbf{E} prior to calculating the dot-product.  We do this to examine the effect of the instantaneous electric fields, due to high frequency ($\gtrsim$10 Hz) waves, on the bulk of the plasma.  The details of these dissipation rates are discussed in the Appendix \ref{app:EnergyDissipation} and the details of our assumptions for calculating $\left\lvert \textbf{j}{\scriptstyle_{o}} \cdot \delta \textbf{E} \right\rvert$ are in Appendix \ref{subapp:Approximations}.

%%  motivate uncertainty calculation for j
\indent  We need to determine the relative uncertainty associated with the assumptions we made to estimate \textbf{j}${\scriptstyle_{o}}$.  This will give us a way to evaluate whether our wave dissipation rates can be trusted because the we expect the uncertainty in $\delta$\textbf{E} to be small by comparison.  We also know that the uncertainty in any given \textbf{B}${\scriptstyle_{o}}$ observation will be relatively small ($\sim$0.1 nT), therefore we can argue that the uncertainty in \textbf{j}${\scriptstyle_{o}}$ will be dominated by uncertainties in V${\scriptstyle_{shn}}$.

%%  uncertainty calculation for j
\indent  To determine the relative impact of these uncertainties, we calculated $\Delta$X${\scriptstyle_{n}}$ using three variances of V${\scriptstyle_{shn}}$, shown in the bottom two panels in Figure \ref{fig:exampleEBSdeltaj}.  The three versions of $\Delta$X${\scriptstyle_{n}}$ were calculated from:  (1) V${\scriptstyle_{shn}}$ $-$ $\sigma{\scriptstyle_{v}}/\sqrt{N}$ (red lines); (2) V${\scriptstyle_{shn}}$ (green lines); and (3) V${\scriptstyle_{shn}}$ $+$ $\sigma{\scriptstyle_{v}}/\sqrt{N}$ (blue lines).  One can see, by examining the bottom two panels in Figure \ref{fig:exampleEBSdeltaj}, that the small deviations in j${\scriptstyle_{y(z)}}$ (red or blue lines) from the values used to create Figure \ref{fig:comparedissrate} (green lines) would not cause significant changes to our estimates for $\left\lvert \textbf{j}{\scriptstyle_{o}} \cdot \delta \textbf{E} \right\rvert$ or $\eta{\scriptstyle_{iaw}} \left\lvert \textbf{j}{\scriptstyle_{o}} \right\rvert^{2}$.  This is because these two quantities are dominated by the very large fluctuations observed in $\delta$\textbf{E} (e.g., see Figures \ref{fig:exampleecdi} -- \ref{fig:exampleSTEREO}).  Therefore, we can safely argue that our estimates for $\left\lvert \textbf{j}{\scriptstyle_{o}} \cdot \delta \textbf{E} \right\rvert$ or $\eta{\scriptstyle_{iaw}} \left\lvert \textbf{j}{\scriptstyle_{o}} \right\rvert^{2}$ can be used as lower bounds for these dissipation rates and safely compared to ($\rho$T)$\Delta \mathfrak{s}$/$\Delta$t.

%%----------------------------------------------------------------------------------------
%%  Subsection:  Approximations
%%----------------------------------------------------------------------------------------
\subsection{Approximations}  \label{subapp:Approximations}
\indent  In this appendix, we will discuss our approximations in our calculation of the current density, \textbf{j}.  We explain why we are able to use the high frequency electric fields with the low frequency magnetic fields to estimate an energy dissipation rate.

\indent  First, we define any quantity observed at $\leq$128 sps as a quasi-static quantity, Q${\scriptstyle_{o}}$, and any quantity observed at $\sim$8192 sps as a fluctuating quantity, $\delta$Q (e.g., note that we have used the notation \textbf{j}${\scriptstyle_{o}}$ throughout, which we justify below).  We then assume that:
\begin{subequations}
  \begin{align}
    \textbf{j} & \approx \textbf{j}{\scriptstyle_{o}} + \delta \textbf{j}  \label{eq:Approximations_0a}  \\
    \textbf{E} & \approx \textbf{E}{\scriptstyle_{o}} + \delta \textbf{E}  \label{eq:Approximations_0b}  \\
    \intertext{and therefore,}
    \textbf{j} \cdot \textbf{E} & \approx \left( \textbf{j}{\scriptstyle_{o}} + \delta \textbf{j} \right) \cdot \left( \textbf{E}{\scriptstyle_{o}} + \delta \textbf{E} \right)  \label{eq:Approximations_0c}  \\
    & = \left( \textbf{j}{\scriptstyle_{o}} \cdot \textbf{E}{\scriptstyle_{o}} \right) + \left( \textbf{j}{\scriptstyle_{o}} \cdot \delta \textbf{E} \right) + \left( \delta \textbf{j} \cdot \textbf{E}{\scriptstyle_{o}} \right) + \left( \delta \textbf{j} \cdot \delta \textbf{E} \right)  \label{eq:Approximations_0d}
  \end{align}
\end{subequations}
where \textbf{E} is the measured electric field.

\indent  We need to justify our approximation of $\left\lvert \textbf{j} \cdot \textbf{E} \right\rvert$ $\sim$ $\left\lvert \textbf{j}{\scriptstyle_{o}} \cdot \delta \textbf{E} \right\rvert$ in our study, which means we need to eliminate everything on the right-hand side of Equation \ref{eq:Approximations_0d} except the second term.  From our results we can make the following arguments, observations, and assumptions.
\begin{enumerate}
  \setlength{\itemsep}{0pt}    %% tighten up spacing between items
  \setlength{\parskip}{0pt}    %% tighten up spacing between items
  %%  Argue:  Jo >> ∂J
  \item  We argue that the largest component of \textbf{j} should be well resolved as gradients in \textbf{B}${\scriptstyle_{o}}$ (i.e., FGM data at $\leq$128 sps) because:
  \begin{enumerate}
    \setlength{\itemsep}{0pt}
    \setlength{\parskip}{0pt}
    \item  we assume the largest component of \textbf{j} arises from the bulk flow relative drifts between oppositely charged particle species;
    \item  these bulk drifts are responsible for the large scale changes in \textbf{B}${\scriptstyle_{o}}$ in the shock ramp and in magnetosonic-whistler waves (e.g., see Figure \ref{fig:exampleBSCfgh});
    \item  our observations show that the majority of the largest changes in \textbf{B}${\scriptstyle_{o}}$ occur at frequencies $<$ 10 Hz because we do not see the corresponding $\delta$\textbf{B} in the SCM data filtered above 10 Hz (e.g., see Figure \ref{fig:exampleEBSjEdjBSC});
    \item  therefore, the associated current densities should have a quasi-static response because we can clearly resolve the main shock structure with $\leq$128 sps FGM data; and
    \item  thus, we conclude that largest the contribution to \textbf{j} should be well resolved in \textbf{B}${\scriptstyle_{o}}$, which we define as \textbf{j}${\scriptstyle_{o}}$.
  \end{enumerate}
  %%  Argue:  neglect |Jo . Eo| (1st term)
  \item  We argue that we can neglect the $\left( \textbf{j}{\scriptstyle_{o}} \cdot \textbf{E}{\scriptstyle_{o}} \right)$ term on the right-hand side of Equation \ref{eq:Approximations_0d} because:
  \begin{enumerate}
    \setlength{\itemsep}{0pt}
    \setlength{\parskip}{0pt}
    \item  we consistently observe E${\scriptstyle_{o}}$ $\ll$ $\delta$E; and
    \item  our calculations consistently show $\left\lvert \textbf{j}{\scriptstyle_{o}} \cdot \textbf{E}{\scriptstyle_{o}} \right\rvert$ $\ll$ $\left\lvert \textbf{j}{\scriptstyle_{o}} \cdot \delta \textbf{E} \right\rvert$.
  \end{enumerate}
  %%  Argue:  neglect |∂J . Eo| and |∂J . ∂E| (3rd and 4th terms)
  \item  We argue that we can neglect the last $\left( \delta \textbf{j} \cdot \textbf{E}{\scriptstyle_{o}} \right)$ and $\left( \delta \textbf{j} \cdot \delta \textbf{E} \right)$ terms on the right-hand side of Equation \ref{eq:Approximations_0d} because:
  \begin{enumerate}
    \setlength{\itemsep}{0pt}
    \setlength{\parskip}{0pt}
    %%  Argue:  ∂E does not contribute to ∂J
    \item  we know that the largest components of $\delta$\textbf{E} are primarily composed of electrostatic waves, therefore $\delta$\textbf{E} should not contribute to $\delta$\textbf{j};
    %%  Argue:  ∂J arises from whistlers
    \item  therefore (and from our first argument regarding \textbf{j}${\scriptstyle_{o}}$), we conclude that $\delta$\textbf{j} must arise from the high frequency electromagnetic whistler mode waves (e.g., see Figure \ref{fig:examplewhistler});
    %%  whistler parameters
    \item  typical amplitudes for whistler mode waves observed by THEMIS are $\delta$B $\sim$ 0.1-1.0 nT and previous studies found wave numbers k$\rho{\scriptstyle_{ce}}$ $\sim$ 0.2-0.8 \citep[e.g.,][]{wilsoniii13a};
    %%  range of ∂J
    \item  thus, assuming $\rho{\scriptstyle_{ce}}$ $\sim$ 1 km and from these wave parameters we can estimate a range of values for $\delta$j $\sim$ k$\delta$B/$\mu{\scriptstyle_{o}}$ $\sim$ 0.02-0.6 $\mu$A m$^{-2}$;
    \item  one can see from comparing j${\scriptstyle_{o}}$ in Figure \ref{fig:exampleEBSdeltaj} with this range of $\delta$j that j${\scriptstyle_{o}}$ $\gg$ $\delta$j;
    %%  Argue:  |∂J . Eo| << |Jo . ∂E|
    \item  therefore, because j${\scriptstyle_{o}}$ $\gg$ $\delta$j and E${\scriptstyle_{o}}$ $\ll$ $\delta$E, we argue that $\left\lvert \delta \textbf{j} \cdot \textbf{E}{\scriptstyle_{o}} \right\rvert$ $\ll$ $\left\lvert \textbf{j}{\scriptstyle_{o}} \cdot \delta \textbf{E} \right\rvert$; and
    %%  Argue:  |∂J . ∂E| << |Jo . ∂E|
    \item  because j${\scriptstyle_{o}}$ $\gg$ $\delta$j and $\delta$E from the whistlers is much smaller than from the electrostatic waves, we argue that $\left\lvert \delta \textbf{j} \cdot \delta \textbf{E} \right\rvert$ $\ll$ $\left\lvert \textbf{j}{\scriptstyle_{o}} \cdot \delta \textbf{E} \right\rvert$.
  \end{enumerate}
\end{enumerate}
\indent  These approximations allow to reduce Equation \ref{eq:Approximations_0d} to $\left\lvert \textbf{j} \cdot \textbf{E} \right\rvert$ $\sim$ $\left\lvert \textbf{j}{\scriptstyle_{o}} \cdot \delta \textbf{E} \right\rvert$, which validates our assumptions in the previous appendix.  Therefore, the contribution to the rate of energy dissipation per unit volume from the electric fields is dominated by the high frequency electrostatic waves that we focus on in this study.

\indent  Since \textbf{j}${\scriptstyle_{o}}$ varies much more slowly than $\delta$\textbf{E}, we can assume \textbf{j}${\scriptstyle_{o}}$ $\sim$ constant over the typical wave periods observed in $\delta$\textbf{E}.  Therefore, we can upsample \textbf{j}${\scriptstyle_{o}}$ to the $\delta$\textbf{E} time steps without significantly compromising our results.

%%----------------------------------------------------------------------------------------
%%  Appendix:  Microscopic Energy Dissipation
%%----------------------------------------------------------------------------------------
\section{Microscopic Energy Dissipation} \label{app:EnergyDissipation}

%%----------------------------------------------------------------------------------------
%%  Subsection:  Introduction and Theory
%%----------------------------------------------------------------------------------------
\subsection{Introduction and Theory}  \label{subapp:IntroductionTheory}

\indent  In this appendix, we analytically define our wave dissipation estimates and how we compare them to our macroscopic energy dissipation estimates.  We provide two different methods for estimating the wave dissipation rate and then discuss the advantages and disadvantages for each.  We estimate the wave dissipation rate in two ways to test the validity of the assumptions required to calculate each.

%%  *eta* calculation --> *eta* |j|^2
\indent  Classical transport theory for an ionized gas was first developed by \citet{spitzer53a}.  In this theory, Coulomb collisions between electrons and ions have an effect that is analogous to an effective friction or drag force resulting in heating.  However, the solar wind and terrestrial bow shock are so tenuous that Coulomb collisions are negligible.  To explain how a shock wave could form in a collisionless medium, \citet{vedenov63} derived a term from quasi-linear theory that is analogous to Boltzmann's collision operator.  Here electromagnetic waves create effective collisions between the wave fields and the particles.  These effective collisions act to reduce the relative drift between electrons and ions that give rise to currents, thus the effect was originally called an \emph{anomalous resistivity}.  If these wave-particle interactions can increase the random kinetic energy (i.e., heat) of the incident particle populations, then they could transform the incident bulk kinetic energy into random kinetic energy irreversibly.  The irreversible change is necessary to increase entropy and initiate the formation of a true shock wave.

\indent  Theory suggests that the most probable candidate in low Mach number quasi-perpendicular shocks is wave-particle interactions \citep[e.g.,][]{tidman71a, treumann09a}, which has been indirectly supported in recent observations \citep[e.g.,][]{wilsoniii07a, wilsoniii10a, wilsoniii12c}.  Previous observations have shown that the ramp region of collisionless shocks are dominated by large amplitude ion-acoustic waves (IAWs) \citep[e.g.,][]{wilsoniii07a}.  \citet[][]{muschietti13a} found that at late times in their simulation, the electron cyclotron drift instability (ECDI) and IAWs had very similar power spectrums (ignoring the peaks due to the Bernstein modes in the ECDI).  The only differences were in the wave polarization and their respective effects on the particle distributions.  The IAWs in their simulation began to form electron phase space holes at later times as well.  In addition, the large currents in the shock ramp, responsible for the jump in B${\scriptstyle_{o}}$, are thought to be the main source of free energy for these waves.  

Therefore, we will use the IAW solution for the effective collision frequency and corresponding resistivity, given by:
\begin{subequations}
  \begin{align}
    \nu{\scriptstyle_{iaw}} & = \omega{\scriptstyle_{pe}} \frac{\varepsilon{\scriptstyle_{o}} \mid \delta E \mid^{2}}{2 n{\scriptstyle_{e}} k{\scriptstyle_{B}} T{\scriptstyle_{e}}}  \label{eq:wavepart_2a}  \\
    \eta{\scriptstyle_{iaw}} & = \frac{ \nu{\scriptstyle_{iaw}} }{\varepsilon{\scriptstyle_{o}} \omega{\scriptstyle_{pe}}^{2}}  \label{eq:wavepart_2b}
  \end{align}
\end{subequations}
where $\left\lvert \delta \textbf{E} \right\rvert$ is the fluctuating electric field amplitude of the wave, and $\nu{\scriptstyle_{iaw}}$ is the effective collision frequency.  Here, $\eta{\scriptstyle_{iaw}}$ represents an effective resistivity, but phenomenologically it arises from an effective drag force caused by $\left\lvert \delta \textbf{E} \right\rvert$ that acts to locally reduce the relative drift between electrons and ions that produces \textbf{j}${\scriptstyle_{o}}$.  Meaning, these waves act to reduce the source of free energy -- the relative drift -- that drives them unstable.

%%  *eta*  =  lower bound
\indent  Note that Equation \ref{eq:wavepart_2a} was derived under the assumption of a weakly turbulent plasma, which is not, in general, accurate.  Recent wave observations \citep{wilsoniii07a, wilsoniii10a, wilsoniii12c} show that wave amplitudes can be very nonlinear, and thus the above equations may underestimate the effects of wave-particle interactions.  In additions, recent Vlasov simulations using realistic mass ratios \citep{petkaki06a, petkaki08a, yoon06a, yoon07a} have observed momentum transfers that are 2-3 orders of magnitude larger than predicted by Equation \ref{eq:wavepart_2b}.  Therefore, we will use our estimates of $\nu{\scriptstyle_{iaw}}$ as a lower bound for the effective resistivity due to wave-particle interactions.

%%  Poynting's Theorem
\indent  Recall from Poynting's theorem that \textbf{\textcolor{Red}{the time rate of change of the energy density of the electromagnetic fields}}\footnote{This can also be treated as the rate of energy transfer per unit volume} $+$ \textbf{\textcolor{Blue}{the rate of electromagnetic energy flux flowing out of a surface}}\footnote{This can also be treated as the power flowing out of a volume through a defined surface} $=$ \textbf{\textcolor{teal}{the energy lost due to momentum transfer between particles and fields}}\footnote{This can also be treated as the rate of work done per unit volume on the charges in the volume element}.  In other words:
\begin{equation}
  \label{eq:quantenergydiss_0}
  \textbf{\textcolor{Red}{ $\partial{\scriptstyle_{t}} \left( \mathcal{W}{\scriptstyle_{B}} + \mathcal{W}{\scriptstyle_{E}} \right)$ }} + \textbf{\textcolor{Blue}{ $\nabla \cdot \delta \textbf{S}$ }} = \textbf{\textcolor{teal}{ $- \textbf{j}{\scriptstyle_{o}} \cdot \delta \textbf{E}$ }}
\end{equation}
If we assume $\delta$\textbf{E} $\approx$ $\left( \overleftrightarrow{\mathbb{\eta}}{\scriptstyle_{_{iaw}}} \cdot \thickspace \textbf{j}{\scriptstyle_{o}} \right)$, then the energy lost can be expressed as:
\begin{subequations}
  \begin{align}
    - \textbf{j}{\scriptstyle_{o}} \cdot \delta \textbf{E} & = - \textbf{j}{\scriptstyle_{o}} \cdot \overleftrightarrow{\mathbb{\eta}}{\scriptstyle_{_{iaw}}} \cdot \textbf{j}{\scriptstyle_{o}}  \label{eq:quantenergydiss_1a}  \\
    & \approx - \left( \eta{\scriptstyle_{\perp}} \mid \textbf{j}{\scriptstyle_{o \perp}} \mid^{2} + \eta{\scriptstyle_{\parallel}} \mid \textbf{j}{\scriptstyle_{o \parallel}} \mid^{2} \right)  \label{eq:currentdensest_1b}  \\
    & \approx - \eta{\scriptstyle_{iaw}} \mid \textbf{j}{\scriptstyle_{o}} \mid^{2}  \label{eq:quantenergydiss_1c}
  \end{align}
\end{subequations}
where $\eta{\scriptstyle_{iaw}}$ is the effective resistivity given by Equation \ref{eq:wavepart_2b}.  Note that $\delta$\textbf{E} and $\delta$\textbf{B} were transformed into the appropriate reference frame prior to the calculation of $\delta$\textbf{S} or either dissipation rate.

%%
%%  pros and cons for each wave dissipation rate estimate...
%%
\indent  We use two methods to estimate the energy dissipation rate due to wave-particle interactions because each has their own advantages and disadvantages.  The assumptions are slightly different for each estimate and therefore, the resulting uncertainties are different.  Some of the disadvantages of using $\eta{\scriptstyle_{iaw}} \left\lvert \textbf{j}{\scriptstyle_{o}} \right\rvert^{2}$ include, but are not limited to:  (1) the calculation of $\nu{\scriptstyle_{iaw}}$ relies upon our assumption of a dispersion relation; (2) this estimate assumes \textbf{E} is parallel to \textbf{j}${\scriptstyle_{o}}$; (3) this estimate assumes \textbf{E} $\approx$ $\overleftrightarrow{\mathbb{\eta}}{\scriptstyle_{_{iaw}}} \cdot \textbf{j}{\scriptstyle_{o}}$; and (4) the calculation of $\nu{\scriptstyle_{iaw}}$ assumes that the fluctuations are quasi-linear which can result in underestimates for the true momentum exchange rates \citep[e.g.,][]{petkaki08a}.  Some of the disadvantages of using $\left\lvert \textbf{j}{\scriptstyle_{o}} \cdot \delta \textbf{E} \right\rvert$ include, but are not limited to:  (1) more reliant upon the accuracy of coordinate basis rotations (e.g., from GSE to NCB); (2) only two components of \textbf{j}${\scriptstyle_{o}}$ can be estimated which may have more of an impact on $\left\lvert \textbf{j}{\scriptstyle_{o}} \cdot \delta \textbf{E} \right\rvert$ than $\eta{\scriptstyle_{iaw}} \left\lvert \textbf{j}{\scriptstyle_{o}} \right\rvert^{2}$; and (3) relies upon the accuracy of the \textbf{E}/$\left\lvert \delta \textbf{E} \right\rvert$.  Correspondingly, each assumption has their respective advantages.  Therefore, we use both methods as a way to test the validity of each.

%%
%%  ratio estimates
%%
\indent  Let us consider Equation \ref{eq:thermodynamics_8a} and assume $\Delta \mathfrak{w}{\scriptstyle_{ext}}$/$\Delta t$ $\rightarrow$ $\left( - \textbf{j}{\scriptstyle_{o}} \cdot \delta \textbf{E} \right)$ $\approx$ $\eta{\scriptstyle_{iaw}} \left\lvert \textbf{j}{\scriptstyle_{o}} \right\rvert^{2}$.  The physical reason for this is that we are assuming that wave energy dissipation is causing the changes in the specific enthalpy per unit time per unit volume ($\Delta \mathfrak{h}$/$\Delta t$) through ohmic dissipation.  Therefore, to quantify the relative impact of the wave energy dissipation rate, we define the following unitless ratios:
\begin{subequations}
  \begin{align}
    \mathcal{R}{\scriptstyle_{\Lambda[\Psi]}} & \equiv \frac{ \left\lvert \textbf{j}{\scriptstyle_{o}} \cdot \delta \textbf{E} \right\rvert }{ \dot{\Lambda} } \left[ \equiv \frac{ \left\lvert \textbf{j}{\scriptstyle_{o}} \cdot \delta \textbf{E} \right\rvert }{ \dot{\Psi} } \right]  \label{eq:quantenergydiss_3a}  \\
    \mathcal{Y}{\scriptstyle_{\Lambda[\Psi]}} & \equiv \frac{ \eta{\scriptstyle_{iaw}} \left\lvert \textbf{j}{\scriptstyle_{o}} \right\rvert^{2} }{ \dot{\Lambda} } \left[ \equiv \frac{ \eta{\scriptstyle_{iaw}} \left\lvert \textbf{j}{\scriptstyle_{o}} \right\rvert^{2} }{ \dot{\Psi} } \right]  \label{eq:quantenergydiss_3b}  \\
    \intertext{where we have used:}
    \dot{\Lambda} & \equiv \dot{\Psi} + C{\scriptstyle_{s}}^{2} \frac{ \Delta \rho }{ \Delta t }  \label{eq:quantenergydiss_3c}  \\
    \dot{\Psi} & \equiv \left( \rho T \right) \frac{ \Delta \mathfrak{s} }{ \Delta t } \text{  .}  \label{eq:quantenergydiss_3d}
  \end{align}
\end{subequations}

%%  physical explanation 1
\indent  We will use these ratios to quantify the relative amount of microscopic electromagnetic energy dissipation to the macroscopic fluid energy dissipation rates.  Note that $\mathcal{R}{\scriptstyle_{\Lambda}}$ and $\mathcal{Y}{\scriptstyle_{\Lambda}}$ include adiabatic compression, which is a reversible process.  Simple MHD theory suggests that a increase(decrease) in Poynting flux should result in a decrease(increase) in kinetic energy flux, which we absorbed into $\Delta \mathfrak{h}$.  The addition of a resistive loss term can cause the change in Poynting flux to result in changes in kinetic energy flux and enthalpy flux \citep[e.g.,][]{birn08a}.  Thus, there is nothing physically inconsistent with $\Delta \mathfrak{h}$/$\Delta t$ $\neq$ 0.  Note that if $\mathcal{R}{\scriptstyle_{\Psi}}$ $=$ $\mathcal{Y}{\scriptstyle_{\Psi}}$, then this implies that $\eta{\scriptstyle_{iaw}} \left\lvert \textbf{j}{\scriptstyle_{o}} \right\rvert^{2}$ $=$ $\left\lvert \textbf{j}{\scriptstyle_{o}} \cdot \delta \textbf{E} \right\rvert$.

%%  physical explanation 2
\indent  If either $\mathcal{R}{\scriptstyle_{\Lambda}}$ or $\mathcal{Y}{\scriptstyle_{\Lambda}}$ are $>$ 1 and $\left( - \textbf{j}{\scriptstyle_{o}} \cdot \delta \textbf{E} \right)$ is positive(negative), then $\Delta \mathfrak{h}$/$\Delta t$ $>$($<$) 0.  It is also possible that there are extra sink(source) terms we have not included in our approximation of $\Delta \mathfrak{w}{\scriptstyle_{ext}}$/$\Delta t$.  However, ratios satisfying $\mathcal{R}{\scriptstyle_{\Lambda[\Psi]}}$ $>$ 1 or $\mathcal{Y}{\scriptstyle_{\Lambda[\Psi]}}$ $>$ 1 correspond to data points where the microscopic electromagnetic energy dissipation exceeds the amount of macroscopic fluid energy dissipation necessary to explain the changes in enthalpy[entropy] across the shock ramp.  Therefore, we will focus on these ratios to show that high frequency electromagnetic waves play a significant role in the global energy budget of low Mach number collisionless shocks.

%%++++++++++++++++++++++++++++++++++++++++++++++++++++++++++++++++++++++++++++++++++++++++
%% Image:  Example BSC with cumulative sum of (-j.∂E)
%%++++++++++++++++++++++++++++++++++++++++++++++++++++++++++++++++++++++++++++++++++++++++
\begin{figure}[!htb]
  \centering
    {\includegraphics[trim = 0mm 0mm 0mm 0mm, clip, width=12cm]
    {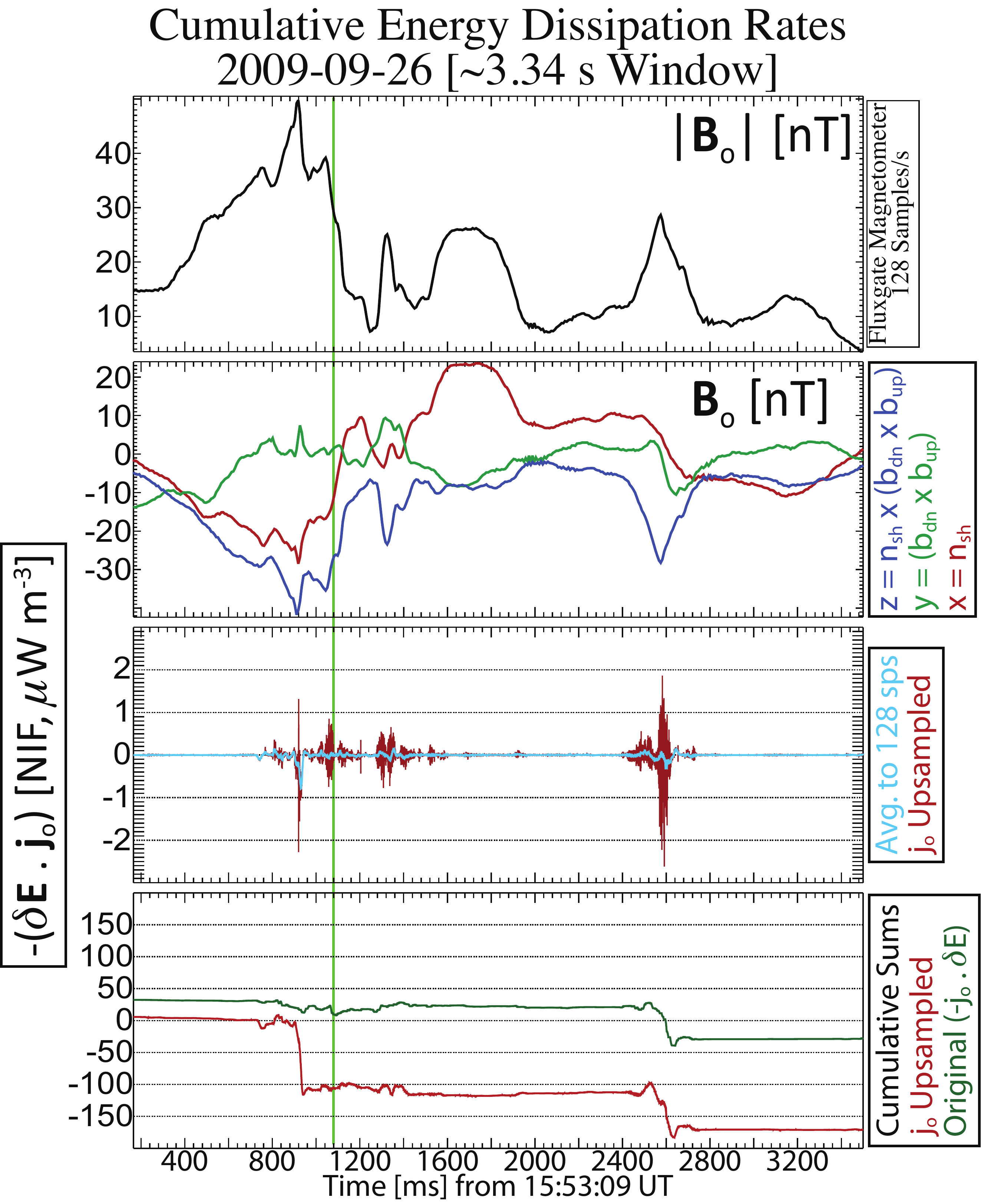}}
    \captionsetup{width=10cm}   %%  adjust caption width
    \caption[Example BSC with cumulative sum of dissipation]{This example shows the same time range as Figure \ref{fig:exampleEBSdeltaj} and the top two panels are the same.  The third panel shows $\left(- \textbf{j}{\scriptstyle_{o}} \cdot \delta \textbf{E} \right)$ at the $\delta$\textbf{E} time steps (red) with corresponding trend line (cyan).  The fourth panel shows the cumulative sum of $\left(- \textbf{j}{\scriptstyle_{o}} \cdot \delta \textbf{E} \right)$ at the $\delta$\textbf{E} time steps (red) and at \textbf{j}${\scriptstyle_{o}}$ time steps (green).}
    \label{fig:exampleEBSCumSumjdE}
\end{figure}
%%++++++++++++++++++++++++++++++++++++++++++++++++++++++++++++++++++++++++++++++++++++++++
%% Image:  Example BSC with cumulative sum of (-j.∂E)
%%++++++++++++++++++++++++++++++++++++++++++++++++++++++++++++++++++++++++++++++++++++++++

%%----------------------------------------------------------------------------------------
%%  Subsection:  Asymmetric Fluctuations
%%----------------------------------------------------------------------------------------
\subsection{Asymmetric Fluctuations}  \label{subapp:AsymmetricFluctuations}

\indent  In this appendix we present an illustrative example that shows there is a net change in $\left(- \textbf{j}{\scriptstyle_{o}} \cdot \delta \textbf{E} \right)$ across the high frequency waves of interest.  Were this not the case, then the waves would not be able to impart momentum/energy to the particles, thus they could not produce a net work on the unit volume.

%%  Figure ∑ (-Jo.∂E):  Introduce
\indent  Figure \ref{fig:exampleEBSCumSumjdE} shows the same time range as Figures \ref{fig:exampleEBSjEdjBSC} and \ref{fig:exampleEBSdeltaj}.  The top two panels are the magnitude of \textbf{B}${\scriptstyle_{o}}$ and its NCB components.  The third panel is the same as the last panel of Figure \ref{fig:exampleEBSjEdjBSC}.  The fourth panel shows the cumulative sum of $\left(- \textbf{j}{\scriptstyle_{o}} \cdot \delta \textbf{E} \right)$ at two different time steps.  The green line was calculated by downsampling $\delta$\textbf{E} to the original \textbf{B}${\scriptstyle_{o}}$ time stamps before calculating $\left(- \textbf{j}{\scriptstyle_{o}} \cdot \delta \textbf{E} \right)$ and then summing.  The red line shows the result when \textbf{j}${\scriptstyle_{o}}$ is upsampled to the $\delta$\textbf{E} time steps before calculating $\left(- \textbf{j}{\scriptstyle_{o}} \cdot \delta \textbf{E} \right)$ (corresponding to red line in third panel).  

%%  Figure ∑ (-Jo.∂E):  Discuss
\indent  The main point of this figure is to show that the high frequency waves that we focus on in this study can produce a net change in $\left(- \textbf{j}{\scriptstyle_{o}} \cdot \delta \textbf{E} \right)$.  This means that the waves have lost electromagnetic energy to the particles, thus dissipating energy.  Figure \ref{fig:exampleEBSCumSumjdE} shows that our main results presented in Figure \ref{fig:comparedissrate} are reliable.  Meaning, the time-average of $\left(- \textbf{j}{\scriptstyle_{o}} \cdot \delta \textbf{E} \right)$ across the waves is not zero so that we have not neglected the restoring effects by using $\left\lvert \textbf{j}{\scriptstyle_{o}} \cdot \delta \textbf{E} \right\rvert$.

\indent  In addition, we examined the cumulative sum of $\delta$\textbf{S} to determine if the waves carried a net energy flux.  The results (not shown) are similar to those found for $\left(- \textbf{j}{\scriptstyle_{o}} \cdot \delta \textbf{E} \right)$, namely, that the waves showed a net change in $\delta$\textbf{S}.  Were these modes pure sinusoidal circularly polarized electromagnetic oscillations, the cumulative sum of $\delta$\textbf{S} would be zero.

\clearpage
%%----------------------------------------------------------------------------------------
%%  Appendix:  Removal of Secondary Ions
%%----------------------------------------------------------------------------------------
\section{Removal of Secondary Ions} \label{app:velocitymomentcorr}
%%  Introduce issue
\indent  Onboard particle distribution moments (or moments calculated from telemetered particle distributions) can suffer from inaccuracies due to spacecraft charging \citep[e.g.,][]{genot04a, geach05a, davis08a}, multiple species \citep[e.g.,][]{paschmann98a}, multiple components \citep[e.g.,][]{wuest07b}, and limited energy ranges (e.g., V${\scriptstyle_{Ti}}$ $\gtrsim$ V${\scriptstyle_{bulk}}$).  Below we discuss how we accounted for these potential inaccuracies when examining the velocity distribution function moments.

%%++++++++++++++++++++++++++++++++++++++++++++++++++++++++++++++++++++++++++++++++++++++++
%% Image:  Raw Moments and Smoothed Field
%%++++++++++++++++++++++++++++++++++++++++++++++++++++++++++++++++++++++++++++++++++++++++
\begin{wrapfigure}{r}{0.425\textwidth}
  \vspace{-15pt}
  \centering
    {\includegraphics[trim = 0mm 0mm 0mm 0mm, clip, width=0.435\textwidth]
    {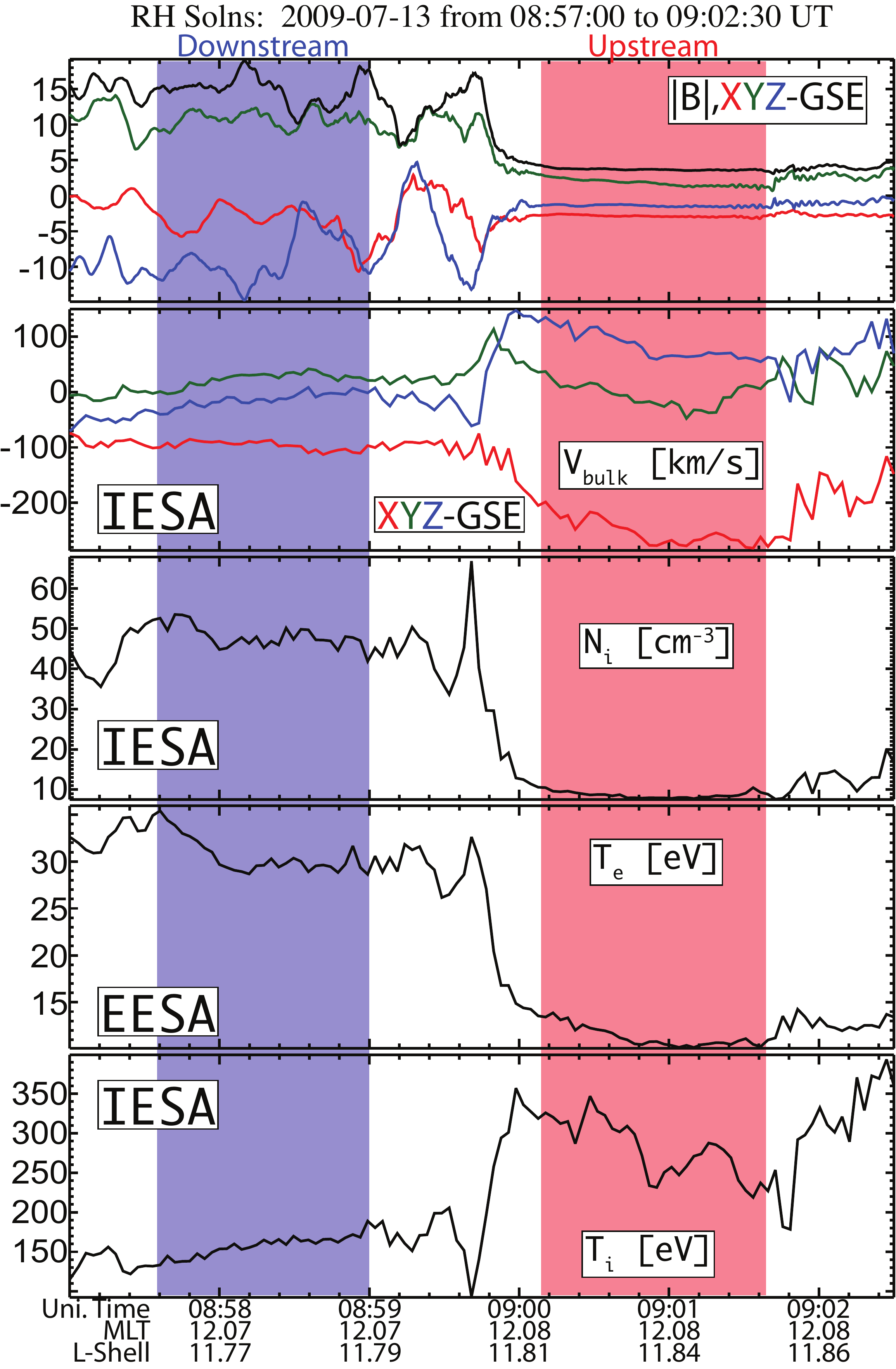}}
    %%  Define new caption setting options
    \captionsetup{width=0.395\textwidth,font=footnotesize,labelfont=bf}
    %%  Define caption
   \caption[Raw Moments and Smoothed Field]{The figure shows an example bow shock crossing with the smoothed \textbf{B}${\scriptstyle_{o}}$ and B${\scriptstyle_{o}}$ (nT, first panel), \textbf{V}${\scriptstyle_{bulk}}$ (km/s, second panel), N${\scriptstyle_{i}}$ (cm$^{-3}$, third panel), T${\scriptstyle_{e}}$ (eV, fourth panel), and T${\scriptstyle_{i}}$ (eV, fifth panel).  The region shaded in blue defines the downstream (i.e., magnetosheath) and red the upstream (i.e., solar wind).  The data was measured by THEMIS-B.}
    \label{fig:rawmoments}
  \vspace{-10pt}
\end{wrapfigure}
%%++++++++++++++++++++++++++++++++++++++++++++++++++++++++++++++++++++++++++++++++++++++++
%% Image:  Raw Moments and Smoothed Field
%%++++++++++++++++++++++++++++++++++++++++++++++++++++++++++++++++++++++++++++++++++++++++
\indent  Figure \ref{fig:rawmoments} shows an outbound bow shock crossing observed by THEMIS-B on 2009-07-13.  The shock ramp center was $\sim$08:59:46 UT and THEMIS-B was at a GSE position of $\sim$$<$$+$11.5, $+$1.0, -2.2$>$ R${\scriptstyle_{E}}$.  Shock waves, by definition, result in an increase in N${\scriptstyle_{i}}$, B${\scriptstyle_{o}}$, T${\scriptstyle_{e}}$, and T${\scriptstyle_{i}}$ on the downstream(shocked) side of the transition region.  The increase in temperature allows V${\scriptstyle_{bulk}}$ to decrease to a subsonic value.   In Figure \ref{fig:rawmoments} we observed an increase in N${\scriptstyle_{i}}$, B${\scriptstyle_{o}}$, and T${\scriptstyle_{e}}$ with corresponding decrease in V${\scriptstyle_{bulk}}$, suggesting this was indeed a shock crossing.  However, one can immediately see that the ions appear to be hotter (5th panel of Figure \ref{fig:rawmoments}) on the upstream(unshocked) rather than the downstream(shocked) side.  Therefore, the decrease in T${\scriptstyle_{i}}$ across the transition region coupled with our knowledge of potential secondary ion contamination (e.g., shock-reflected ions) led us to examine the ion velocity moments in greater detail.

%%  Introduce DFs
\indent  It is well known that the terrestrial bow shock is capable of producing multiple populations of reflected ions \citep[e.g.,][]{bonifazi81a, bonifazi81b, fuselier86a}.  Therefore, we initially assumed that these secondary ions were responsible for some fraction of the error in the ion velocity moments.  To test whether ion beams were contaminating the upstream ion velocity moments, we examined the entire ion velocity distribution functions in three different reference frames and three different projections.  To do this, we converted our observations to phase(velocity) space densities, translated the data into the new reference frame, rotated the data into physically significant coordinate basis (discussed below), and projected the triangulated results onto the three planes comprising this new coordinate basis.

\begin{figure}[!htb]
  \vspace{-5pt}
  \centering
  {\includegraphics[trim = 0mm 0mm 0mm 0mm, clip, width=0.85\textwidth]
  {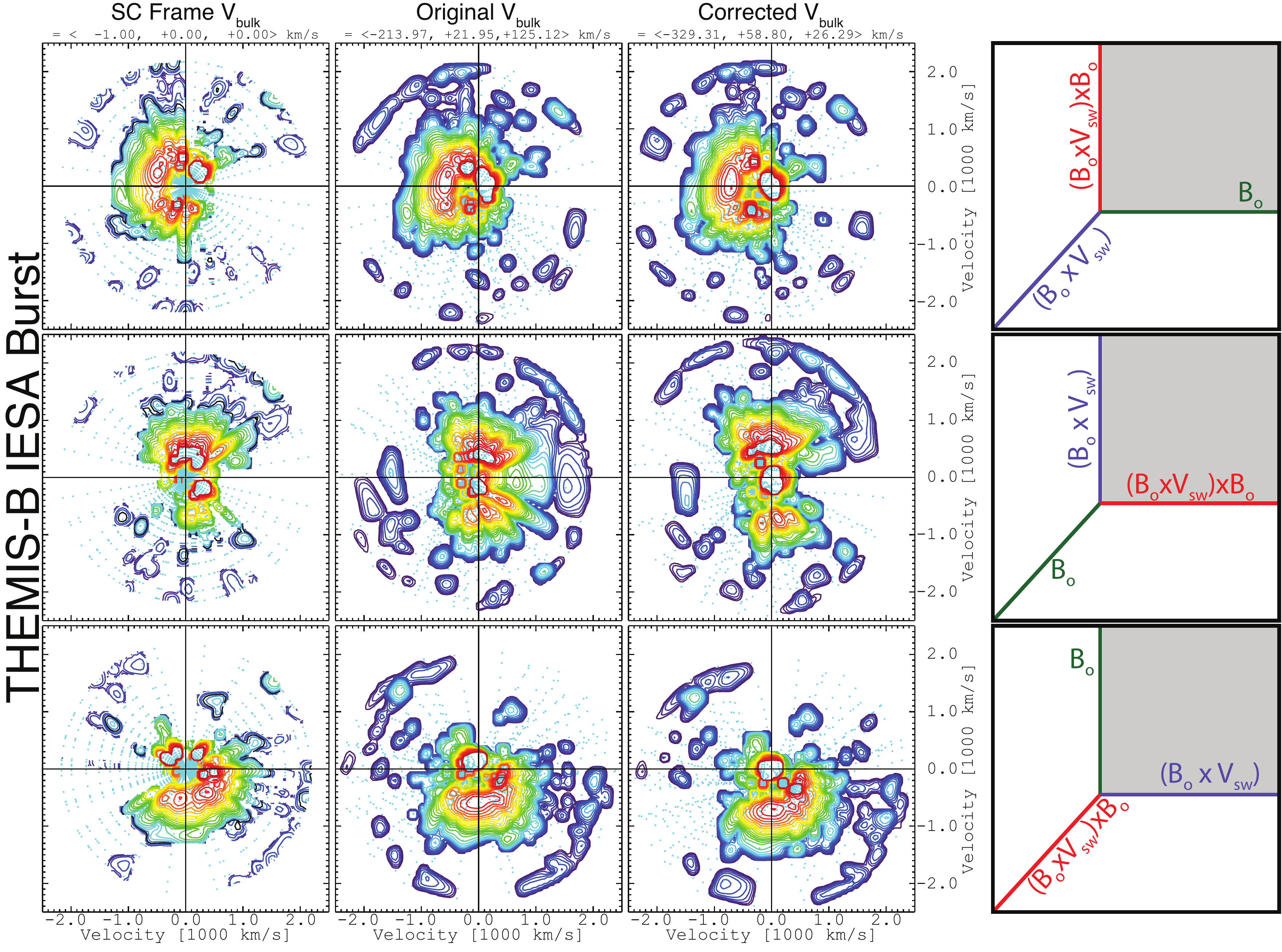}}
  %%  Define new caption setting options
  \captionsetup{width=0.85\textwidth,font=footnotesize,labelfont=bf}
  \caption[SC Frame vs. Original vs. Corrected DF in 3 Planes]{An example THEMIS IESA Burst particle velocity distribution observed upstream of a bow shock crossing.  The three columns of contour plots correspond to three different rest frames defined by \textbf{V}${\scriptstyle_{bulk}}$ at the top of each column.  Each contour plot shows contours of constant phase space density (uniformly scaled from 1$\times$10$^{-14}$ to 1$\times$10$^{-8}$ s$^{3}$cm$^{-3}$km$^{-3}$, where red is high) versus velocity projected onto three different planes defined by the shaded region in the coordinate axes shown in right-hand column.  The velocity axes range from $\pm$2500 km/s and the crosshairs show the location of the origin.}
  \label{fig:exampledf3planes}
  \vspace{-15pt}
\end{figure}

%%  Introduce Coordinates and Reference Frames
\indent  Figure \ref{fig:exampledf3planes} shows contours of constant phase space density projected onto three planes (rows) and in three reference frames (first three columns).  The first reference frame (first column) shown is the spacecraft frame.  The second reference frame (second column) was defined by the level-2 average bulk flow velocity, \textbf{V}${\scriptstyle_{bulk}}$.  The third reference frame (third column) shows the distribution plotted in the ion core bulk flow rest frame.  The physically significant coordinate basis we used was defined by the value used for \textbf{V}${\scriptstyle_{bulk}}$ and the observed quasi-static magnetic field, \textbf{B}${\scriptstyle_{o}}$ (shaded planes in fourth column).  These distributions do not assume gyrotropy.

%%  Introduce Vbulk correction
\indent  This adjustment to \textbf{V}${\scriptstyle_{bulk}}$ that resulted in the third column was determined by plotting the distributions in a manner similar to that shown in Figure \ref{fig:exampledf3planes} and then iteratively changing our estimate for \textbf{V}${\scriptstyle_{bulk}}$ until the core of the distribution was centered on the origin in all three planes.  We made this adjustment by starting in the second reference frame (second column) and adjusting \textbf{V}${\scriptstyle_{bulk}}$ until the peak of the distribution rested at the origin (third column of Figure \ref{fig:exampledf3planes}).  Then we found the difference between this reference frame and the level-2 estimate for \textbf{V}${\scriptstyle_{bulk}}$ to get our adjustment.  It is important to note that when plotting particle velocity distributions in this manner that the reference frame and coordinate basis can strongly influence the results, as shown in Figure \ref{fig:exampledf3planes}.  For instance, in the spacecraft frame of reference (first column), the secondary population is not obviously a field-aligned beam.  However, in the core bulk flow rest frame (third column), the signature of a strong field-aligned beam is obvious.  We observed this beam for several minutes upstream of this shock ramp.

%%++++++++++++++++++++++++++++++++++++++++++++++++++++++++++++++++++++++++++++++++++++++++
%% Image:  Original vs. Corrected DF in 3 Planes
%%++++++++++++++++++++++++++++++++++++++++++++++++++++++++++++++++++++++++++++++++++++++++
%%++++++++++++++++++++++++++++++++++++++++++++++++++++++++++++++++++++++++++++++++++++++++
%% Image:  Original vs. Corrected DF in 3 Planes
%%++++++++++++++++++++++++++++++++++++++++++++++++++++++++++++++++++++++++++++++++++++++++
\indent    Some previous studies have plotted foreshock ion velocity distributions in the spacecraft frame, however Figure \ref{fig:exampledf3planes} provides an example of why this can lead to confusion or a misinterpretation of the results.  Note that the plane of projection in Figure \ref{fig:exampledf3planes} is dependent upon the definition of \textbf{V}${\scriptstyle_{bulk}}$.  If the estimate for \textbf{V}${\scriptstyle_{bulk}}$ or the plane of projection is inaccurate, then the projected distribution can be misleading.  For instance, when comparing the results for the first plane (top row), one can see that the core of the second projection looks far more anisotropic than the core in the third projection.  If one examines the bottom row of distributions, one can see surprisingly different results between the first and third columns.  The apparent discrepancies are a consequence of the plotting routines, not a characteristic of the distribution.  For more examples of these types of plots, see \citet{wilsoniii09a, wilsoniii10a, wilsoniii12c}.

\indent  After adjusting \textbf{V}${\scriptstyle_{bulk}}$, we created a mask to eliminate all secondary ion populations.  We also removed ions within a small cone around the sun direction to reduce the effects of ``UV contamination.''  An example can be observed in the third distribution of the top row as the intense, narrow beam-like feature in the third quadrant near $\sim$500 km/s.  We only applied these masks to those distributions with well defined secondary ion populations.  After removing the secondary ions, we re-calculated the particle velocity moments.  Therefore, these re-calculated velocity moments will reflect only the core of the ion distribution function.

%%++++++++++++++++++++++++++++++++++++++++++++++++++++++++++++++++++++++++++++++++++++++++
%% Image:  Original vs. Corrected Moments
%%++++++++++++++++++++++++++++++++++++++++++++++++++++++++++++++++++++++++++++++++++++++++
\begin{figure}[!htb]
  \centering
  {\includegraphics[trim = 0mm 0mm 0mm 0mm, clip, width=0.75\textwidth]
  {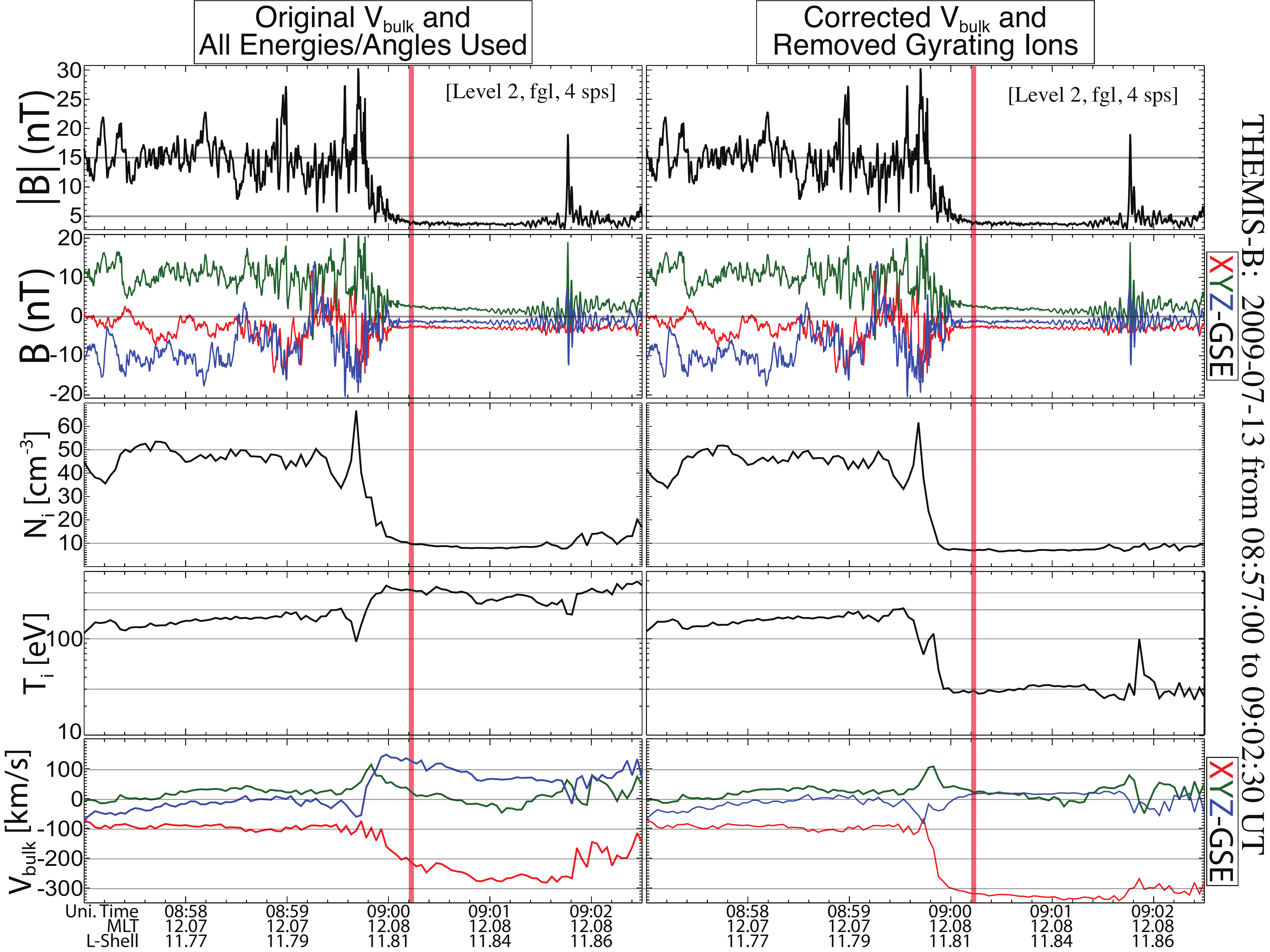}}
  %%  Define new caption setting options
  \captionsetup{width=0.70\textwidth,font=footnotesize,labelfont=bf}
  %%  Define caption
  \caption[Original vs. Corrected Moments]{Comparison between the level-2 moments (left-hand column) and the estimates for the core only (right-hand column).  The panels are as follows:  B${\scriptstyle_{o}}$ (nT, first panel), \textbf{B}${\scriptstyle_{o}}$ (nT, second panel), N${\scriptstyle_{i}}$ (cm$^{-3}$, third panel), T${\scriptstyle_{i}}$ (eV, fourth panel), and \textbf{V}${\scriptstyle_{bulk}}$ (km/s, fifth panel).  The magnetic fields are identical in each column.  The horizontal gray lines in each panel are to help the reader compare differences between the two results.}
  \label{fig:corrected}
\end{figure}
%%++++++++++++++++++++++++++++++++++++++++++++++++++++++++++++++++++++++++++++++++++++++++
%% Image:  Original vs. Corrected Moments
%%++++++++++++++++++++++++++++++++++++++++++++++++++++++++++++++++++++++++++++++++++++++++

\indent  We show an example comparison between the original level-2 velocity moments and the core only estimates in Figure \ref{fig:corrected}.  The time range shown is the same as in Figure \ref{fig:rawmoments}.  The red shaded region corresponds to the time range of the ion velocity distribution shown in Figure \ref{fig:exampledf3planes}.  The results for T${\scriptstyle_{i}}$ and \textbf{V}${\scriptstyle_{bulk}}$ show dramatic differences, whereas N${\scriptstyle_{i}}$ shows only minor differences in the upstream.  The most important observations are that T${\scriptstyle_{i}}$ now increases across the shock ramp and that $\mid$\textbf{V}${\scriptstyle_{bulk}}$$\mid$ shows differences of up to $\sim$30\%.  Not only did we observe large differences in $\mid$\textbf{V}${\scriptstyle_{bulk}}$$\mid$, one can see that there are significant changes in the flow direction as well.

%%  FABs and SLAMS
\indent  There is a large amplitude magnetic fluctuations observed near $\sim$09:02 UT, which causes a deflection of the core \textbf{V}${\scriptstyle_{bulk}}$.  We observed heating of the core ions and electrons near this structure as well.  A comparison between the N${\scriptstyle_{i}}$ plots in Figure \ref{fig:corrected} shows that the secondary ions were enhanced near this structure.  When we examined the ion distributions (e.g., see Figure \ref{fig:exampledf3planes}) we observed enhanced field-aligned and gyrophase-bunched ions near these fluctuations, consistent with recent observations \citep[][]{wilsoniii13b}.

%%  Discuss why N2/N1 > 4
\indent  Upon examination of the results shown in Table \ref{tab:shockparams}, one can see that a few events have N${\scriptstyle_{i2}}$/N${\scriptstyle_{i1}}$ $>$ 4.  There are many possible explanations for this, including but not limited to:  (1) N${\scriptstyle_{i1}}$ does not include reflected particles; (2) the spacecraft velocity had a significant component along the shock plane; (3) time-variations due to single spacecraft observations; (4) inaccuracies in velocity moments due to V${\scriptstyle_{Ti}}$ $\gtrsim$ V${\scriptstyle_{bulk}}$; (5) etc.  Note that events with N${\scriptstyle_{i2}}$/N${\scriptstyle_{i1}}$ $>$ 4 are only in violation of a time-stationary neutral fluid approximations.  If, for instance, the bow shock were being increasingly compressed as the spacecraft passed through the downstream region, one might expect N${\scriptstyle_{i2}}$/N${\scriptstyle_{i1}}$ $>$ 4.

\end{document}